\documentclass[apj]{emulateapj}
\usepackage{amsmath}
\pdfoutput=1
\usepackage{color}
\newcommand{\etal}{et~al.\/}
\newcommand{\ie}{i.e.\/}
\newcommand{\eg}{e.g.\/}
\newcommand{\etc}{etc.\/}
\newcommand{\kms}{km\,s$^{-1}$}
\newcommand{\msun}{M$_\odot$}
\newcommand{\lsun}{L$_\odot$}

\shorttitle{Star Forming Galaxy Thermometry}
\shortauthors{Mangum \etal}

\begin{document}
\title{Ammonia Thermometry of Star Forming Galaxies}

\author{Jeffrey G.~Mangum}
\affil{National Radio Astronomy Observatory, 520 Edgemont Road,
  Charlottesville, VA  22903, USA}
\email{jmangum@nrao.edu}

\author{Jeremy Darling}
\affil{Center for Astrophysics and Space Astronomy, Department of Astrophysical
and Planetary Sciences, Box 389, University of Colorado, Boulder, CO 
80309-0389, USA}
\email{jdarling@origins.colorado.edu}

\author{Christian Henkel\altaffilmark{1}}
\affil{Max-Planck-Institut f\"ur Radioastronomie, Auf dem H\"ugel
  69, 53121 Bonn, Germany}
\altaffiltext{1}{Also Astronomy Department, Faculty of Science, King Abdulaziz
  University, P.~O.~Box 80203, Jeddah, Saudi Arabia}
\email{chenkel@mpifr-bonn.mpg.de}

\author{Karl M.~Menten}
\affil{Max-Planck-Institut f\"ur Radioastronomie, Auf dem H\"ugel
  69, 53121 Bonn, Germany}
\email{kmenten@mpifr-bonn.mpg.de}

\author{Meredith MacGregor\altaffilmark{2}}
\affil{Harvard University, Department of Astronomy, 60
  Garden Street, Cambridge, MA 02138}
\altaffiltext{2}{Also National Radio Astronomy Observatory, 520
  Edgemont Road, Charlottesville, VA  22903, USA}
\email{mmacgreg@fas.harvard.edu}

\author{Brian E.~Svoboda\altaffilmark{2,3}}
\affil{Steward Observatory, University of Arizona, 933 North Cherry
  Avenue, Tucson, AZ  85721, USA}
\altaffiltext{3}{Also Western Washington University, Department of
  Physics and Astronomy, 516 High Street, Bellingham, WA 98225-9164}
\email{svobodb@email.arizona.edu}

\and

\author{Eva Schinnerer}
\affil{Max-Planck-Institut f\"ur Astronomie, K\"onigstuhl 17, 69117
  Heidelberg, Germany}
\email{schinner@mpia.de}

\begin{abstract}
With a goal toward deriving the physical conditions in external
galaxies, we present a study of the ammonia (NH$_3$) emission and
absorption in a sample of star forming systems.  
Using the unique sensitivities to kinetic temperature
afforded by the excitation characteristics of several inversion
transitions of NH$_3$, we have continued our characterization of
the dense gas in star forming galaxies by measuring the
kinetic temperature in a sample of 23 galaxies and one galaxy offset
position selected for their high infrared luminosity.  We derive
kinetic temperatures toward 13 galaxies, 9 of which possess multiple
kinetic temperature and/or velocity 
components.  Eight of these galaxies exhibit kinetic temperatures
$>100$\,K, which are in many cases at least a factor of two larger than
kinetic temperatures derived previously.  Furthermore, the derived
kinetic temperatures in our galaxy sample, which are in many cases at
least a factor of two larger than derived dust temperatures, point to
a problem with the common assumption that dust and gas kinetic
temperatures are equivalent.  As previously suggested, the use of dust
emission at wavelengths greater than 160\,$\mu$m to derive dust
temperatures, or dust heating from older stellar populations, may be
skewing derived dust temperatures in these galaxies to lower values.
We confirm the detection of high-excitation OH $^2\Pi_{3/2}$\,J=9/2
absorption toward Arp\,220 \citep{Ott2011}.  We also report the first
detections of non-metastable NH$_3$ inversion transitions toward
external galaxies in the (2,1) (NGC\,253, NGC\,660, IC\,342, and
IC\,860), (3,1), (3,2), (4,3), (5,4) (all in NGC\,660) and (10,9)
(Arp\,220) transitions.
\end{abstract}


\section{Introduction}
\label{intro}

The molecular mass in external galaxies can be well determined by
measuring the distribution of Carbon Monoxide (CO) emission
\citep[see][]{Young1991}.  These extragalactic CO 
measurements have yielded a detailed picture of the molecular mass in
many systems.  Interpretation of these CO measurements is
limited, though, as the emission from the CO molecule is generally
dominated by radiative transfer effects, including high 
optical depth.  CO is not a reliable monitor of the spatial density
and kinetic temperature in star formation regions.  Emission line
measurements from less-abundant molecules provide a more sensitive
diagnostic of the spatial density and kinetic temperature of the dense
gas in galaxies (for example, \cite{Henkel1991} for a review,
\cite{Gao2004a} (HCN), \cite{Nguyen1992} (HCO$^+$), \cite{Meier2005}
and \cite{Lindberg2011} (HC$_3$N), \cite{Mauersberger2003} (NH$_3$),
\cite{Mangum2008,Mangum2013} (H$_2$CO)).

Results from a survey of a sample of mainly nearby galaxies
\citep{Mangum2008,Mangum2013} have shown that Formaldehyde (H$_2$CO) is
a reliable and accurate density probe for extragalactic environments
\textit{where the kinetic temperature is known}.  The derivation of
n(H$_2$) in our sample of star forming galaxies relies upon assumed
kinetic temperatures.  The inversion transitions 
of NH$_3$ and the rotational transitions of H$_2$CO possess very
similar excitation conditions, thus likely trace similar dense gas
environments.  Using the unique 
sensitivities to kinetic temperature afforded by the excitation
characteristics of up to 14 different inversion transitions of
NH$_3$, we have continued our characterization of
the dense gas in star forming galaxies by measuring the
kinetic temperature in a sample of galaxies selected for their high
infrared luminosity.  In \S\ref{NH3Probe} we discuss the specific
properties of the NH$_3$ molecule which make it a good probe of
kinetic temperature. \S\ref{Observations} presents our observation
summary; \S\ref{Results} our NH$_3$ measurement 
results; \S\ref{Analysis} analyses of our NH$_3$ measurements,
employing Large Velocity Gradient (LVG) model fits.  In
\S\ref{Arp220OH} we discuss our detection of the OH 
$^2\Pi_{3/2}$\,J=9/2 transition toward Arp\,220.  In \S\ref{NGC660NH3}
we present an analysis of the 
remarkable NH$_3$ absorption spectra measured toward NGC\,660.  In
\S\ref{NH3Spatial} we discuss the spatial extent of the NH$_3$
emitting gas.  In \S\ref{HighTK} we study the connection between the
high kinetic temperatures derived and models of the heating processes in
star forming environments.  In \S\ref{DustandGas} we discuss the
disagreement between dust and gas kinetic temperatures.
\S\ref{Conclusions} presents our conclusions.

\section{Ammonia as a Kinetic Temperature Probe}
\label{NH3Probe}

Ammonia (NH$_3$) is a proven and unbiased tracer of the high density
regions within molecular clouds in a variety of galactic and
extragalactic environments
\citep[see][]{Walmsley1983,Mauersberger2003}.  Because NH$_3$ is a
symmetric top molecule (energy levels defined by quantum numbers (J,K)),
exchange of population between the K-ladders 
within a given symmetry state (ortho or para) occurs only via
collisional processes.  The relative intensity of these rotational
ground-state inversion transitions for different K-ladders then
provides a direct measurement of the kinetic temperature.

\begin{figure*}
\centering
\includegraphics[scale=0.55]{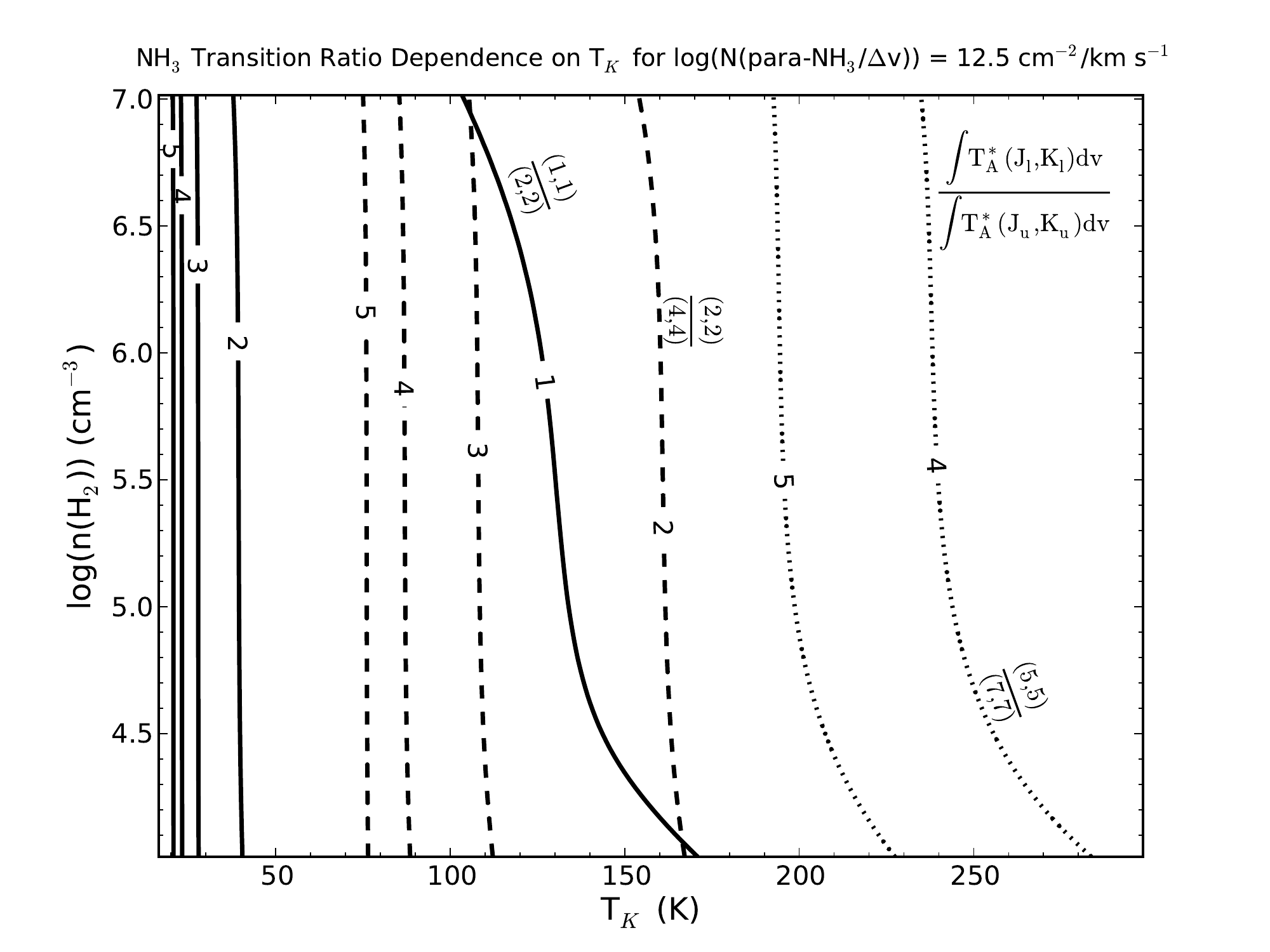}\\
\includegraphics[scale=0.55]{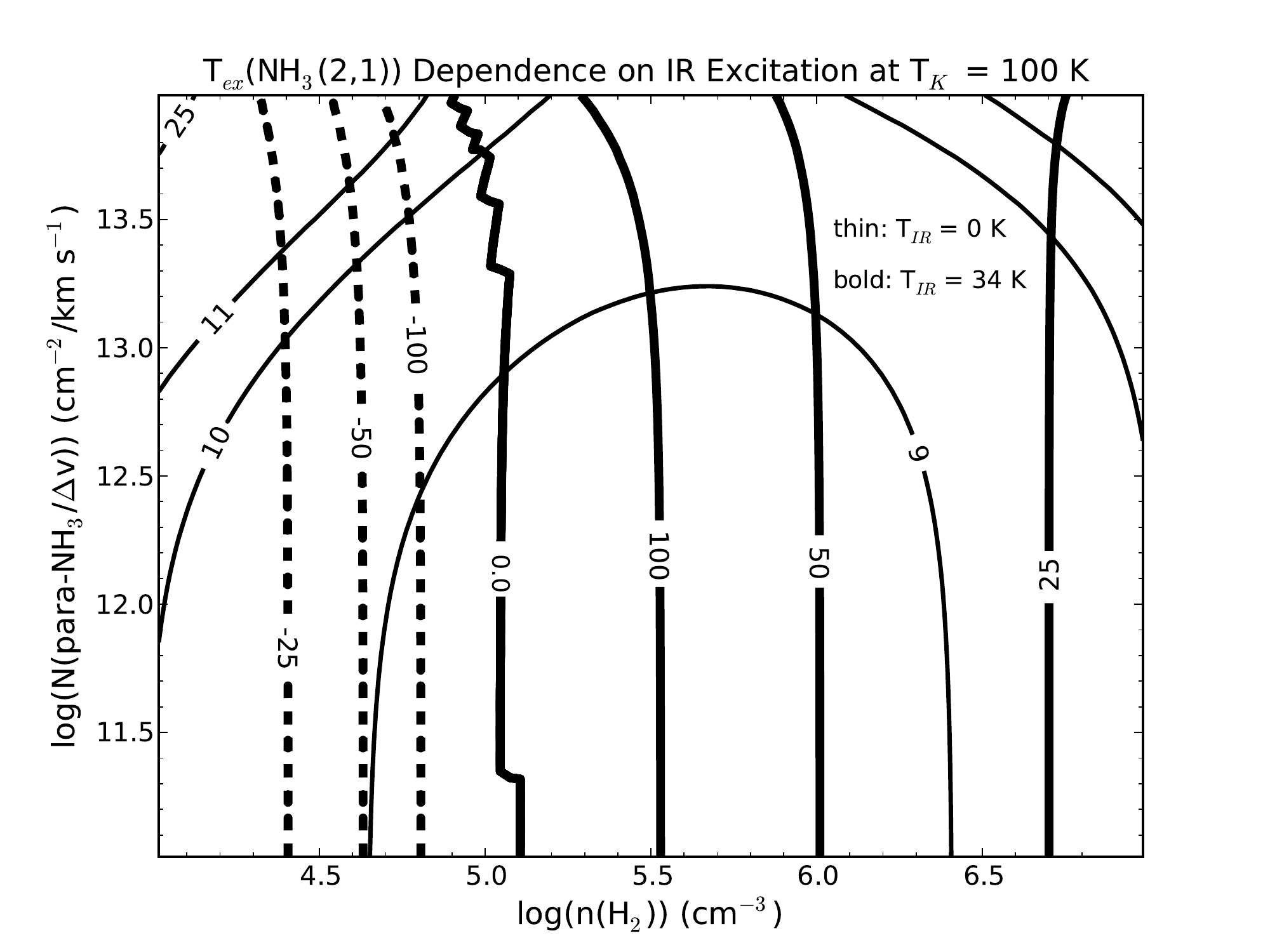}
\caption{\textit{Top:} Example LVG-modelled kinetic temperature
  sensitivity for the NH$_3$ (1,1)/(2,2) (solid contours), (2,2)/(4,4)
  (dashed contours), and (5,5)/(7,7) (dotted contours) transition
  ratios.  Note how the lower-excitation (1,1)/(2,2) ratio 
  is linear with kinetic temperature up to $\sim 40$ K, while the
  higher-excitation (2,2)/(4,4) and (5,5)/(7,7) ratios are relatively
  linear up to $\sim 150$ and $\sim 250$ K,
  respectively. \textit{Bottom:} Example LVG-modelled NH$_3$ (2,1) 
  excitation temperature as a function of spatial density and
  para-NH$_3$ column density per unit linewidth at T$_K = 100$\,K.
  Thin contours indicate T$_{ex}$ for no infrared radiation field,
  while bold contours indicate T$_{ex}$ when T$_{IR} = 34$\,K
  (appropriate for NGC\,253 and IC\,342; see \S\ref{Comparison}).  For
  n(H$_2$) $\lesssim 10^5$ the (2,1) transition is inverted
  (\ie\ masing) when subjected to a strong infrared radiation field.}
\label{fig:NH3Plots}
\end{figure*}

The inversion transitions of NH$_3$ at 23--27 GHz have been
used to measure the kinetic temperature in both cool (T$_K \simeq 20$
K) and warm (T$_K \simeq 300$ K) galactic and extragalactic star
formation environments.  For example, \cite{Hermsen1988} compared
measurements of 32 inversion transitions of NH$_3$ ranging
in energy from 23 to 1250 K ((J,K) = (1,1) to (10,8)) toward the
Orion-KL star formation region.  The wide span of NH$_3$ excitation
afforded by these measurements allowed for a very detailed
characterization of the kinetic temperature within the structural
components of the Orion-KL star formation environment.  A similar
study of the NH$_3$ (1,1) through (9,9) transitions toward NGC\,253,
IC\,342, and Maffei\,2 by \cite{Mauersberger2003} revealed warm (T$_K$
= 100--140 K) gas toward NGC\,253, IC\,342, and Maffei\,2 and cooler
gas (T$_K$ = 60 K) toward M\,82.  A demonstration of the kinetic
temperature sensitivity of the NH$_3$ transitions is shown in the
upper panel of Figure~\ref{fig:NH3Plots}.  In this demonstration, the
(1,1)/(2,2) 
line ratio monitors the lower kinetic temperatures ($\lesssim 40$ K),
while the (2,2)/(4,4) and (5,5)/(7,7) ratios monitor the higher
kinetic temperatures ($\lesssim 150$ K and 250 K, respectively).

In addition to the line-intensity comparative measurements of the
metastable (J = K) NH$_3$ inversion transitions which yield kinetic
temperature, non-metastable (J$\not=$K) inversion transitions can be
used to trace the infrared radiation environment in star formation
regions.  As 
originally noted by \cite{Morris1973}, for n(H$_2$) $\lesssim 10^7$
cm$^{-3}$ and no nearby source of infrared radiation the excitation
temperatures of the non-metastable transitions remain very low.  
This low excitation temperature is due to the rapid spontaneous
decay from the (J+1,J) states into the (J,J) states (A$_{ij}$ $\sim
0.013$\,s$^{-1}$ for the (2,1)$-$(1,1) rotational transition).  In the
presence of a source of infrared flux, \cite{Morris1973} showed that
the level populations of the (2,1) and (3,2) non-metastable inversion
transitions become inverted.  This behavior also holds true for the
other non-metastable NH$_3$ transitions.
This sensitivity to background infrared radiation makes
the (2,1) transition a sensitive probe of the radiation environment
within star formation regions.  Figure~\ref{fig:NH3Plots} (bottom)
shows the LVG-model predicted excitation 
temperature of the (2,1) transition within typical physical conditions
for our star forming galaxy sample (assuming no background continuum).
When an infrared radiation field is introduced the excitation
temperature becomes negative (\ie\ masing) for n(H$_2$) $\lesssim 10^5$
cm$^{-3}$, leading to emission in the (2,1) transition.

\section{Observations}
\label{Observations}

The measurements reported here were made using the National Radio
Astronomy Observatory (NRAO\footnote{The National Radio Astronomy
  Observatory is a facility of the National Science Foundation
  operated under cooperative agreement by Associated Universities,
  Inc.}) Green Bank Telescope (GBT) during thirty-four observing
sessions from 2009/03/18 through 2012/03/04.  Single-pointing
measurements were obtained of the NH$_3$ transitions listed in
frequency setup ``A50'' of Table~\ref{tab:frequencies} toward a sample
of 23 infrared-luminous and/or star forming galaxies
(Table~\ref{tab:galaxies}) which exhibit H$_2$CO emission or absorption
\citep{Mangum2008,Mangum2013}.  Additional measurements of the
generally higher-excitation and ortho-NH$_3$ transitions listed in
the other frequency setups described in Table~\ref{tab:frequencies}
were made toward NGC\,253, NGC\,660, IC\,342, NGC\,3079, IC\,860, and
Arp\,220.  For galaxies with expected large linewidths
(IR\,01418+1651, IR\,15107+0724, and Arp\,220) we used the 200\,MHz
bandwidth configurations listed in Table~\ref{tab:frequencies}.
Table~\ref{tab:galaxies} lists the observation frequency setup(s) used
for each galaxy.

\begin{deluxetable*}{llll} 
\tablewidth{0pt}
\tablecolumns{4}
\tablecaption{Measured Transition Properties and Observing Configurations\label{tab:frequencies}}
\tablehead{
\colhead{Transition} & 
\colhead{Frequency} & 
\colhead{E$_{upper}$} &
\colhead{Frequency Setup\tablenotemark{a}} \\
& \colhead{(MHz)} & \colhead{(K)} &
}
\startdata
NH$_3$ (3,1) & 22234.5630 & 167.3 & J50 \\
NH$_3$ (5,4) & 22653.0720 & 344.5 & J50 \\
NH$_3$ (4,3) & 22688.3490 & 239.0 & J50 \\
NH$_3$ (3,2) & 22834.1820 & 151.3 & J50 \\
NH$_3$ (2,1) & 23098.8190 & 80.0  & A50 \\
NH$_3$ (1,1) & 23694.4955 & 22.7  & A50,C200 \\
NH$_3$ (2,2) & 23722.6333 & 63.9  & A50,C200 \\
OH $^2\Pi_{3/2}$ J=9/2 F=$4-4$ & 23817.6153 & 512.1 & D200 \\
OH $^2\Pi_{3/2}$ J=9/2 F=$5-5$ & 23826.6211 & 512.1 & D200 \\
NH$_3$ (3,3) & 23870.1292 & 123.1 & B50,D200 \\
NH$_3$ (4,4) & 24139.4163 & 200.1 & A50,C200 \\
NH$_3$ (10,9) & 24205.2870 & 1137.6 & C200 \\
NH$_3$ (5,5) & 24532.9887 & 295.0 & B50,E200 \\
NH$_3$ (6,6) & 25056.0250 & 407.8 & B50,D200,G200 \\
NH$_3$ (7,7) & 25715.1820 & 538.3 & B50,E200,F200,I50 \\
NH$_3$ (8,8) & 26518.9810 & 687.7 & F200,G200,H200,I50 \\
NH$_3$ (9,9) & 27477.9430 & 853.1 & H200,I50 \\
HC$_3$N $3-2$ & 27294.2950 & 2.6 & I50 \\
\enddata
\tablenotetext{a}{Nomenclature: Letter designates a specific
  spectrometer configuration which includes the associated transitions
  for a given bandwidth.  For example, ``A50'' for the spectrometer
  configuration that includes the NH$_3$ (2,1), (1,1), (2,2), and
  (4,4) transitions measured using a 50\,MHz bandwidth.}
\end{deluxetable*} 

All measurements utilized the dual-beam nodding ($\theta_{B} =
30^{\prime\prime}$ with beam separation = $179^{\prime\prime}$) technique
afforded by the dual- and multi-beam dual-polarization K-band receiver
systems at the GBT.  Measurements made before 2011/10/23 used the
facility dual-beam dual-polarization receiver system, while those made
thereafter used beams 2 and 3 of the 7-channel K-band Focal Plane
Array (KFPA).  Spectra were produced from these measurements using the
GBT spectrometer configured to yield simultaneous measurement of 4
(frequency setups A50, B50, J50, and I50) or 2 (all other frequency
setups) spectral windows each with 50 or 200\,MHz of bandwidth sampled
by 4096 channels.  This correlator configuration produced a spectral
channel width of 12.2 and 48.8\,kHz (0.16 and 0.62\,\kms\ at 23.7 GHz).

\begin{deluxetable*}{lccccccl} 
\tablewidth{0pt}
\tablecolumns{8}
\tablecaption{Extragalactic Ammonia Target List\label{tab:galaxies}}
\tablehead{
\colhead{Source} & 
\colhead{$\alpha$} & 
\colhead{$\delta$} & 
\colhead{v$_{hel}$\tablenotemark{a}} & 
\colhead{D$_L$\tablenotemark{b}} &
\colhead{T$_{dust}$\tablenotemark{c}} &
\colhead{L$_{IR}$\tablenotemark{d}} &
\colhead{Obs Frequency Setups\tablenotemark{e}} \\
& \colhead{(J2000)} &\colhead{(J2000)}& 
\colhead{(km s$^{-1}$)} & \colhead{(Mpc)} & \colhead{(K)} &
\colhead{($\times 10^{10}$\,\lsun)} &
}
\startdata
NGC\,253       & 00:47:33.1  & $-$25:17:18 & 251    & $3.44\pm0.24$  &
34 & $3.41\pm0.24$ & A,B,I, and J50 \\
NGC\,598       & 01:33:54.0  & $+$30:40:07 & $-$179 & $0.88\pm0.28$ &
\nodata & $0.13\pm0.05$ & A50 \\ 
NGC\,660       & 01:43:01.7  & $+$13:38:36 & 856    & $12.2\pm0.9$  &
37 & $3.03\pm0.22$ & A,B, and J50\\
IR\,01418+1651 & 01:44:30.5  & $+$17:06:09 & 8101   & $109.7\pm7.7$ &
\nodata & $36.33\pm2.72$ & C200\\
NGC\,891       & 02:22:33.4  & $+$42:20:57 & 529    & $9.43\pm0.66$  &
28 & $2.25\pm0.16$ & A50 \\
Maffei\,2      & 02:41:55.1  & $+$59:36:15 & $-$17  & $3.11\pm0.23$ &
\nodata & $0.31\pm0.05$ & A50 \\ 
NGC\,1144      & 02:55:12.2  & $-$00:11:01 & 8750   & $115.3\pm8.1$ &
32 & $23.86\pm2.16$ & A50,C200\\ 
NGC\,1365      & 03:33:36.4  & $-$36:08:25 & 1652   & $21.5\pm1.5$  &
32 & $14.31\pm1.00$ & A50 \\ 
IC\,342        & 03:46:49.7  & $+$68:05:45 & 31     & $3.82\pm0.27$  &
30 & $1.01\pm0.07$ & A50,B50 \\ 
NGC\,2146      & 06:18:37.7  & $+$78:21:25 & 918    & $16.7\pm1.2$  &
38 & $12.56\pm1.36$ & A50\\ 
M\,82          & 09:55:52.2  & $+$69:40:47 & 203    & $5.92\pm0.42$  &
45 & $15.69\pm1.11$  & A50 \\ 
M\,82SW        & 09:55:50.0  & $+$69:40:43 & 203    & \ldots  & \ldots
& $15.69\pm1.11$ & A50  \\
NGC\,3079      & 10:01:57.8  & $+$55:40:47 & 1150   & $20.7\pm1.5$  &
32 & $6.93\pm0.50$ & A,B,I, and J50, D200 \\
NGC\,3628      & 11:20:17.2  & $+$13:35:20 & 847    & $8.5\pm0.6$  &
30 & $1.27\pm0.09$ & A50 \\
NGC\,3690      & 11:28:32.2  & $+$58:33:44 & 3121   & $48.5\pm3.4$  &
\nodata & $76.96\pm5.40$ & A50 \\
NGC\,4214      & 12:15:39.2  & $+$36:19:37 & 291    & $3.75\pm0.27$ &
\nodata & $0.07\pm0.005$ & A50 \\
NGC\,4418      & 12:26:54.6  & $-$00:52:39 & 2040   & $34.7\pm2.4$ &
\nodata & $14.08\pm0.98$ & A50 \\
Mrk\,231       & 12:56:14.2  & $+$56:52:25 & 12642  & $178.1\pm12.5$ &
\nodata & $319.46\pm22.45$ & A50 \\
IC\,860        & 13:15:04.1  & $+$24:37:01 & 3866   & $53.8\pm3.8$  &
\nodata & $11.95\pm0.93$ & A50,B50 \\ 
M\,83          & 13:37:00.9  & $-$29:51:57 & 518    & $4.03\pm0.28$  &
31 & $1.57\pm0.11$ & A50 \\
IR\,15107+0724 & 15:13:13.1  & $+$07:13:27 & 3897   & $61.9\pm4.4$  &
\nodata & $19.98\pm1.44$ & C200 \\
Arp\,220       & 15:34:57.1  & $+$23:30:11 & 5434   & $82.9\pm5.8$  &
44 & $167.11\pm11.70$ & C,D,E,F,G, and H200 \\
NGC\,6946      & 20:34:52.3  & $+$60:09:14 & 48     & $5.49\pm0.39$  &
30 & $1.51\pm0.11$ & A50 \\
NGC\,6951      & 20:37:14.1  & $+$66:06:20 & 1425   & $24.3\pm1.7$ &
\nodata & $3.74\pm0.26$ & A50 \\
\enddata
\tablenotetext{a}{Heliocentric velocity drawn from the
  literature.}
\tablenotetext{b}{NED Hubble flow distance corrected for Virgo
  cluster, Great Attractor, and Shapley supercluster.  For NGC\,598 no
  Hubble flow distance was available, so NED ``redshift-independent'' distance
assumed.} 
\tablenotetext{c}{From \citet{Gao2004b}, who used IRAS 60 and
  100$\mu$m dust continuum emission ratios with an assumed dust
  emissivity $\propto\nu^{-\beta}$ with $\beta = 1.5$.}
\tablenotetext{d}{~Luminosities from \citet{Sanders2003}, derived from
  IRAS fluxes over 8 to 1000\,$\mu$m.}
\tablenotetext{e}{See Table~\ref{tab:frequencies} for transitions measured.}
\end{deluxetable*} 

To calibrate all of our measurements to a uniform intensity scale we
used the following measurements:
\begin{itemize}
\item One or more of the standard flux calibration sources 0137+331
  (3C\,48), 0139+413 (3C\,84), 1331+305 (3C\,286) were measured during
  each observing run. 
\item An estimate of the atmospheric opacity during each of our
  observing runs was derived based on atmospheric model
  calculations using ambient pressure, temperature, and relative
  humidity measurements.  At 23 GHz we found that the zenith opacity
  $\tau_0 \sim 0.04 - 0.10$ during our observations.  Assuming
  elevation $\gtrsim$ 30 degrees implies that the opacity correction
  to our measured amplitudes is $\lesssim 1.22$.
\end{itemize}
Combined with a calculation of the aperture efficiency using the
following GBT-recommended formula:

\begin{equation}
\label{eq:etaa}
\eta_A = 0.72\exp\left[-\left(\alpha\nu(GHz)\right)^2\right]
\end{equation}

\noindent{where} $\alpha = 0.0163$ for measurements made on or before
2009/02/06, while $\alpha = 0.0145$ for measurements made after
2009/02/06, the conversion between antenna temperature corrected for
atmospheric attenuation (T$^*_A$) and flux density (S) was calculated
using:

\begin{equation}
\label{eq:kperjy}
\frac{T^*_A}{S} = 2.846\eta_A\exp(-\tau_0\csc(EL))~K/Jy
\end{equation}

Fluxes for our calibrator sources were calculated using the
measurements and frequency-dependent relations derived by
\cite{Ott1994}.  The $T^*_A/S$ values for our thirty-four observing
sessions which comprise our NH$_3$ measurements ranged from
$0.91\pm0.07$ K/Jy to $1.92\pm0.07$ K/Jy.  With a calculated value for
$T^*_A/S$ appropriate for the measurements made in each observing run
we then normalized all of our NH$_3$ measurements to a common
$T^*_A/S$ 
value of 1.88 K/Jy, thus allowing us to average spectra for a given galaxy
over multiple observing runs.  This calibration sequence assumes
point-source emission.  To calibrate to a main beam brightness
temperature scale, one can use $\eta_{mb} \simeq 1.32 \eta_A$ with
brightness temperature defined as $T_{mb} \simeq
\frac{T^*_A}{\eta_{mb}}$.  GBT amplitude calibration is
reported to be accurate to 10-15\% at all frequencies, limited mainly
by temporal drifts in the noise diodes used as amplitude
calibration standards.

\section{Results}
\label{Results}

\begin{figure}
\centering
\includegraphics[trim=20mm 20mm 20mm 30mm, clip, scale=0.40]{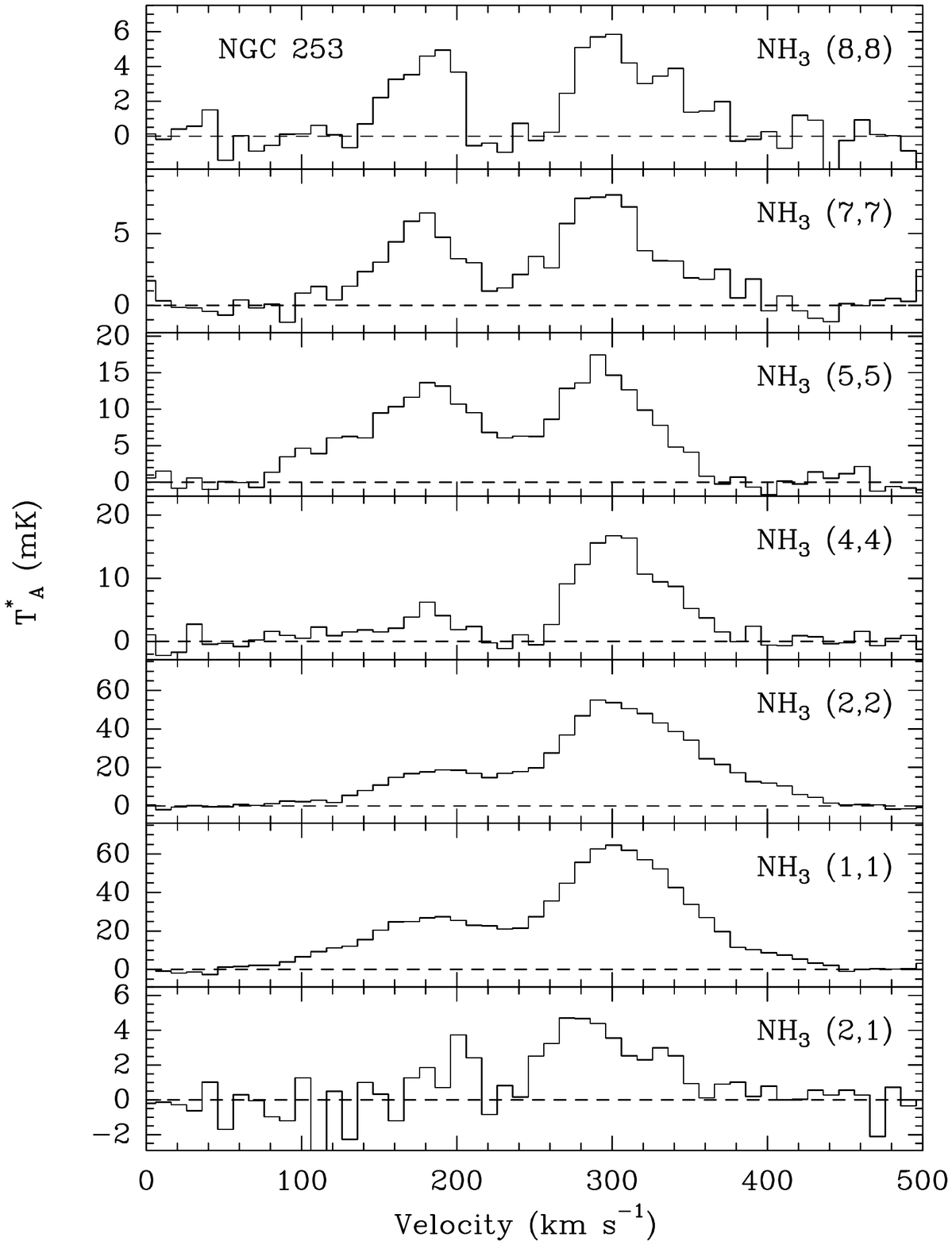}
\includegraphics[trim=20mm 20mm 20mm 30mm, clip, scale=0.40]{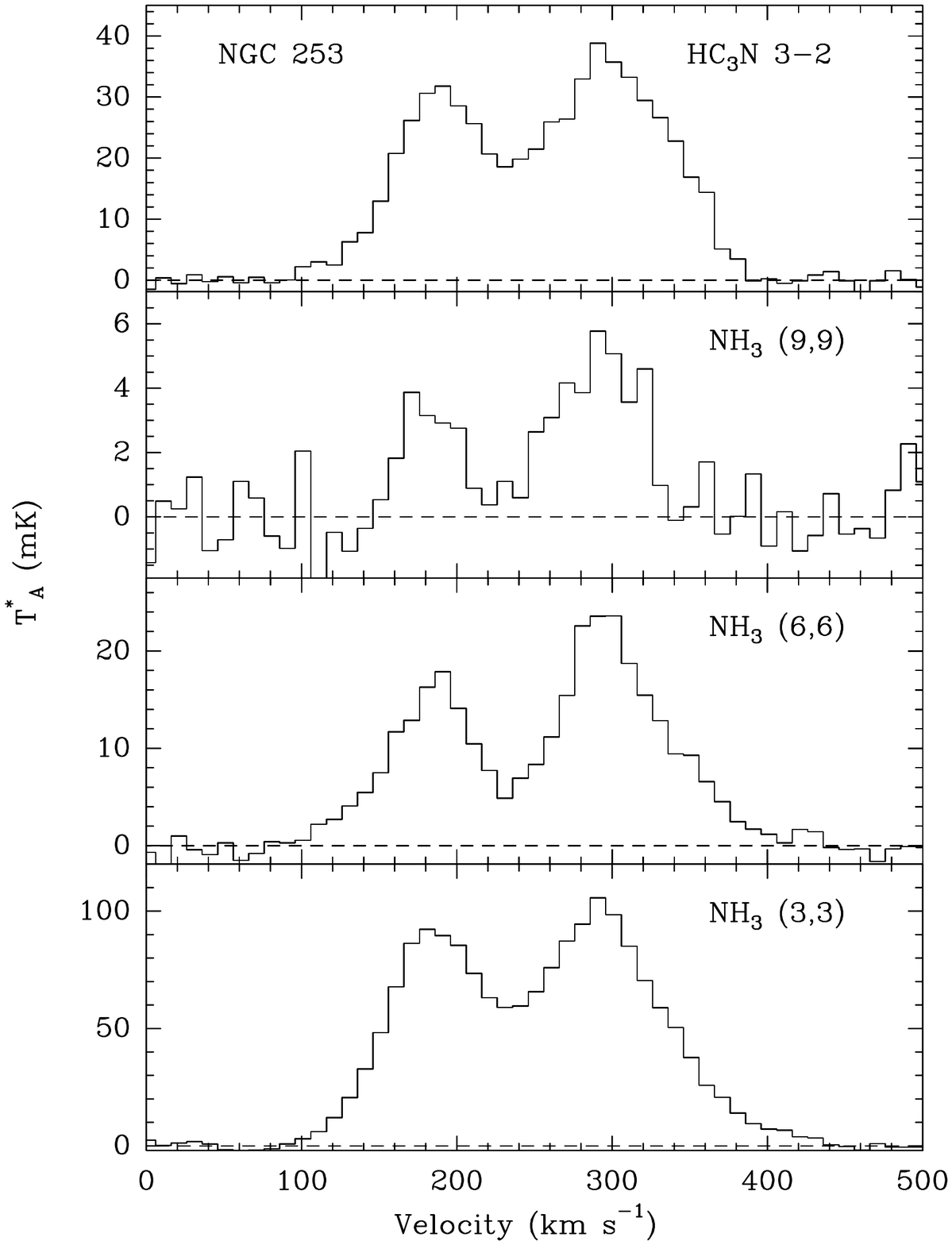}
\caption{NH$_3$ and HC$_3$N J=$3-2$ spectra of NGC\,253.  The top
  panel shows spectra of para-NH$_3$, while the bottom panel lists
  ortho-NH$_3$ and HC$_3$N J=$3-2$ spectra.}
\label{fig:NGC253NH3Spec}
\end{figure}

\begin{figure*}
\centering
\includegraphics[trim=20mm 20mm 20mm 30mm, clip, scale=0.30]{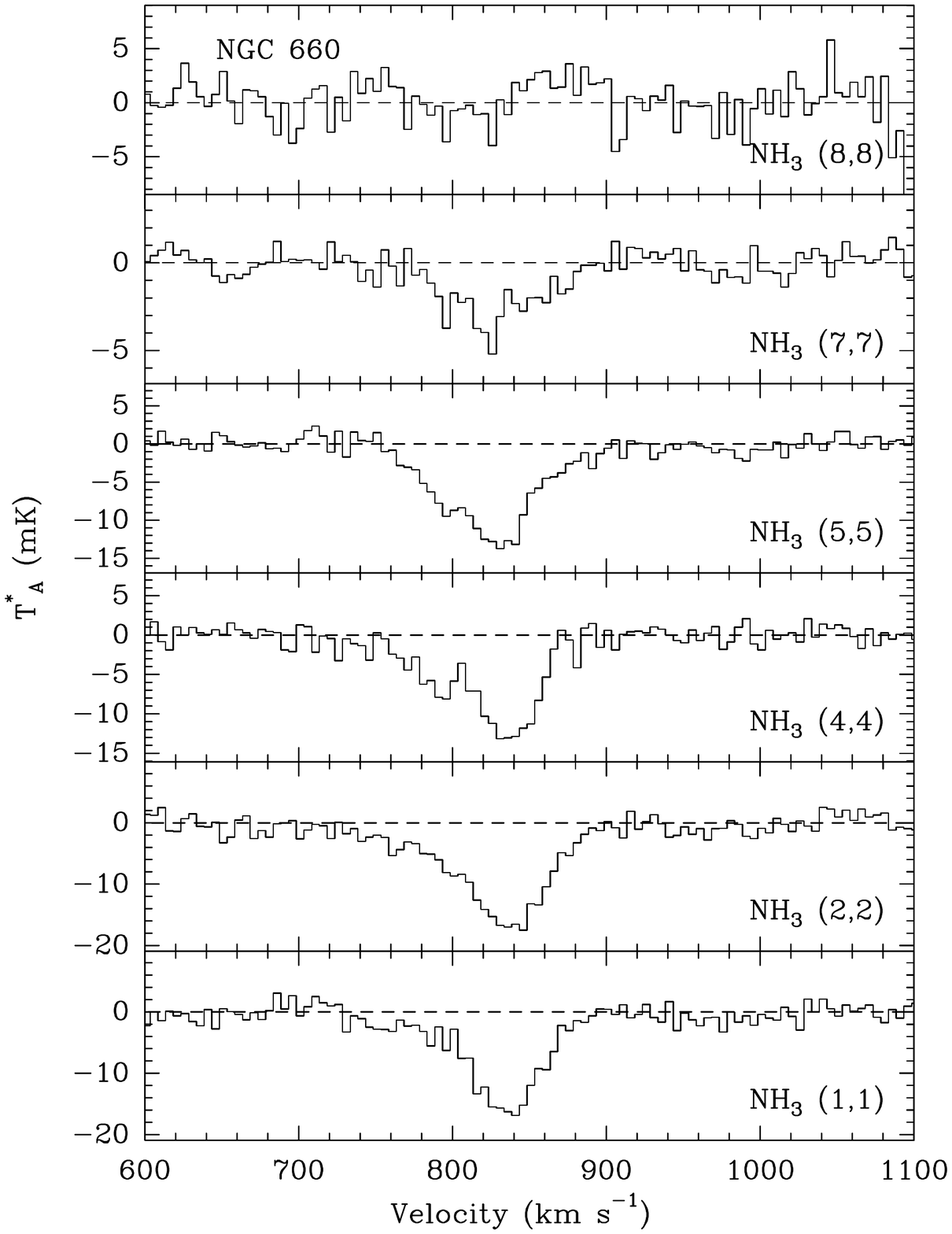}
\includegraphics[trim=20mm 20mm 20mm 30mm, clip, scale=0.30]{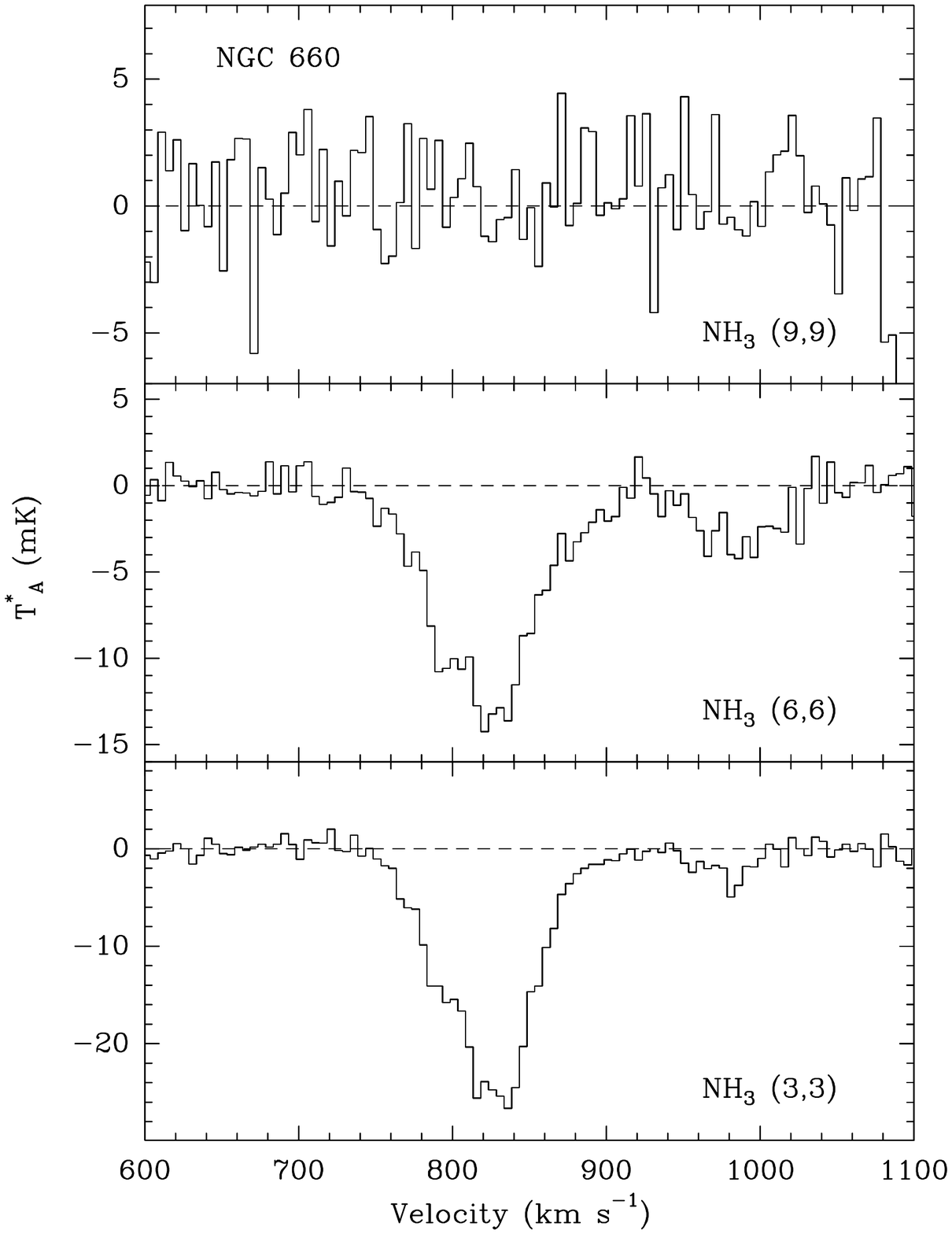}
\includegraphics[trim=20mm 20mm 20mm 30mm, clip, scale=0.30]{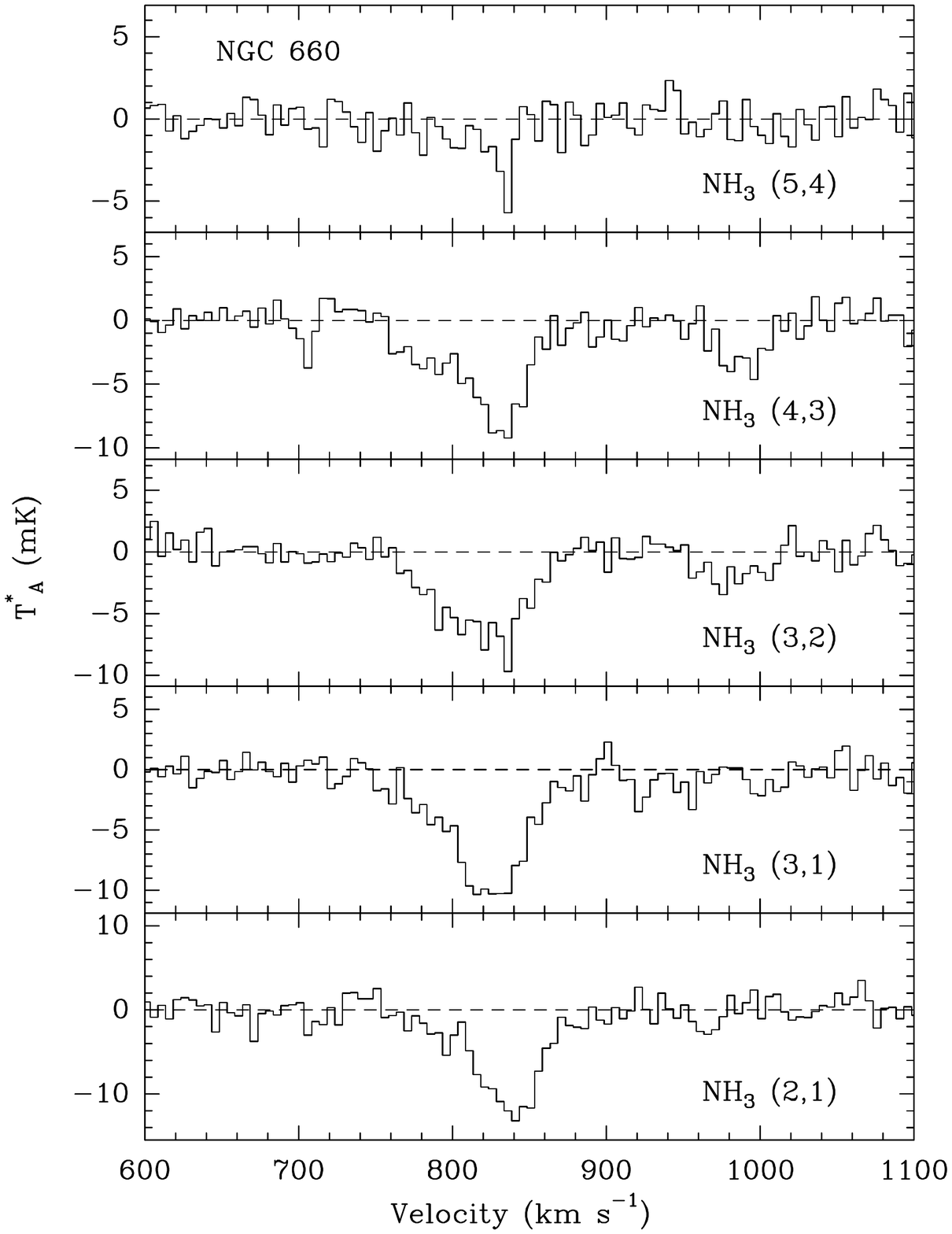}
\caption{NH$_3$ spectra of NGC\,660.  The left, middle, and right
  panels show spectra of metastable (J=K) para-NH$_3$, metastable
  ortho-NH$_3$, and non-metastable (J$\neq$K) NH$_3$, respectively.}
\label{fig:NGC660NH3spec}
\end{figure*}

\begin{figure*}
\centering
\includegraphics[trim=20mm 15mm 20mm 30mm, clip, scale=0.30]{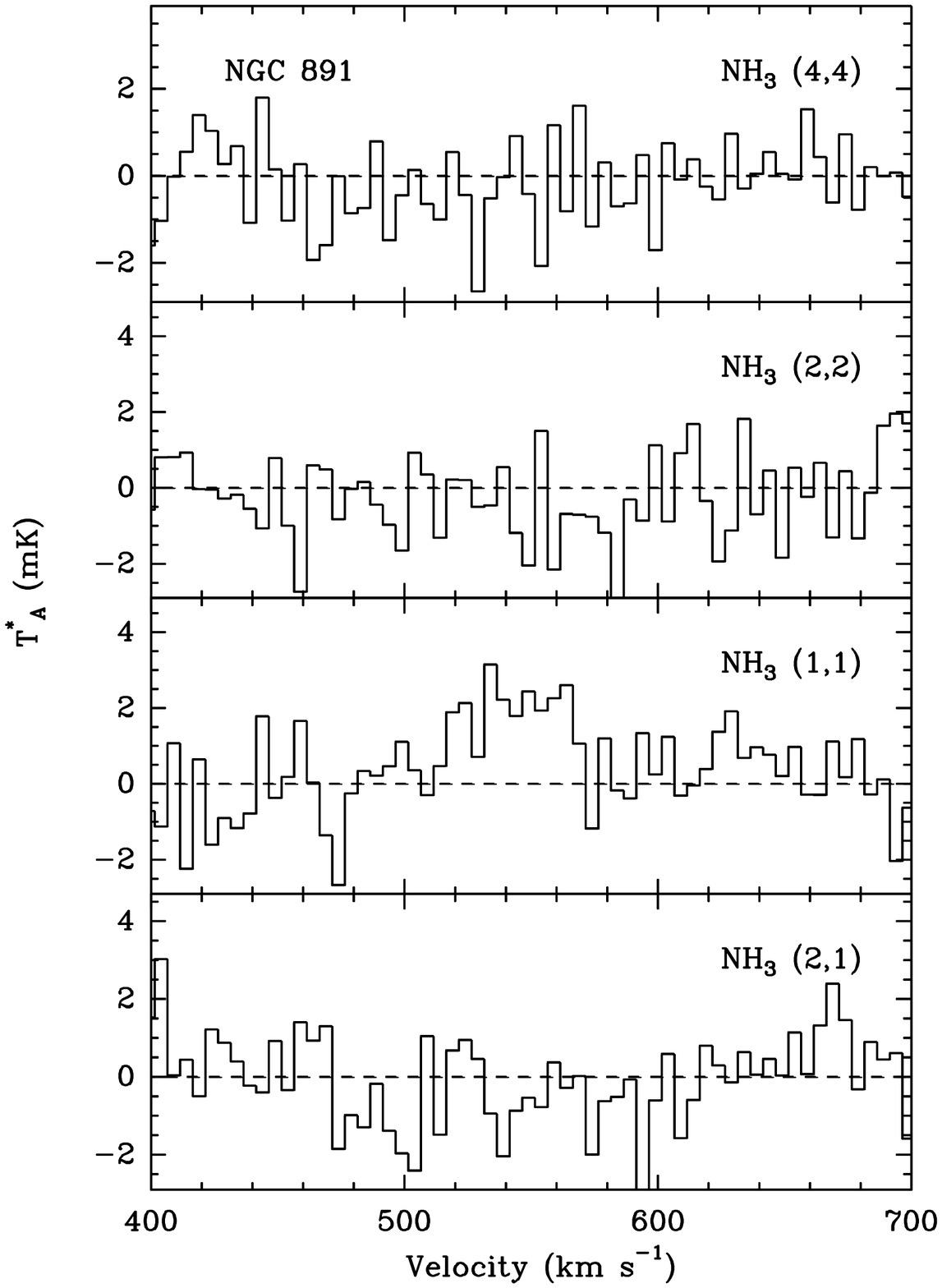}
\includegraphics[trim=20mm 15mm 20mm 30mm, clip, scale=0.31]{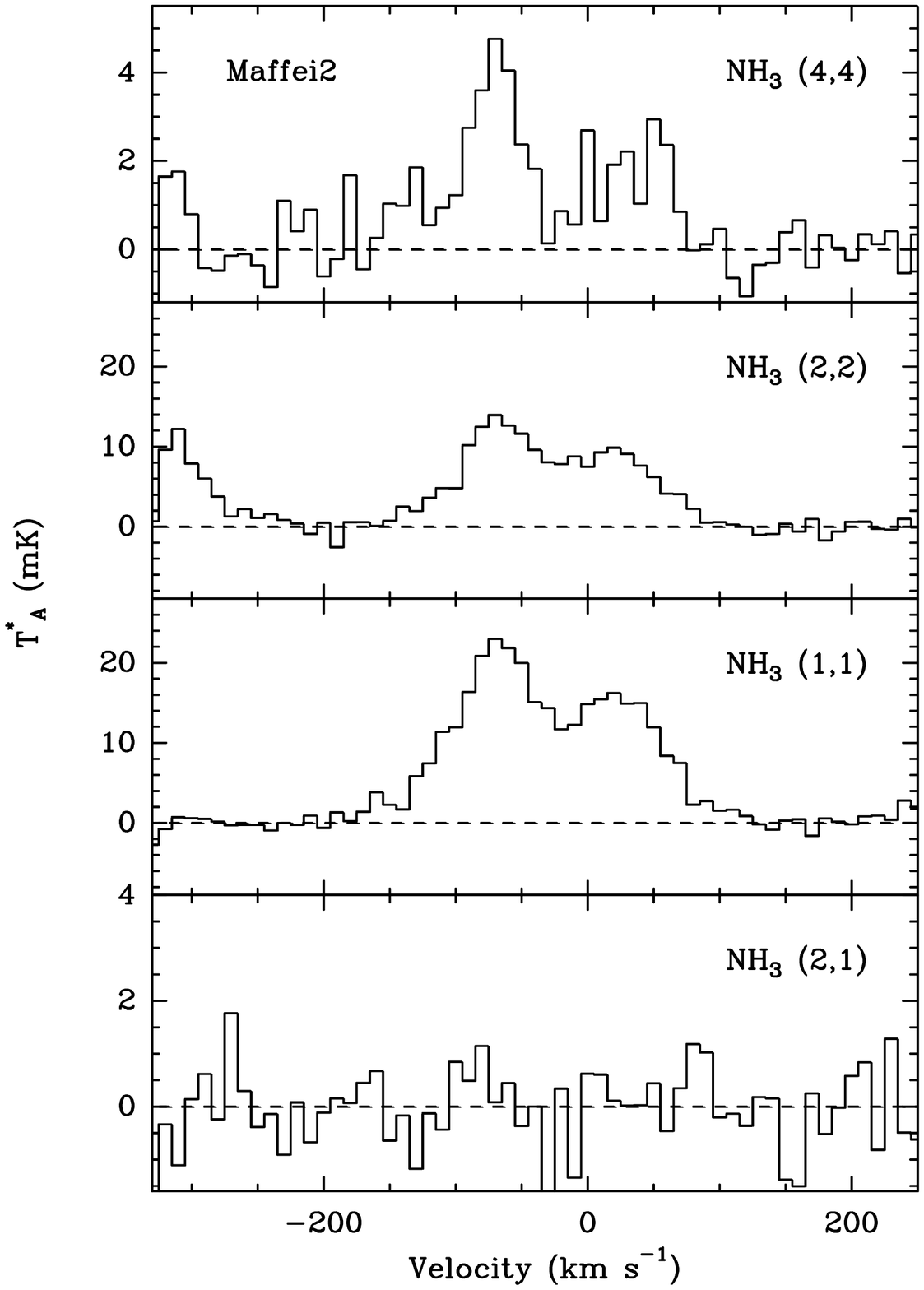}
\includegraphics[trim=20mm 15mm 20mm 30mm, clip, scale=0.30]{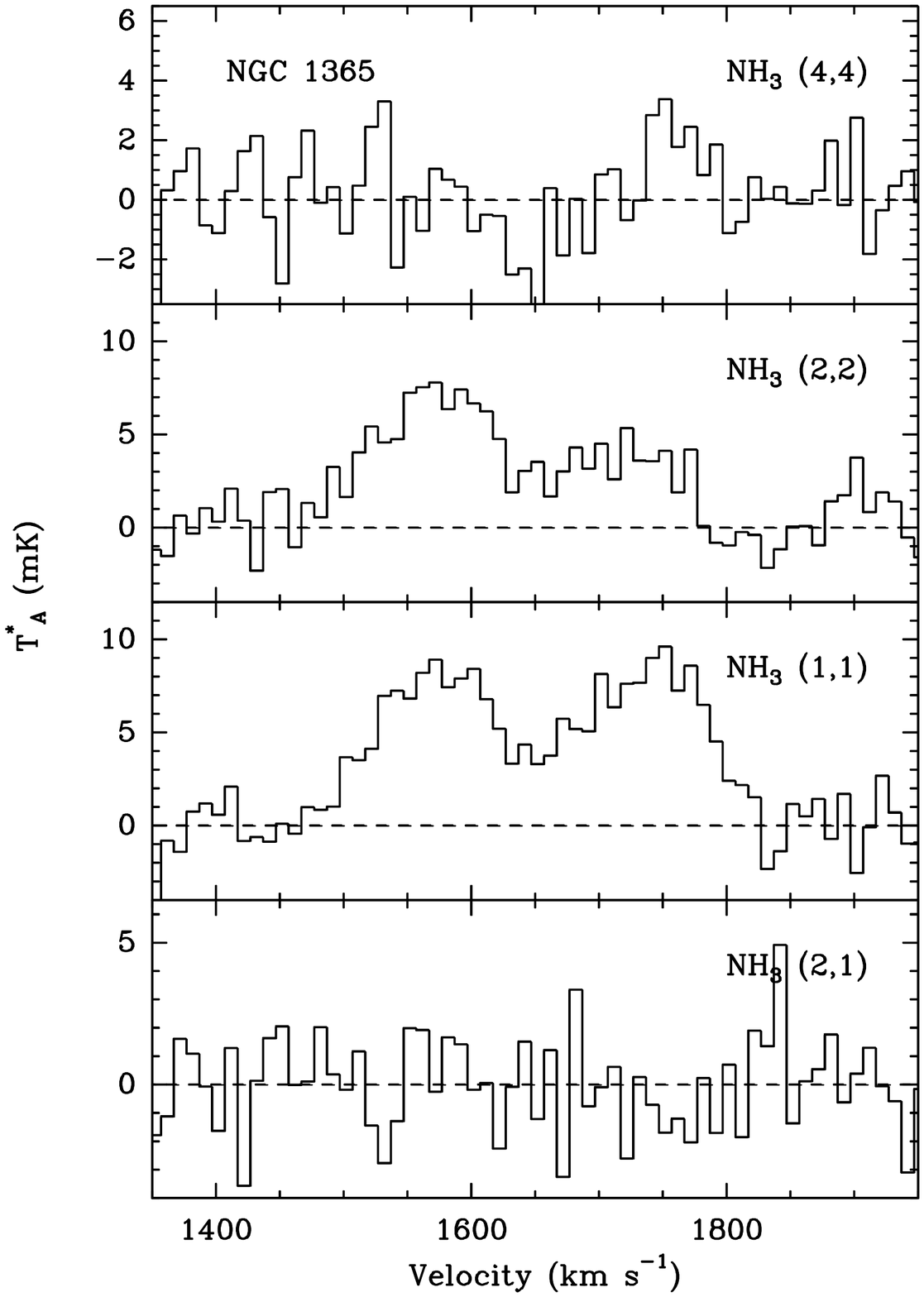}
\caption{NH$_3$ spectra of NGC\,891 (left), Maffei\,2 (middle), and
  NGC\,1365 (right).}
\label{fig:NGC891Maffei2NGC1365NH3Spec}
\end{figure*}

\begin{figure*}
\centering
\includegraphics[trim=20mm 15mm 20mm 30mm, clip, scale=0.45]{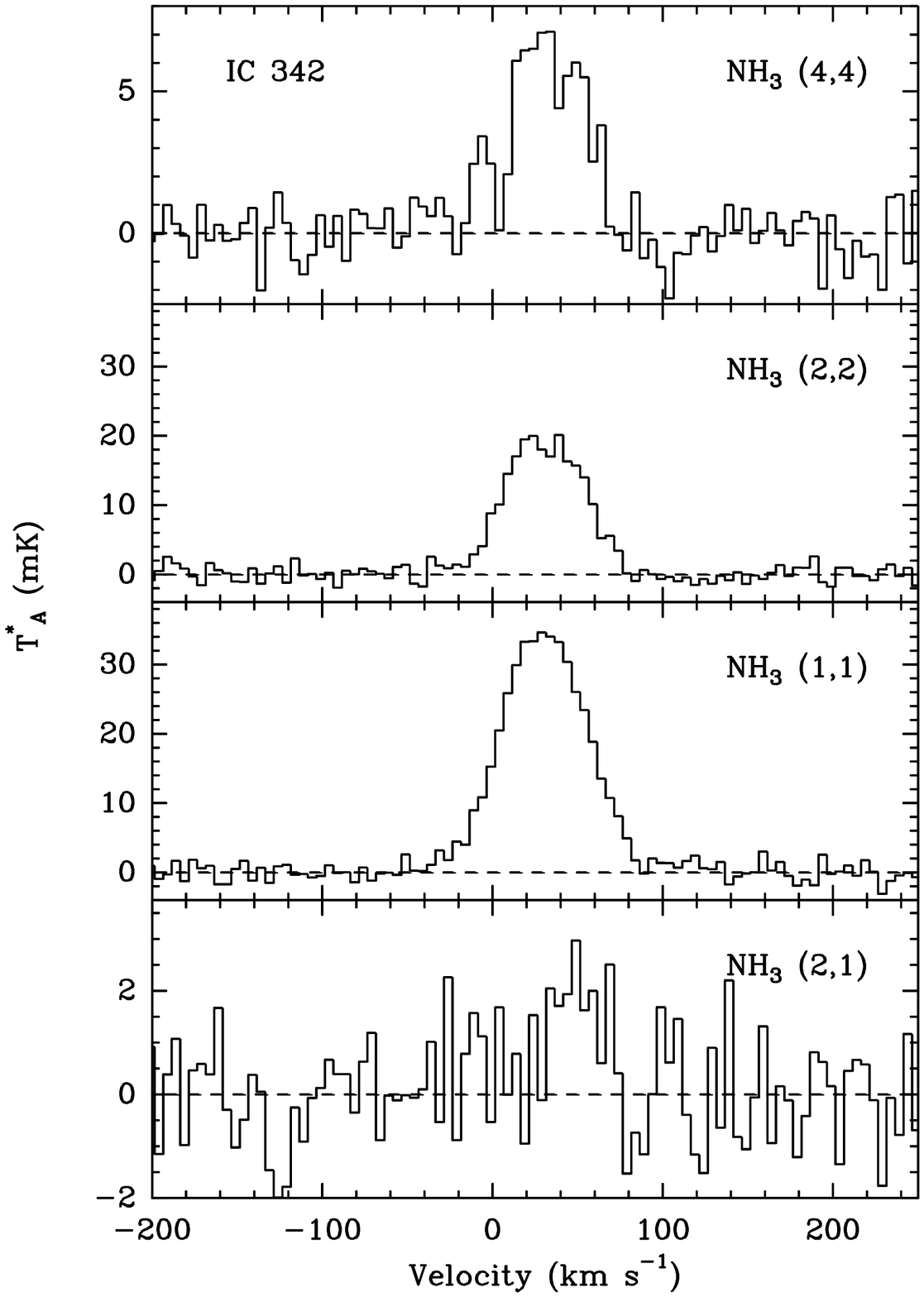}
\includegraphics[trim=20mm 15mm 20mm 30mm, clip, scale=0.45]{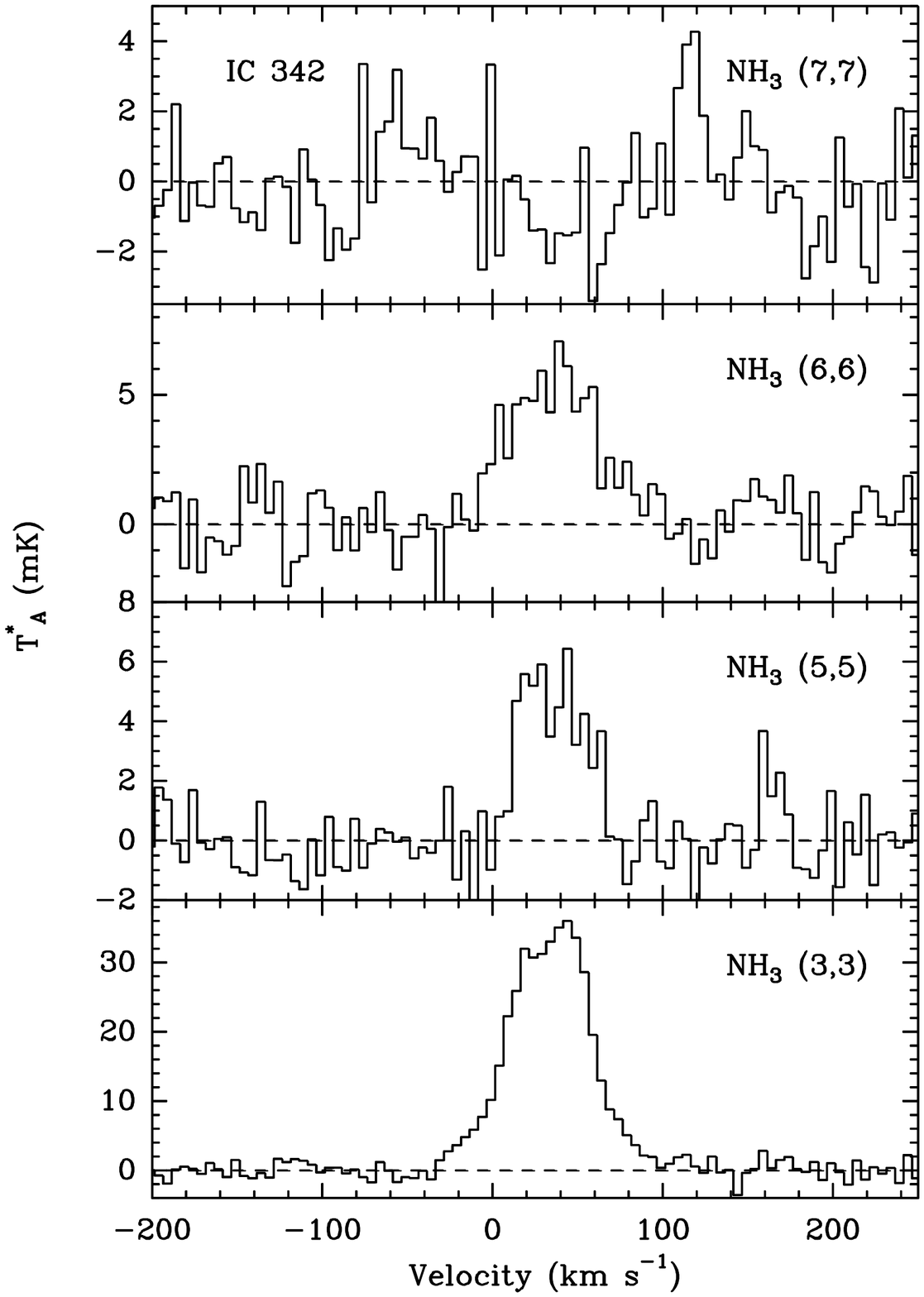}
\caption{NH$_3$ spectra of IC\,342.}
\label{fig:IC342NH3Spec}
\end{figure*}

\begin{figure*}
\centering
\includegraphics[trim=20mm 15mm 20mm 30mm, clip, scale=0.45]{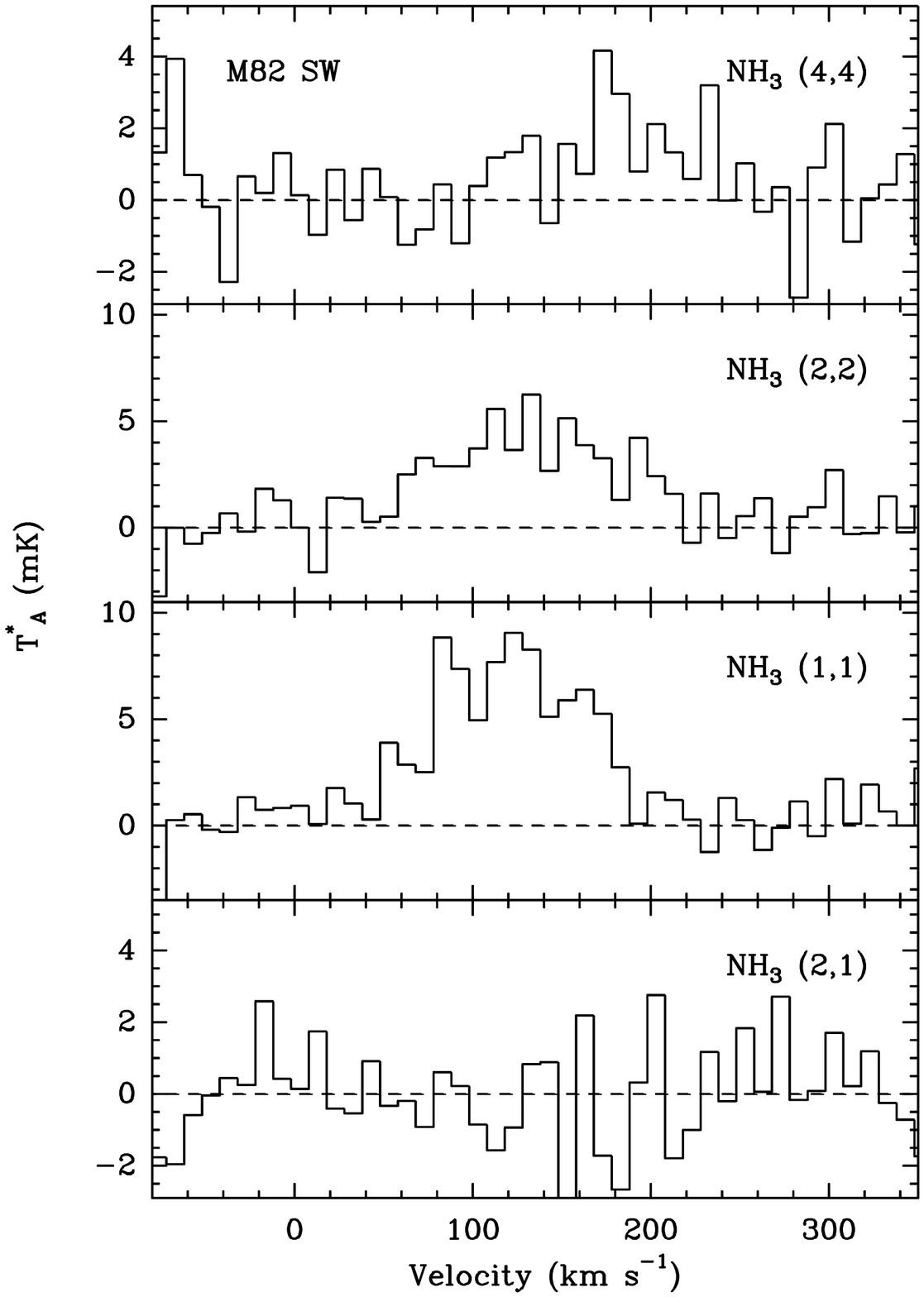}
\includegraphics[trim=20mm 15mm 20mm 30mm, clip, scale=0.45]{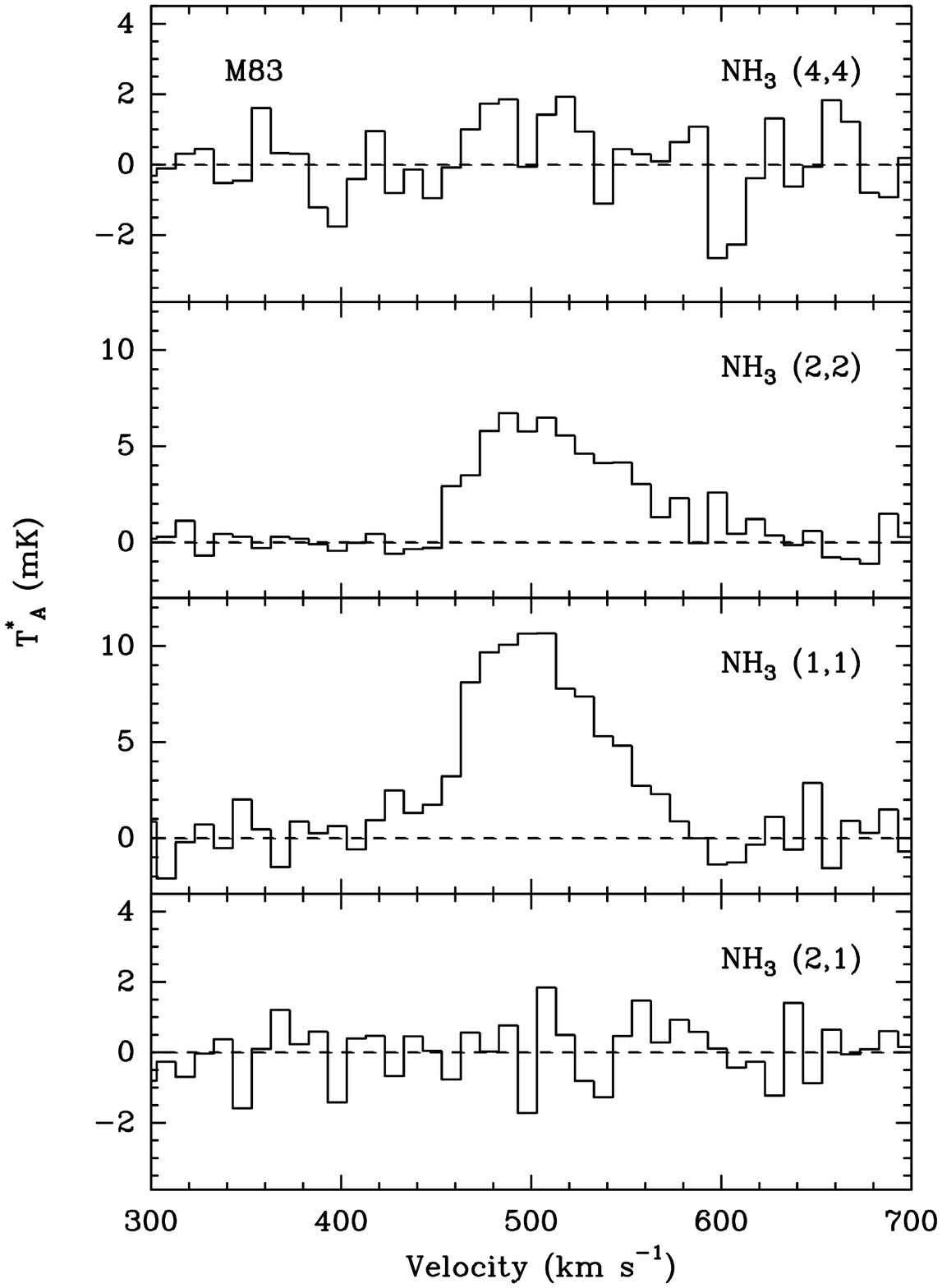}
\caption{NH$_3$ spectra of M\,82SW (left) and M\,83 (right).}
\label{fig:M82SWM83NH3Spec}
\end{figure*}

\begin{figure*}
\centering
\includegraphics[trim=20mm 15mm 20mm 30mm, clip, scale=0.45]{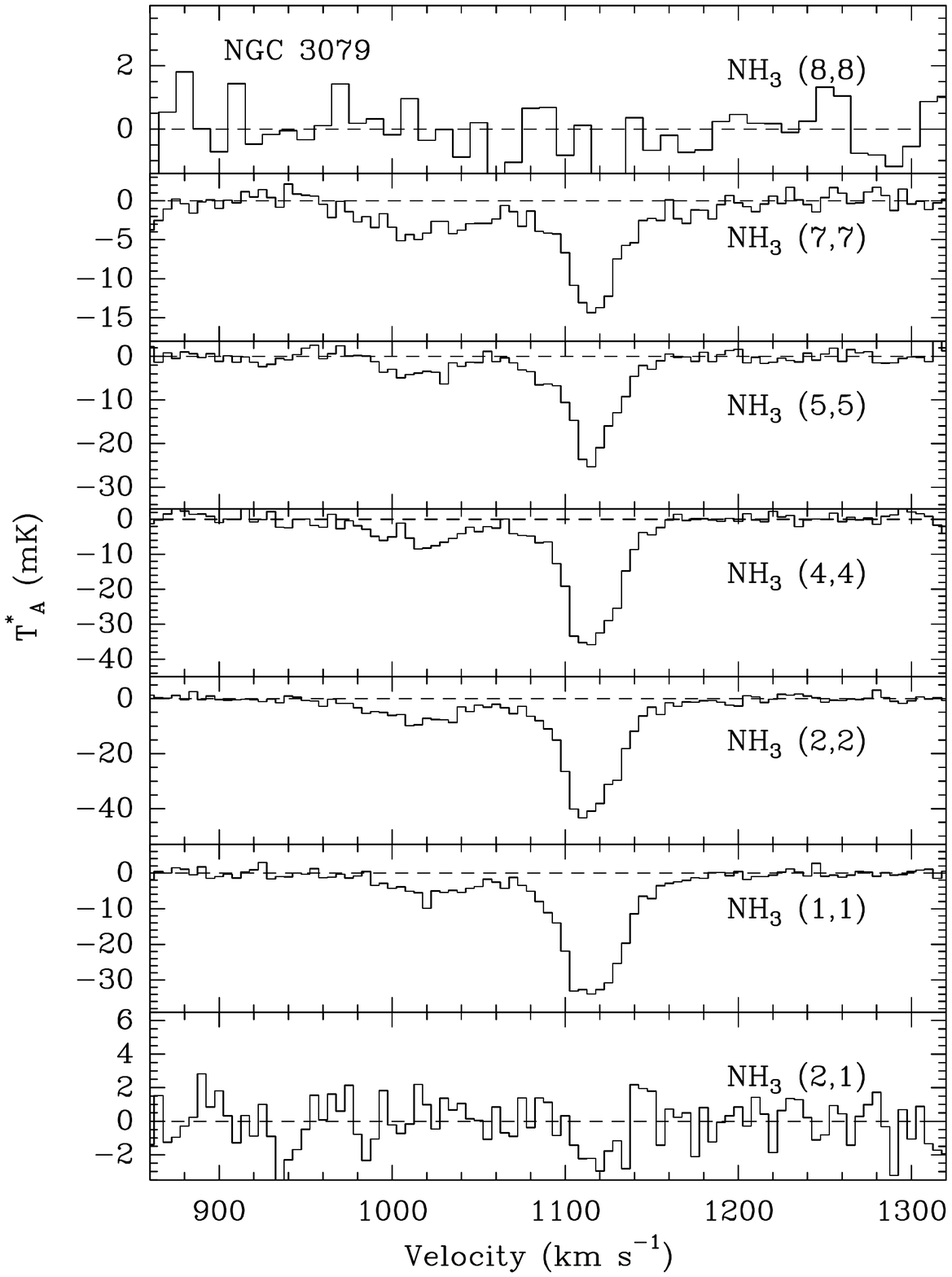}
\includegraphics[trim=20mm 15mm 20mm 30mm, clip, scale=0.45]{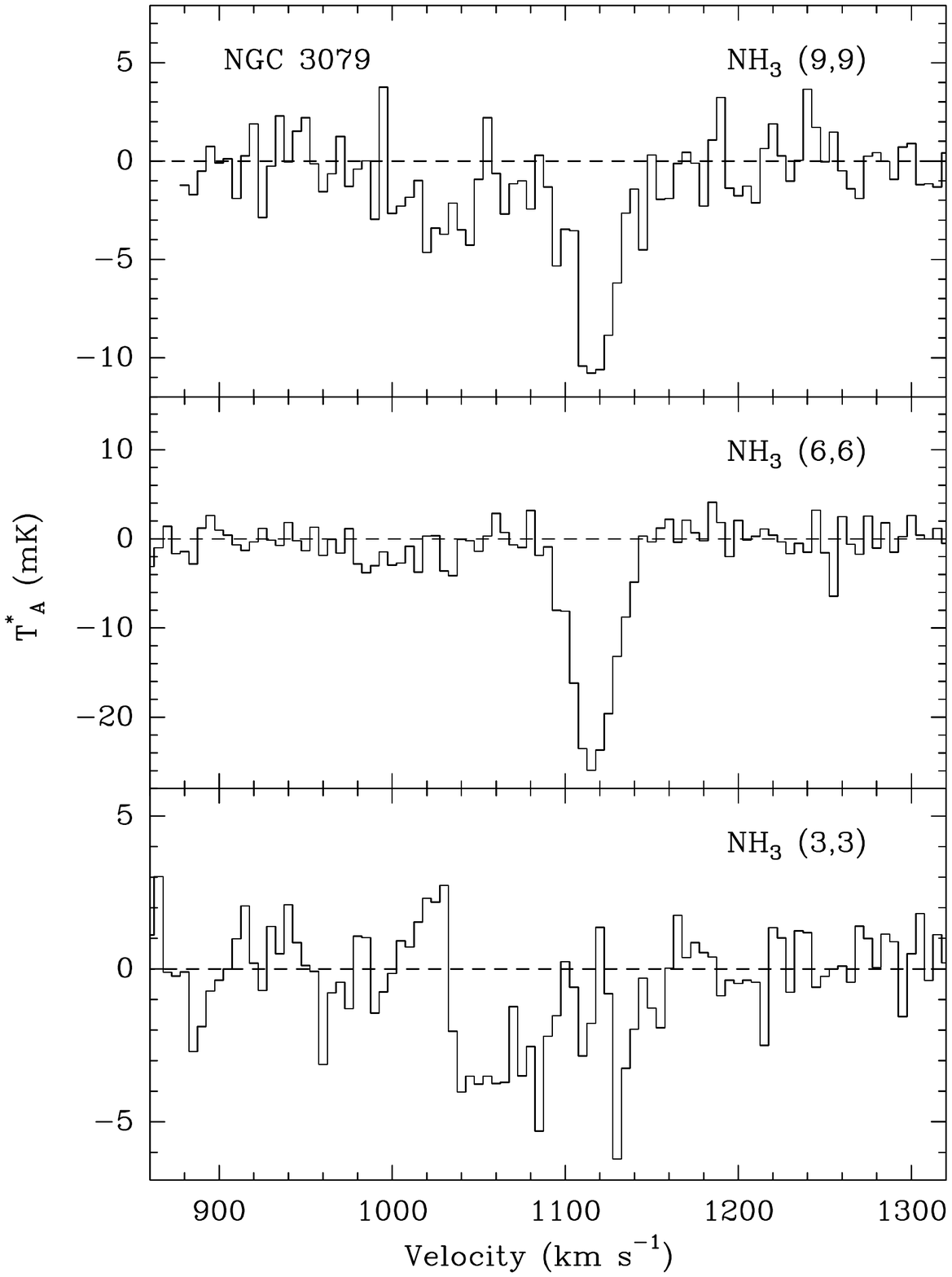}
\caption{Para-NH$_3$ (left panel) and ortho-NH$_3$ (right panel)
  spectra of NGC\,3079.}
\label{fig:NGC3079NH3Spec}
\end{figure*}

\begin{figure*}
\centering
\includegraphics[trim=20mm 15mm 20mm 30mm, clip, scale=0.45]{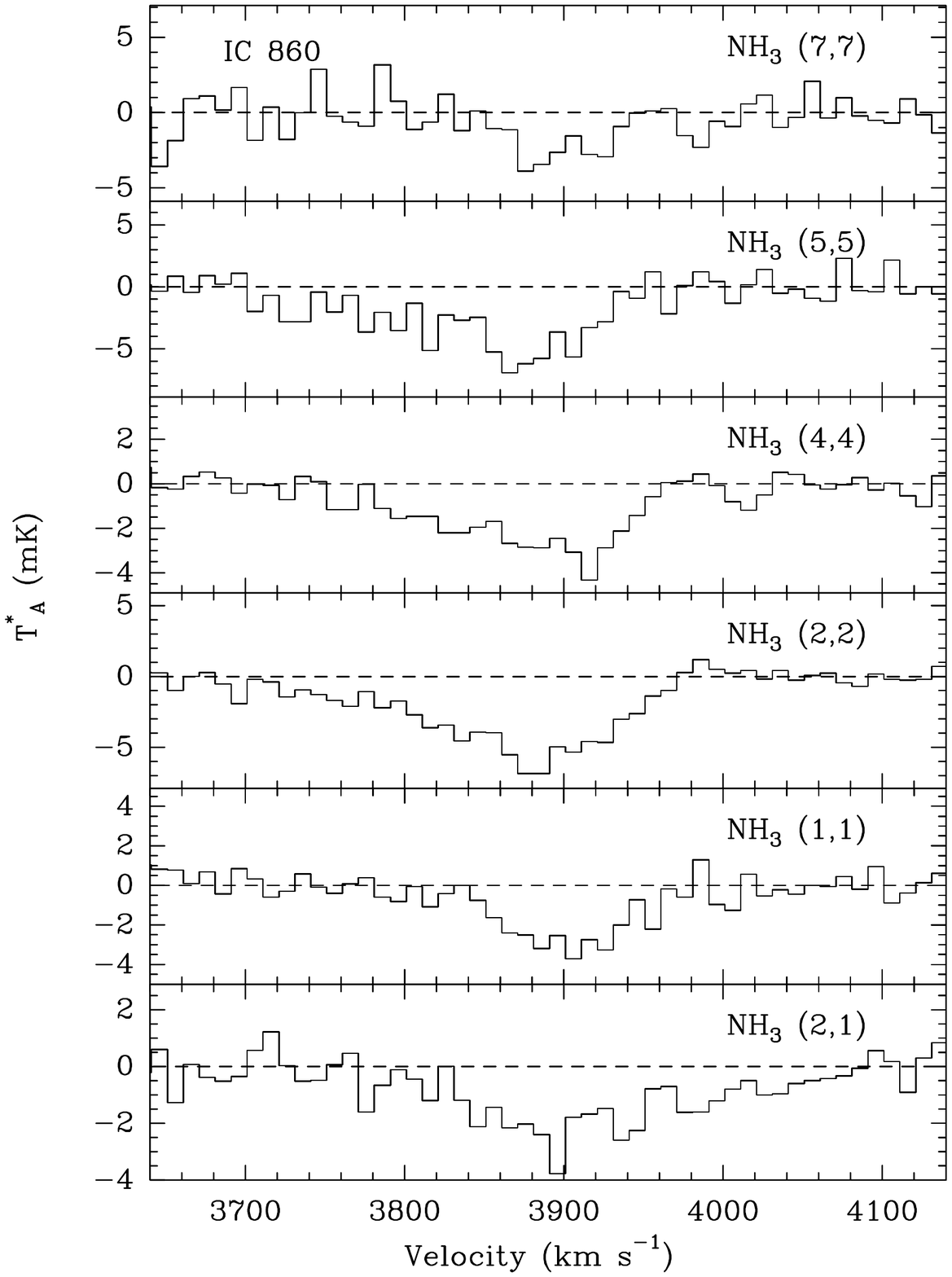}
\includegraphics[trim=20mm 15mm 20mm 30mm, clip, scale=0.45]{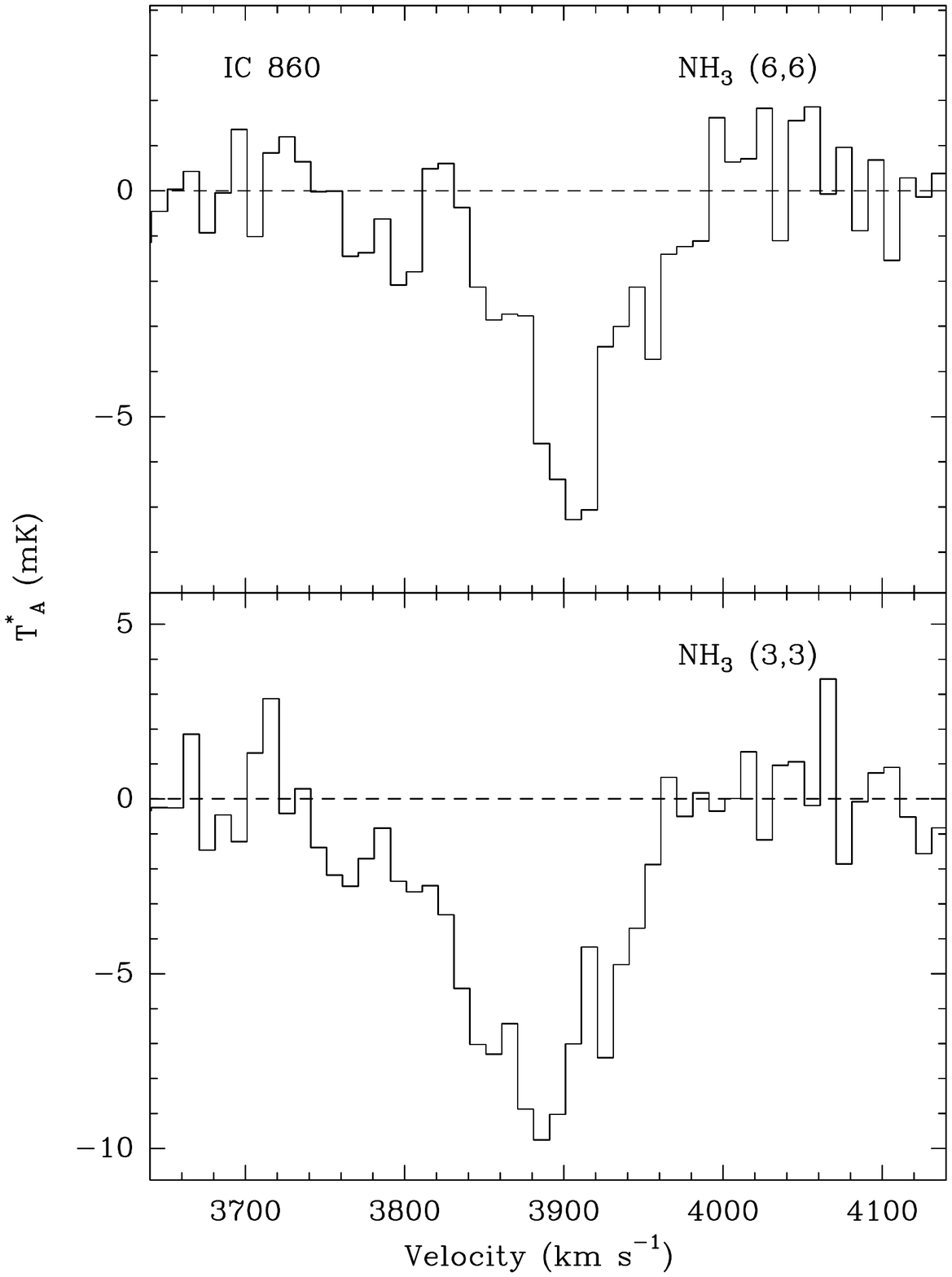}
\caption{Para-NH$_3$ (left panel) and ortho-NH$_3$ (right panel)
  spectra of IC\,860.}
\label{fig:IC860NH3Spec}
\end{figure*}

\begin{figure*}
\centering
\includegraphics[trim=20mm 15mm 20mm 30mm, clip, scale=0.45]{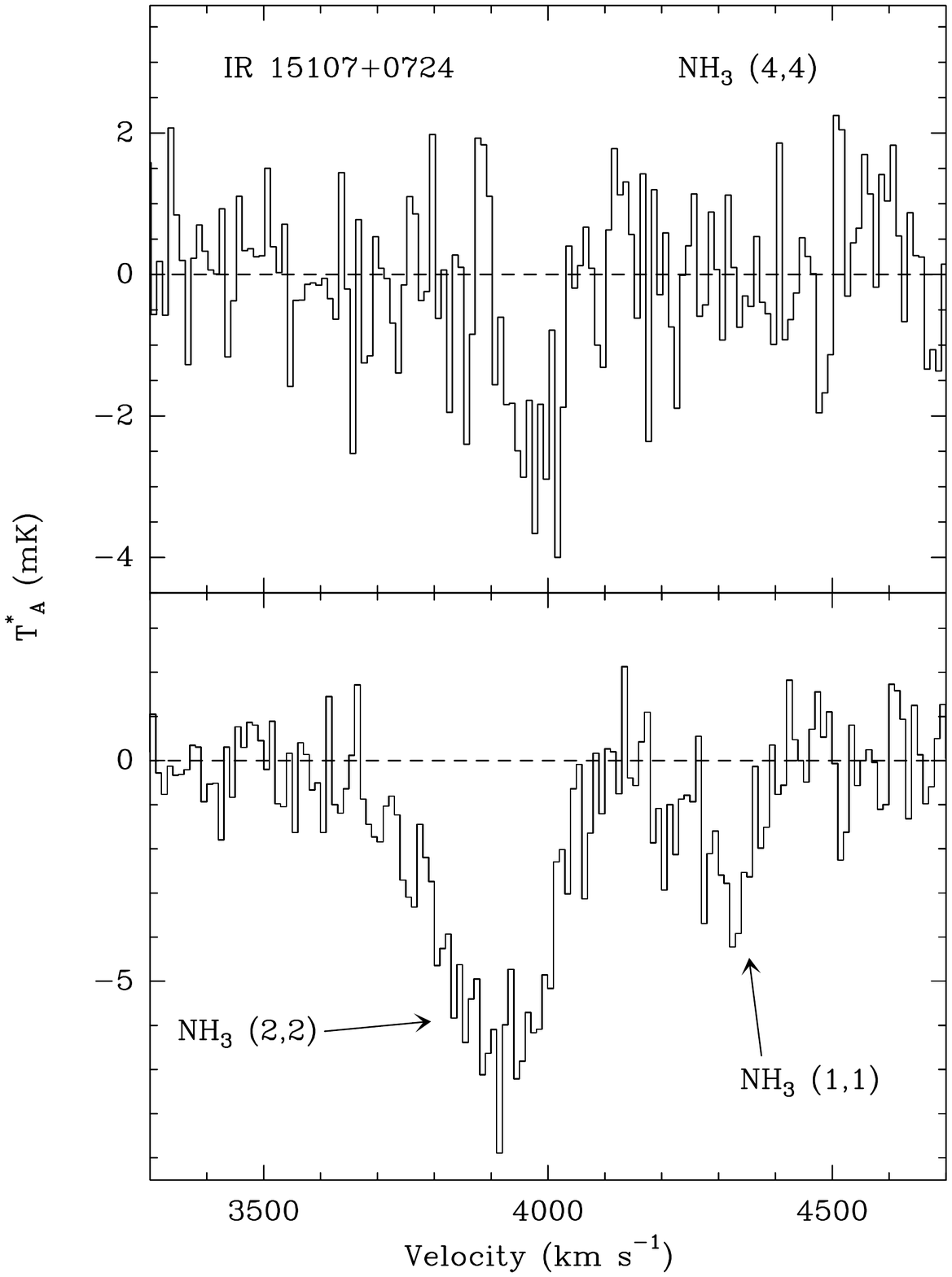}
\includegraphics[trim=20mm 15mm 20mm 30mm, clip, scale=0.45]{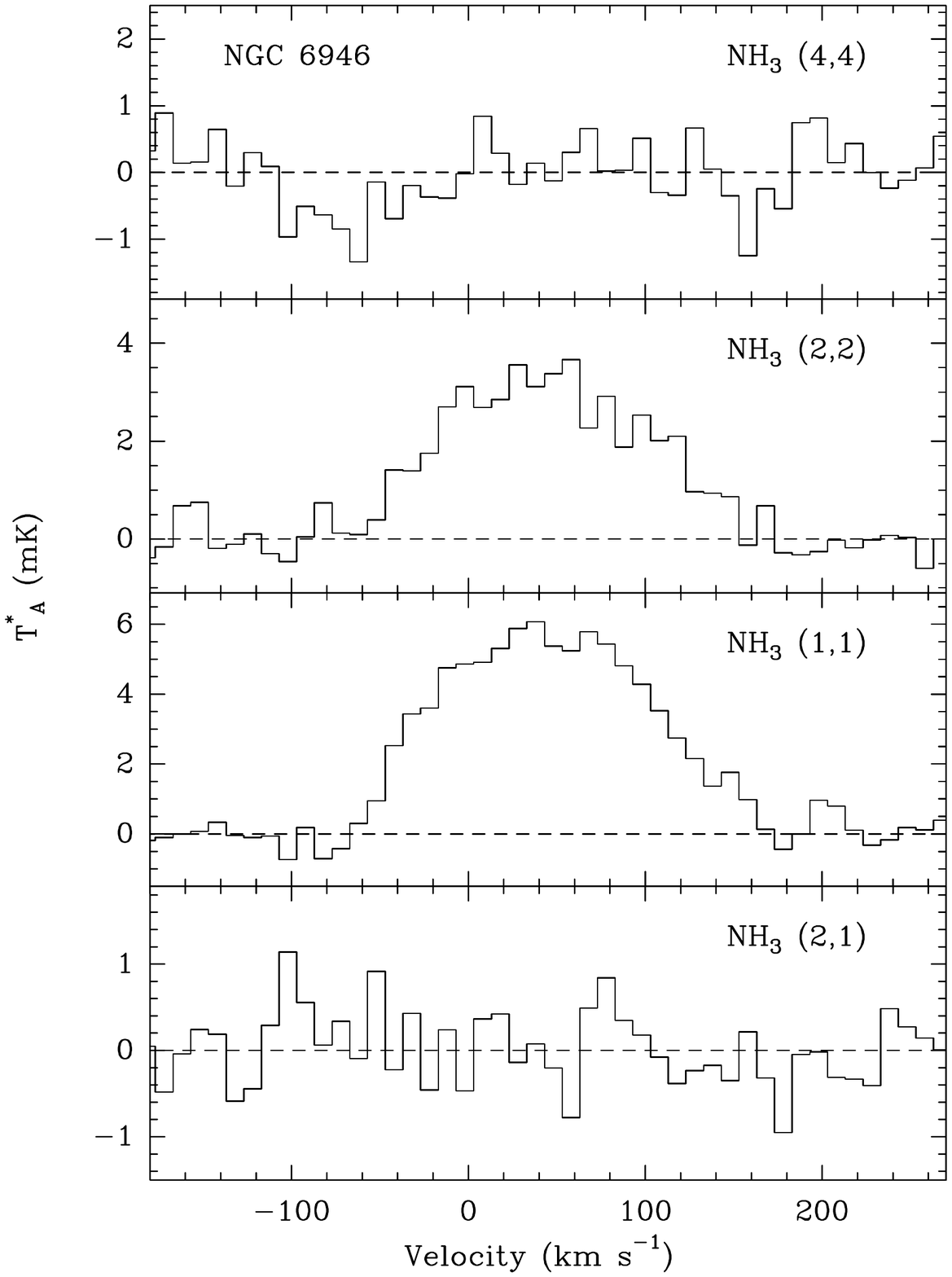}
\caption{NH$_3$ spectra of IR\,15107+0724 (left) and NGC\,6946
  (right).  For the IR\,15107+0724 NH$_3$ (1,1) and (2,2) spectrum the
velocity scale is referenced to the NH$_3$ (2,2) transition frequency.}
\label{fig:IR15107NGC6946NH3pec}
\end{figure*}

\begin{figure*}
\centering
\includegraphics[trim=20mm 15mm 20mm 30mm, clip, scale=0.45]{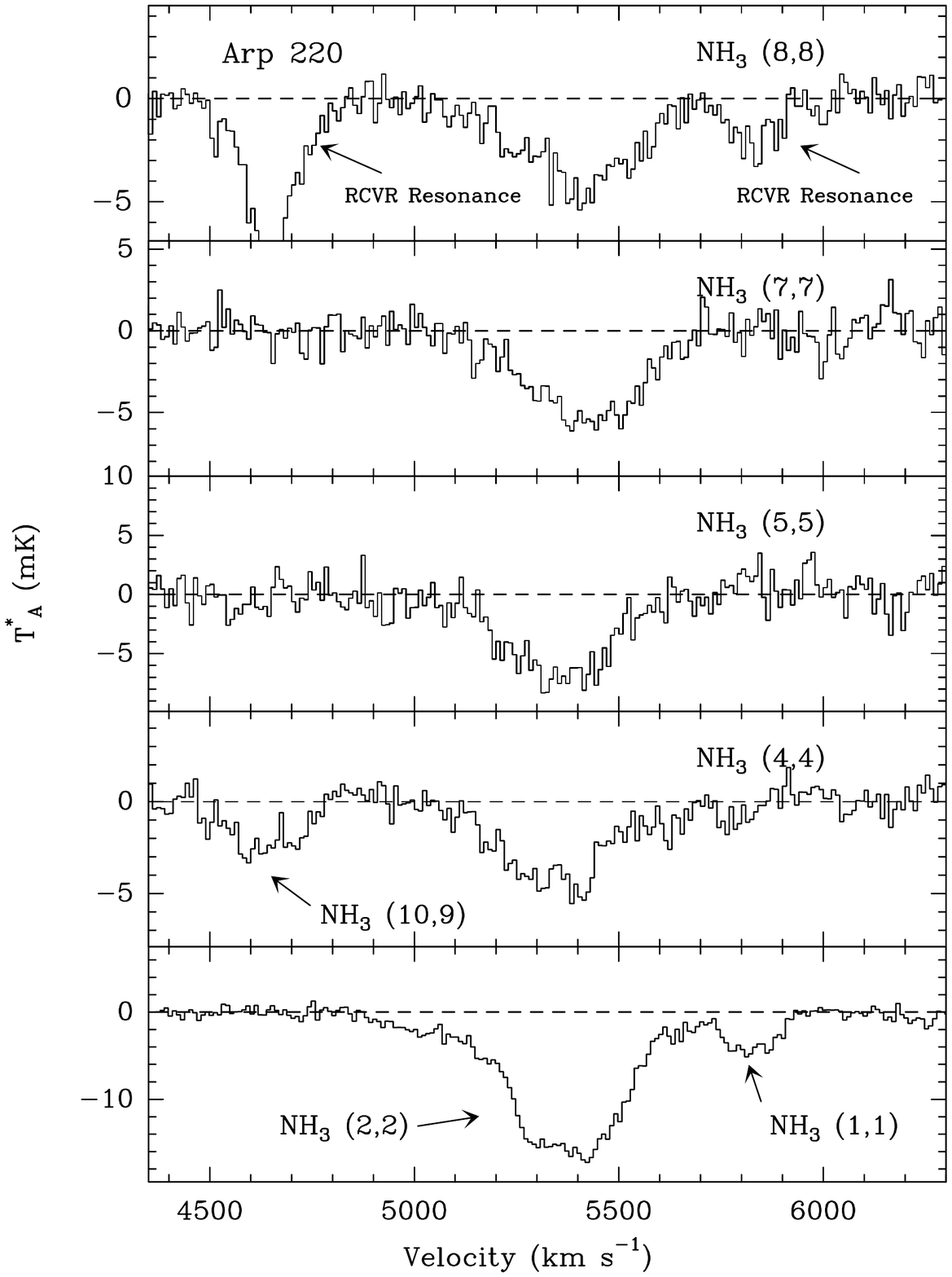}
\includegraphics[trim=20mm 15mm 20mm 30mm, clip, scale=0.45]{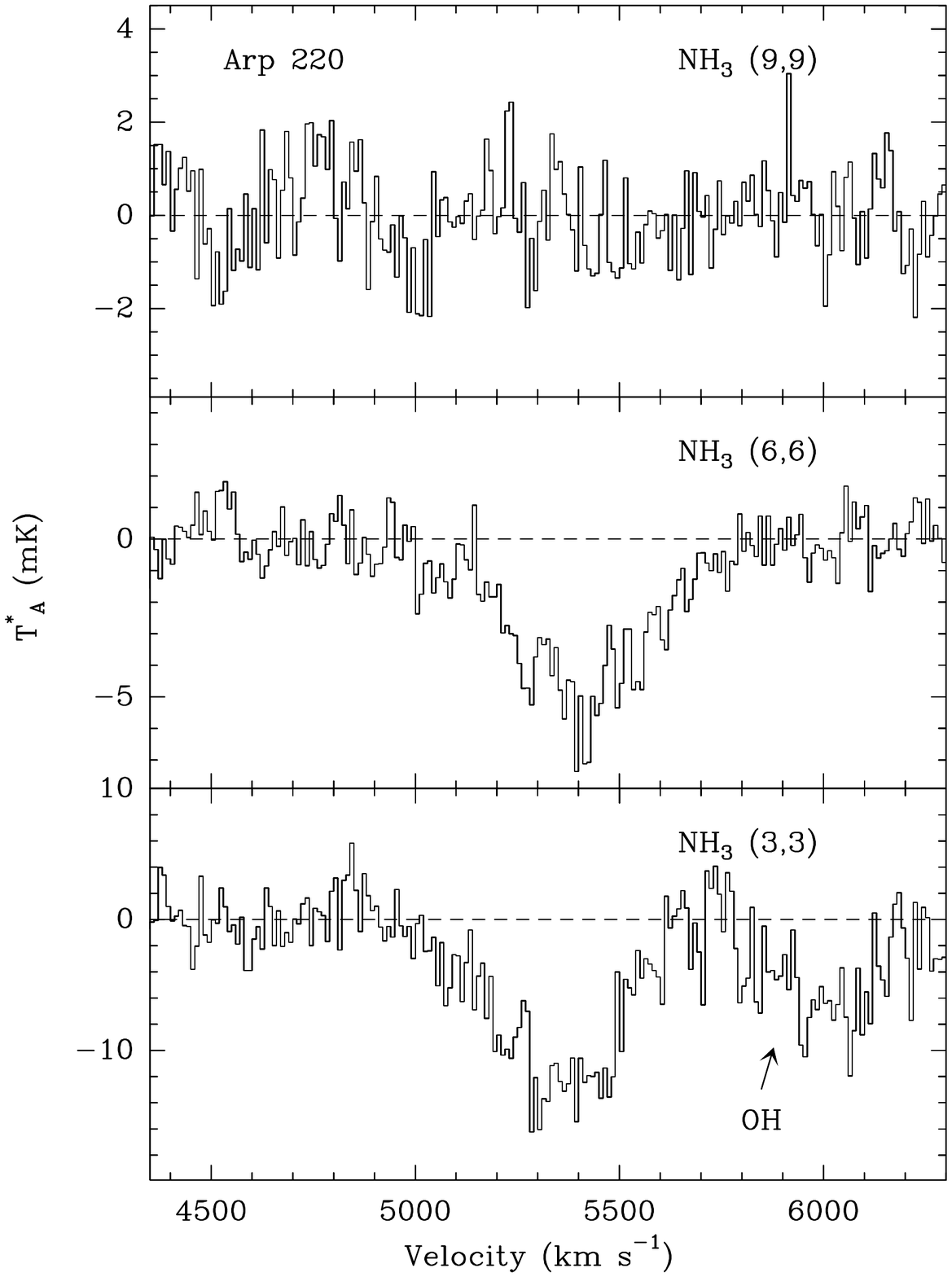}
\caption{Para-NH$_3$ (left panel) and ortho-NH$_3$ (right panel)
  spectra of Arp\,220. For the NH$_3$ (1,1) and (2,2) spectrum the
velocity scale is referenced to the NH$_3$ (2,2) transition frequency.}
\label{fig:Arp220NH3Spec}
\end{figure*}

In the following all spectra have been smoothed to either 5 or
10\,km\,s$^{-1}$ to increase the signal-to-noise ratio of individual
channels in these measurements.
Table~\ref{tab:nh3results} lists the peak intensity, velocity, line width,
and integrated intensity derived from both direct channel-by-channel
integration of the FWZI line profiles and Gaussian fits for those galaxies
with detected NH$_3$ emission or absorption.  The measured band-center
continuum level derived from the zero-level offset of our
spectroscopic measurements is also listed.  Spectra for the detected
galaxies are displayed in Figures~\ref{fig:NGC253NH3Spec} through
\ref{fig:Arp220NH3Spec}.  These spectra include the first
  detections of non-metastable NH$_3$ inversion transitions toward
  external galaxies in the (2,1) (NGC\,253, NGC\,660, IC\,342, and
  IC\,860), (3,1), (3,2), (4,3), (5,4) (all in NGC\,660) and (10,9)
  (Arp\,220) transitions.

%
%
\begin{deluxetable*}{lllrrrrr} 
\tablewidth{500pt}
\tablecolumns{8}
\tablecaption{Extragalactic NH$_3$ Measurements\tablenotemark{a}\label{tab:nh3results}}
\tablehead{
\colhead{Source} & 
\colhead{Transition} & 
\colhead{Fit\tablenotemark{b}} &
\colhead{$T^*_A$} & 
\colhead{v$_{hel}$\tablenotemark{c}} & 
\colhead{FW(HM/ZI)\tablenotemark{d}} & 
\colhead{$\int$T$^*_A$dv\tablenotemark{e}} &
\colhead{S$_{cont}$}\\
&&& \colhead{(mK)} & 
\colhead{(km s$^{-1}$)} & 
\colhead{(km s$^{-1}$)} & 
\colhead{(mK km s$^{-1}$)} &
\colhead{(mJy)}
}
\startdata
NGC\,253	      & (1,1) & D,10  & \nodata & 256.3 & 382.6 &
9347.2(92.9) & 467 \\ 
              &       & G1,10 & 26.2(1.5) & 176.1(2.2) & 93.9(5.5) &
2614.5(69.1) & \\
              &       & G2,10 & 63.0(1.5) & 306.6(0.9) & 97.4(2.4) &
6538.9(70.4) & \\

              & (2,2) & D,10  & \nodata & 263.5 & 368.2 & 7536.5(58.6) & \\
              &       & G1,10 & 15.2(1.0) &
176.1(0.0)\tablenotemark{f} & 93.9(0.0)\tablenotemark{f} & 1550.9(44.4) & \\
              &       & G2,10 & 51.5(1.0) & 309.9(1.4) & 108.7(3.8) &
5964.3(47.8) & \\ 

              & (3,3) & D,10 & \nodata & 260.3 & 373.3 & 15495.0(82.3) & \\ 	

              &       & G1,10 & 89.0(1.4) & 182.7(1.0) & 69.4(2.2) &
6578.6(53.3) & \\                                     
              &       & G2,10 & 97.1(1.4) & 291.6(1.1) & 100.8(2.9) &
10412.0(64.2) & \\

              & (4,4) & D,10  & \nodata & 237.4 & 328.5 & 1445.8(42.3) & \\
              &       & G1,10 & 3.1(0.7) & 176.1(0.0)\tablenotemark{f}
& 93.9(0.0)\tablenotemark{f} & 310.9(34.0) & \\
              &       & G2,10 & 16.5(0.7) & 305.0(1.7) & 63.3(3.8) &
1112.9(27.9) & \\

              & (5,5) & D,10  & \nodata & 220.5 & 302.4 & 1485.1(52.8) & \\
              &       & G1,10 & 12.1(1.0) & 177.7(2.3) & 102.2(6.5) &
1316.4(46.1) & \\
              &       & G2,10 & 15.5(1.0) & 296.2(1.5) & 66.6(3.5) &
1099.9(37.2) & \\

              & (6,6) & D,10  & \nodata & 246.0 & 335.8 & 3081.1(63.0) & \\
              &       & G1,10 & 16.4(1.1) & 183.1(1.4) & 63.5(3.7) &
1110.5(41.1) & \\
              &       & G2,10 & 22.2(1.1) & 298.5(1.2) & 81.2(3.3) &
1920.7(46.5) & \\

              & (7,7) & D,10  & \nodata & 241.8 & 315.0 & 929.0(34.5) & \\
              &       & G1,10 & 6.0(0.6) & 176.5(2.1) & 55.8(5.1) &
355.3(21.8) & \\
              &       & G2,10 & 7.4(0.60 & 295.1(1.9) & 75.7(5.6) &
593.5(25.4) & \\

              & (8,8) & D,10  & \nodata & 172.8 & 84.2 & 613.6(22.9) & \\
              &       & G1,10 & 5.0(0.8) & 180.0(2.9) & 42.2(5.6) &
225.4(24.4) & \\
              &       & G2,10 & 5.3(0.8) & 305.5(3.6) & 67.8(7.9) &
384.9(30.9) & \\
				      	
              & (9,9) & D,10  & \nodata & 235.5 & 211.4 & 543.7(45.6) & \\
              &       & G1,10 & 3.7(1.0) & 181.9(2.6) & 42.0(5.6) &
165.3(30.5) & \\
              &       & G2,10 & 5.3(1.0) & 290.6(2.3) & 65.4(5.4) &
366.6(38.0) & \\

              & (2,1) & D,10  & \nodata & 229.2 & 324.9 & 432.3(33.2) & \\
              &       & G2,10 & 4.2(0.8) & 290.9(8.0) & 86.0(26.4) &
381.8(33.2) & \\

              & (3,1) & \nodata & (4.4) & \nodata & \nodata & \nodata
& \\

              & (3,2) & \nodata & (1.3) & \nodata & \nodata & \nodata
& \\

              & (4,3) & \nodata & (1.4) & \nodata & \nodata & \nodata
& \\

              & (5,4) & \nodata & (1.6) & \nodata & \nodata & \nodata
& \\

              & HC$_3$N $3-2$ & D,10  & \nodata & 241.1 & 299.6 &
5859.2(47.1) & \\
              &                 & G1,10 & 30.3(0.9) & 187.2(1.2) &
69.1(0.0) & 2228.1(33.9) & \\
              &                 & G2,10 & 36.2(0.9) & 298.8(1.2) &
95.9(2.9) & 3698.6(39.9) & \\

NGC\,598      & (1,1) & D,10  & (1.1) & \nodata & \nodata & \nodata & 13 \\ 
              & (2,2) & D,10  & (1.3) & \nodata & \nodata & \nodata & \\ 
              & (4,4) & D,10  & (1.3) & \nodata & \nodata & \nodata & \\ 
              & (2,1) & D,10  & (1.2) & \nodata & \nodata & \nodata & \\

NGC\,660      & (1,1) & D,5 & \nodata & 811.0 & 175.1 & $-1102.7$(37.5) & 78 \\
              &       & G1,5 & -2.8(1.3) & 769.3(9.7) & 59.2(18.7) &
$-$176.9(32.7) & \\ 
              &       & G3,5 & $-$16.6(1.3) & 835.8(1.4) & 49.1(3.0) &
$-$865.2(29.8) & \\
              &       & G4,5 & $-$1.6(1.3) & 974.6(9.6) & 59.8(17.9) &
$-$102.7(32.9) & \\

              & (2,2) & D,5  & \nodata & 800.9 & 212.2 & $-1384.6$(46.6) & \\
              &       & G1,5 & $-3.5$(1.4) &
765.0(0.0)\tablenotemark{f} & 64.7(15.1) & $-243.7$(38.6) & \\
              &       & G2,5 & $-2.2$(1.4) &
800.0(0.0)\tablenotemark{f} & 26.5(18.7) & $-62.5$(24.7) & \\
              &       & G3,5 & $-17.1$(1.4) & 837.9(2.0) & 53.8(2.7) &
$-978.1$(35.2) & \\ 
              &       & G4,5 & $-1.4$(1.4) & 975.3(15.5) & 82.3(34.7) &
$-122.6$(43.5) & \\ 

              & (3,3) & D,5  & \nodata & 821.6 & 187.8 & $-1953.1$(25.9) & \\
              &       & G1,5 & \nodata & \nodata & \nodata & \nodata & \\
              &       & G2,5 & $-7.9$(0.8) & 785.0(1.8) & 33.3(3.9) &
$-280.4$(16.0) & \\
              &       & G3,5 & $-20.3$(0.8) & 831.2(1.2) & 54.6(2.0) &
$-1519.0$(20.5) & \\
              &       & G4,5 & $-3.2$(0.8) & 979.0(2.7) & 37.4(7.1) &
$-126.7$(17.0) & \\

              & (4,4) & D,5  & \nodata & 786.1 & 205.9 & $-863.9$(32.3) & \\
              &       & G1,5 & $-1.7$(1.0) &
765.0(0.0)\tablenotemark{f} & 88.1(43.7) & $-157.1$(31.7) & \\ 
              &       & G2,5 & $-5.4$(1.0) &
795.0(0.0)\tablenotemark{f} & 36.7(9.7) & $-210.7$(20.4) & \\ 
              &       & G3,5 & $-13.6$(1.0) & 838.8(1.3) & 36.3(2.5) &
$-525.6$(20.3) & \\

              & (5,5) & D,5  & \nodata & 825.3 & 178.3 & $-1042.4$(28.2) & \\
              &       & G1,5 & \nodata & \nodata & \nodata & \nodata & \\
              &       & G2,5 & $-4.0$(1.0) & 786.2(2.6) & 30.1(4.9) &
$-129.3$(17.7) & \\
              &       & G3,5 & $-12.8$(1.0) & 829.1(1.7) & 57.4(4.4) &
$-545.0$(24.4) & \\
              &       & G4,5 & $-1.4$(1.0) & 988.8(6.7) & 40.5(28.7) &
$-41.9$(20.5) & \\

              & (6,6) & D,5  & \nodata & 888.1 & 324.8 & $-1283.0$(31.0) & \\
              &       & G1,5 & \nodata & \nodata & \nodata & \nodata & \\
              &       & G2,5 & $-6.4$(0.8) & 792.0(0.0)\tablenotemark{f} & 30.0(0.0)\tablenotemark{f} &
$-204.9$(14.1) & \\
              &       & G3,5 & $-13.1$(0.8) & 830.0(0.0)\tablenotemark{f} & 55.0(0.0)\tablenotemark{f} &
$-769.5$(19.1) & \\
              &       & G4,5 & $-3.6$(0.8) & 985.9(3.7) & 58.8(8.7) &
$-226.6$(19.8) & \\

\enddata
\end{deluxetable*} 

\addtocounter{table}{-1}
\begin{deluxetable*}{lllrrrrr} 
\tablewidth{500pt}
\tablecolumns{8}
\tablecaption{\vspace{-8.3pt} \hspace{90pt} --- {\it Continued}}
\tablehead{
\colhead{Source} & 
\colhead{Transition} & 
\colhead{Fit\tablenotemark{b}} &
\colhead{$T^*_A$} & 
\colhead{v$_{hel}$\tablenotemark{c}} & 
\colhead{FW(HM/ZI)\tablenotemark{d}} & 
\colhead{$\int$T$^*_A$dv\tablenotemark{e}} &
\colhead{S$_{cont}$}\\
&&& \colhead{(mK)} & 
\colhead{(km s$^{-1}$)} & 
\colhead{(km s$^{-1}$)} & 
\colhead{(mK km s$^{-1}$)} &
\colhead{(mJy)}
}
\startdata
NGC\,660      & (7,7) & D,5  & \nodata & 837.0 & 141.1 & $-277.2$(18.1) & \\
              &       & G1,5 & \nodata & \nodata & \nodata & \nodata & \\
              &       & G2,5 & \nodata & \nodata & \nodata & \nodata & \\
              &       & G3,5 & $-3.4$(0.7) & 825.6(3.1) & 66.1(7.1) &
$-241.5$(18.6) & \\
              &       & G4,5 & \nodata & \nodata & \nodata & \nodata & \\

              & (8,8) & D,5  & (2.2) & \nodata & \nodata & \nodata & \\

              & (9,9) & D,5  & (3.5) & \nodata & \nodata & \nodata & \\

              & (2,1) & D,5  & \nodata & 828.0 & 143.3 & $-686.4$(37.0) & \\
              &       & G1,5 & \nodata & \nodata & \nodata & \nodata & \\
              &       & G2,5 & $-2.8$(1.4) & 789.5(4.6) & 27.4(17.2) &
$-82.5$(24.3) & \\ 
              &       & G3,5 & $-12.7$(1.4) & 838.1(1.4) & 42.7(3.7) &
$-577.9$(30.3) & \\ 
              &       & G4,5 & $-1.3$(1.4) &
970.0(0.0)\tablenotemark{f} & 40.0(0.0)\tablenotemark{f} &
$-55.9$(29.4) & \\

              & (3,1) & D,5  & \nodata & 839.5 & 132.0 & $-782.7$(27.2) & \\
              &       & G1,5 & \nodata & \nodata & \nodata & \nodata & \\
              &       & G2,5 & $-2.9$(1.0) & 789.0(0.0)\tablenotemark{f} & 35.0(0.0)\tablenotemark{f} & $-108.0$(36.0) & \\
              &       & G3,5 & $-10.6$(1.0) & 827.7(1.6) & 46.7(4.5) &
$-527.5$(42.0) & \\
              &       & G4,5 & \nodata & \nodata & \nodata & \nodata & \\

              & (3,2) & D,5  & \nodata & 845.8 & 325.0 & $-563.9$(26.4) & \\
              &       & G1,5 & \nodata & \nodata & \nodata & \nodata & \\
              &       & G2,5 & $-3.9$(1.0) & 789.0(0.0)\tablenotemark{f} & 35.0(0.0)\tablenotemark{f} & $-143.6$(24.0) & \\ 
              &       & G3,5 & $-7.3$(1.0) & 829.1(1.7) & 44.4(4.0) &
$-346.7$(27.0) & \\
              &       & G4,5 & $-2.5$(1.0) & 980.8(3.7) & 39.1(6.5) &
$-105.0$(18.0) & \\

              & (4,3) & D,5  & \nodata & 854.7 & 325.0 & $-646.0$(29.1) & \\
              &       & G1,5 & \nodata & \nodata & \nodata & \nodata & \\
              &       & G2,5 & $-3.9$(1.1) & 789.0(0.0)\tablenotemark{f} & 35.0(0.0)\tablenotemark{f} & $-146.2$(29.0) & \\
              &       & G3,5 & $-9.0$(1.1) & 831.9(1.7) & 34.5(4.1) & $-329.3$(33.0) & \\
              &       & G4,5 & $-3.9$(1.1) & 988.0(3.7) & 32.2(7.9) & $-133.0$(29.0) & \\

              & (5,4) & D,5  & \nodata & 830.0 & 95.0 & $-148.3$(21.6) & \\
              &       & G1,5 & \nodata & \nodata & \nodata & \nodata & \\
              &       & G2,5 & \nodata & \nodata & \nodata & \nodata & \\
              &       & G3,5 & $-2.4$(1.0) & 830.0(0.0)\tablenotemark{f} & 40.0(0.0)\tablenotemark{f} & $-102.4$(27.0) & \\
              &       & G4,5 & \nodata & \nodata & \nodata & \nodata & \\

              & HC$_3$N $3-2$ & D,5  & (2.7) & \nodata & \nodata & \nodata & \\

IR\,01418$+$1651& (1,1) & D,10  & (1.6) & \nodata & \nodata & \nodata &
13 \\ 
               & (2,2) & D,10  & (1.6) & \nodata & \nodata & \nodata & \\ 
               & (4,4) & D,10  & (1.5) & \nodata & \nodata & \nodata & \\ 
               & (2,1) & D,10  & (1.6) & \nodata & \nodata & \nodata & \\

NGC\,891      & (1,1) & D,5  & \nodata & 540.5 & 94.4 & 124.3(24.3) & 44 \\
              &       & G1,5 & 2.5(1.1) & 543.9(3.4) & 45.3(7.7) &
120.2(25.0) & \\ 
              & (2,2) & D,5  & (1.1) & \nodata & \nodata & \nodata & \\ 
              & (4,4) & D,5  & (1.0) & \nodata & \nodata & \nodata & \\ 
              & (2,1) & D,5  & (1.1) & \nodata & \nodata & \nodata & \\

Maffei\,2     & (1,1) & D,10  & \nodata & $-32.0$ & 328.8 & 3068.5(36.8) &
35 \\
              &       & G1,10 & 21.5(0.6) & $-68.5$(1.7) &
82.0(4.5) & 1878.0(27.6) & \\
              &       & G2,10 & 15.8(0.6) & 27.4(2.2) &
70.6(5.0) & 1269.0(25.6) & \\

              & (2,2) & D,10  & \nodata & $-28.3$ & 251.8 & 1779.8(45.0) & \\
              &       & G1,10 & 12.9(0.9) & $-68.1$(2.0) &
69.5(4.8) & 955.9(35.5) & \\
              &       & G2,10 & 9.3(0.9) & 21.2(3.1) &
78.0(6.6) & 771.4(37.6) & \\

              & (4,4) & D,10  & \nodata & $-34.8$ & 247.4 & 398.6(29.4) & \\
              &       & G1,10 & 4.4(0.6) & $-70.6$(2.2) & 47.2(6.5) &
221.9(19.3) \\
              &       & G2,10 & 2.2(0.6) & 34.6(5.9) & 73.9(14.1) &
398.6 & \\ 

              & (2,1) & D,10  & (0.7) & \nodata & \nodata & \nodata & \\

NGC\,1144     & (1,1) & D,10  & (1.7) & \nodata & \nodata & \nodata & 42 \\ 
              & (2,2) & D,10  & (1.6) & \nodata & \nodata & \nodata & \\ 
              & (4,4) & D,10  & (1.7) & \nodata & \nodata & \nodata & \\ 
              & (2,1) & D,10  & (2.7) & \nodata & \nodata & \nodata & \\

NGC\,1365     & (1,1) & D,10  & \nodata & 1646.0 & 363.2 & 1972.7(80.0) & 40 \\
              &       & G1,10 & 8.5(1.3) & 1573.6(2.7) &
112.0(7.0) & 1014.0(66.6) & \\
              &       & G2,10 & 8.8(1.3) & 1739.3(2.6) &
110.5(6.8) & 1040.9(66.2) & \\

              & (2,2) & D,10  & \nodata & 1630.1 & 302.2 & 1255.7(68.9) & \\
              &       & G1,10 & 7.4(1.3) & 1572.7(3.6) &
116.1(9.8) & 918.4(64.0) & \\
              &       & G2,10 & 4.4(1.3) & 1721.8(5.5) &
85.0(10.5) & 397.2(55.0) & \\

              & (4,4) & D,10  & \nodata & 1753.9 & 122.1 & 157.3(66.6) & \\
              &       & G1,10 & 1.7(1.5) & 1758.6(10.6) &
85.0(0.0)\tablenotemark{f} & 157.3(66.6) & \\

              & (2,1) & D,10  & (1.6) & \nodata & \nodata & \nodata & \\

IC\,342       & (1,1) & D,5  & \nodata & 18.3 & 186.4 & 2191.8(37.3) & 75 \\
              &       & G1,5 & 33.3(1.2) & 25.3(2.8) &
51.1(3.3) & 1810.6(29.2) & \\
              &       & G2,5 & 9.1(1.2) & 55.3(4.0) &
36.7(5.2) & 354.2(24.8) & \\

              & (2,2) & D,5  & \nodata & 22.7 & 150.7 & 1135.9(28.6) & \\
              &       & G1,5 & 17.7(1.0) & 19.1(1.7) &
36.7(2.5) & 691.7(21.2) & \\
              &       & G2,5 & 13.0(1.0) & 48.6(1.9) &
31.4(2.7) & 434.8(19.6) & \\

\enddata
\end{deluxetable*} 

\addtocounter{table}{-1}
\begin{deluxetable*}{lllrrrrr} 
\tablewidth{500pt}
\tablecolumns{8}
\tablecaption{\vspace{-8.3pt} \hspace{90pt} --- {\it Continued}}
\tablehead{
\colhead{Source} & 
\colhead{Transition} & 
\colhead{Fit\tablenotemark{b}} &
\colhead{$T^*_A$} & 
\colhead{v$_{hel}$\tablenotemark{c}} & 
\colhead{FW(HM/ZI)\tablenotemark{d}} & 
\colhead{$\int$T$^*_A$dv\tablenotemark{e}} &
\colhead{S$_{cont}$}\\
&&& \colhead{(mK)} & 
\colhead{(km s$^{-1}$)} & 
\colhead{(km s$^{-1}$)} & 
\colhead{(mK km s$^{-1}$)} &
\colhead{(mJy)}
}
\startdata
IC\,342       & (4,4) & D,5  & \nodata & 28.6 & 98.4 & 371.7(21.2) & \\
              &       & G1,5 & 6.1(1.0) & 25.8(2.2) &
41.3(4.5) & 267.5(21.0) & \\
              &       & G2,5 & 2.8(1.0) & 52.1(3.8) &
31.0(0.0)\tablenotemark{f} & 92.4(19.0) & \\

              & (2,1) & D,5  & \nodata & 43.9 & 76.4 & 81.1(19.1) & \\
              &       & G2,5 & 2.4(1.0) & 48.7(2.9) &
31.7(5.7) & 81.1(19.1) & \\

              & (3,3) & D,5  & \nodata & 32.4 & 149.6 & 2106.4(35.5) & \\
              &       & G1,5 & 26.1(1.3) & 20.9(1.1) &
41.9(3.7) & 1165.7(28.0) & \\
              &       & G2,5 & 24.7(1.3) & 48.7(1.0) &
33.5(3.0) & 880.0(25.0) & \\

              & (5,5) & D,5  & \nodata & 39.0 & 81.6 & 244.2(22.1) & \\
              &       & G1,5 & 5.8(1.1) & 25.3(0.0)\tablenotemark{f} &
28.7(3.9) & 175.7(19.7) & \\
              &       & G2,5 & 3.7(1.1) & 55.3(0.0)\tablenotemark{f} &
23.1(4.2) & 92.0(17.7) & \\

              & (6,6) & D,5  & \nodata & 45.0 & 125.22 & 375.5(30.3) & \\
              &       & G1,5 & 3.4(1.2) & 20.9(0.0)\tablenotemark{f} &
49.2(8.6) & 176.3(28.6) & \\
              &       & G2,5 & 3.7(1.2) & 48.7(0.0)\tablenotemark{f} &
52.6(8.9) & 207.5(29.5) & \\

              & (7,7) & D,5  & (1.5) & \nodata & \nodata & \nodata & \\

NGC\,2146     & (1,1) & D,10  & (0.8) & \nodata & \nodata & \nodata &
130 \\ 
              & (2,2) & D,10  & (0.8) & \nodata & \nodata & \nodata & \\ 
              & (4,4) & D,10  & (0.9) & \nodata & \nodata & \nodata & \\ 
              & (2,1) & D,10  & (0.8) & \nodata & \nodata & \nodata & \\

M\,82         & (1,1) & D,10  & (3.7) & \nodata & \nodata & \nodata &
731 \\ 
              & (2,2) & D,10  & (3.0) & \nodata & \nodata & \nodata & \\ 
              & (4,4) & D,10  & (2.8) & \nodata & \nodata & \nodata & \\ 
              & (2,1) & D,10  & (2.5) & \nodata & \nodata & \nodata & \\

M\,82SW       & (1,1) & D,10  & \nodata & 118.1 & 150.7 & 840.1(52.1) & 776 \\
              &       & G1,10 & 8.0(1.1) & 119.9(2.8) &
104.5(6.5) & 888.2(52.1) & \\

              & (2,2) & D,10  & \nodata & 130.7 & 177.9 & 561.4(71.5) & \\
              &       & G1,10 & 4.7(1.7) & 132.2(6.3) &
122.1(15.3) & 607.5(88.9) & \\
              & (4,4) & D,10  & (1.0) & \nodata & \nodata & \nodata & \\

              & (2,1) & D,10  & (1.3) & \nodata & \nodata & \nodata & \\

NGC\,3079     & (1,1) & D,5  & \nodata & 1087.8 & 229.2 & $-1941.5$(40.2) &
174 \\
              &       & G1,5 & $-6.4$(1.2) & 1024.4(2.7) &
56.0(6.9) & $-381.0$(29.8) & \\
              &       & G2,5 & $-34.8$(1.2) & 1115.9(0.4) &
40.8(1.1) & $-1512.3$(25.5) & \\

              & (2,2) & D,5  & \nodata & 1077.7 & 260.0 & $-2326.1$(39.0) & \\
              &       & G1,5 & $-8.0$(1.1) & 1017.4(2.7) &
65.9(7.2) & $-563.5$(29.5) & \\
              &       & G2,5 & $-42.4$(1.1) & 1114.7(0.4) &
37.0(1.0) & $-1673.0$(22.1) & \\

              & (4,4) & D,5  & \nodata & 1069.7 & 218.6 & $-1831.2$(43.4) & \\
              &       & G1,5 & $-6.5$(1.3) & 1019.9(3.6) &
64.5(10.0) & $-447.3$(35.3) & \\
              &       & G2,5 & $-36.6$(1.3) & 1115.3(0.5) &
35.1(1.2) & $-1366.6$(26.1) & \\

              & (2,1) & D,5  & (1.5) & \nodata & \nodata & \nodata & \\

              & (3,3) & D,5  & (1.2) & \nodata & \nodata & \nodata & \\

              & (5,5) & D,5  & \nodata & 1073.2 & 193.6 & $-1031.1$(39.4) & \\
              &       & G1,5 & $-4.6$(1.3) & 1016.9(3.1) & 44.5(7.5) &
$-216.2$(28.4) & \\
              &       & G2,5 & $-22.5$(1.3) & 1115.2(0.5) & 33.9(1.5)
& $-812.1$(24.7) & \\

              & (6,6) & D,5  & \nodata & 1058.4	& 192.5 & $-830.9$(54.6) & \\
              &       & G1,5 & $-2.5$(1.8) & 1000.5(8.9) & 59.7(17.9)
& $-161.5$(45.7) & \\
              &       & G2,5 & $-25.7$(1.8) & 1116.4(0.5) & 28.3(1.2)
& $-775.2$(31.5) & \\

              & (7,7) & D,5  & \nodata & 1069.9 & 228.6 & $-884.8$(33.2) & \\
              &       & G1,5 & $-4.1$(1.0) & 1021.1(6.9) & 79.5(17.7)
& $-347.1$(29.4) & \\
              &       & G2,5 & $-13.6$(1.0) & 1116.5(1.4) & 36.6(4.0)
& $-529.7$(20.0) & \\

              & (8,8) & D,5  & (0.8) & \nodata & \nodata & \nodata & \\

              & (9,9) & D,5  & \nodata & 1074.3 & 187.0 & $-532.8$(43.5) & \\
              &       & G1,5 & $-3.6$(1.4) & 1028.4(5.7) & 40.6(16.9)
& $-155.4$(30.4) & \\
              &       & G2,5 & $-10.4$(1.4) & 1117.3(1.7) & 31.0(5.2)
& $-342.4$(26.5) & \\

              & HC$_3$N $3-2$ & D,5  & (1.4) & \nodata & \nodata
& \nodata & \\

              & (3,1) & D,5  & (1.4) & \nodata & \nodata & \nodata & \\

              & (4,3) & D,5  & (1.2) & \nodata & \nodata & \nodata & \\

              & (5,4) & D,5  & (1.1) & \nodata & \nodata & \nodata & \\

              & (3,2) & D,5  & (1.1) & \nodata & \nodata & \nodata & \\

NGC\,3628     & (1,1) & D,10  & (0.9) & \nodata & \nodata & \nodata &
138 \\ 
              & (2,2) & D,10  & (1.1) & \nodata & \nodata & \nodata & \\ 
              & (4,4) & D,10  & (1.1) & \nodata & \nodata & \nodata & \\ 
              & (2,1) & D,10  & (1.3) & \nodata & \nodata & \nodata & \\

NGC\,3690     & (1,1) & D,10  & (2.1) & \nodata & \nodata & \nodata &
147 \\ 
              & (2,2) & D,10  & (1.6) & \nodata & \nodata & \nodata & \\ 
              & (4,4) & D,10  & (1.6) & \nodata & \nodata & \nodata & \\ 
              & (2,1) & D,10  & (1.8) & \nodata & \nodata & \nodata & \\

NGC\,4214     & (1,1) & D,10  & (1.7) & \nodata & \nodata & \nodata &
147 \\ 
              & (2,2) & D,10  & (1.3) & \nodata & \nodata & \nodata & \\ 
              & (4,4) & D,10  & (1.4) & \nodata & \nodata & \nodata & \\ 
              & (2,1) & D,10  & (1.5) & \nodata & \nodata & \nodata & \\ 

\enddata
\end{deluxetable*} 

\addtocounter{table}{-1}
\begin{deluxetable*}{lllrrrrr} 
\tablewidth{500pt}
\tablecolumns{8}
\tablecaption{\vspace{-8.3pt} \hspace{90pt} --- {\it Continued}}
\tablehead{
\colhead{Source} & 
\colhead{Transition} & 
\colhead{Fit\tablenotemark{b}} &
\colhead{$T^*_A$} & 
\colhead{v$_{hel}$\tablenotemark{c}} & 
\colhead{FW(HM/ZI)\tablenotemark{d}} & 
\colhead{$\int$T$^*_A$dv\tablenotemark{e}} &
\colhead{S$_{cont}$}\\
&&& \colhead{(mK)} & 
\colhead{(km s$^{-1}$)} & 
\colhead{(km s$^{-1}$)} & 
\colhead{(mK km s$^{-1}$)} &
\colhead{(mJy)}
}
\startdata
NGC\,4418     & (1,1) & D,10  & (0.6) & \nodata & \nodata & \nodata &
147 \\ 
              & (2,2) & D,10  & (0.6) & \nodata & \nodata & \nodata & \\ 
              & (4,4) & D,10  & (0.5) & \nodata & \nodata & \nodata & \\ 
              & (2,1) & D,10  & (0.4) & \nodata & \nodata & \nodata & \\

Mrk\,231      & (1,1) & D,10  & (2.6) & \nodata & \nodata & \nodata &
173 \\ 
              & (2,2) & D,10  & (2.3) & \nodata & \nodata & \nodata & \\ 
              & (4,4) & D,10  & (2.6) & \nodata & \nodata & \nodata & \\ 
              & (2,1) & D,10  & (2.6) & \nodata & \nodata & \nodata & \\

IC\,860        & (1,1) & D,10  & \nodata & 3894.2 & 237.9 & $-305.4$(24.1) &
27 \\
              &       & G1,10 & $-3.3$(0.5) & 3901.6(5.9) &
86.3(14.6) & $-301.9$(21.8) & \\

              & (2,2) & D,10  & \nodata & 3837.8 & 316.9 & $-829.8$(22.6) & \\
              &       & G1,10 & $-5.8$(0.4) & 3874.0(3.8) &
129.7(9.6) & $-797.4$(21.7) & \\

              & (4,4) & D,10  & \nodata & 3846.3 & 262.2 & $-430.8$(25.7) & \\
              &       & G1,10 & $-3.0$(0.5) & 3881.0(6.1) &
136.4(15.7) & $-430.8$(25.7) & \\

              & (2,1) & D,10  & \nodata & 3935.2 & 339.0 & $-390.5$(38.8) & \\
              &       & G1,10 & $-2.2$(0.7) & 3908.1(7.8) &
168.0(20.2) & $-397.0$(40.9) & \\

              & (3,3) & D,10  &	\nodata & 3851.5 & 247.4 & $-938.6$(63.8) & \\ 
              &       & G1,10 & $-8.8$(1.3) & 3881.7(2.5) & 103.4(5.8)
& $-965.2$(32.2) & \\
              &       & G2,10 & $-2.4$(1.3) & 3761.6(4.8) & 27.9(10.8)
& $-71.0$(61.9) & \\

              & (5,5) & D,10  & \nodata & 3829.5 & 236.2 & $-529.2$(140.3) & \\
              &       & G1,10 & $-5.3$(2.9) & 3864.8(6.6) & 119.4(16.8) & $-671.0$(149.6) & \\
              &       & G2,10 & $-2.6$(2.9) & 729.3(6.2) &
28.0(0.0)\tablenotemark{f} & $-78.7$(72.3) & \\

              & (6,6) & D,10  & \nodata & 3913.3 & 193.2 & $-538.4$(46.4) & \\
              &       & G1,10 & $-6.4$(1.1) & 3904.9(3.0) & 75.9(7.4)
& $-513.5$(43.7) & \\

              & (7,7) & D,10  & \nodata & 3908.4 & 181.2 & $-194.4$(54.5) & \\
              &       & G1,10 & $-3.3$(1.3) & 3894.9(6.9) & 65.8(15.1)
& $-220.2$(49.3) & \\

M\,83         & (1,1) & D,10  & \nodata & 505.6 & 194.8 & 897.4(46.4) & 76 \\
              &       & G1,10 & 10.6(1.1) & 500.1(1.7) & 79.3(4.2) &
897.4(46.4) & \\ 

              & (2,2) & D,10  & \nodata & 533.5 & 181.1 & 606.6(37.1) & \\
              &       & G1,10 & 6.3(0.8) & 507.6(3.5) &
90.5(9.2) & 606.6(37.1) & \\

              & (4,4) & D,10  & (1.1) & \nodata & \nodata & \nodata & \\

              & (2,1) & D,10  & (0.8) & \nodata & \nodata & \nodata & \\

IR\,15107+0724 & (1,1) & D,10  & \nodata & 3974.6 & 386.8 & $-502.4$(55.9) & 15.4 \\
               &       & G1,10 & $-2.1$(1.0) & 3946.2(17.5) & 223.0(0.0)\tablenotemark{f} & $-490.0$(74.0) & \\

               & (2,2) & D,10  & \nodata & 3900.0 & 400.0 & $-1638.6$(67.7) & \\
               &       & G1,10 & $-6.8$(1.0) & 3905.6(5.4) &
223.1(12.9) & $-1597.5$(80.0) & \\    			    	

               & (4,4) & D,10  & \nodata & 3989.8 & 281.1 & $-240.9$(55.1) & \\
               &       & G1,10 & $-1.5$(1.0) & 3976.5(25.9) & 223.0(0.0)\tablenotemark{f} & $-351.4$(83.0) & \\

Arp\,220      & (1,1) & D,10  & \nodata & 5310.0 & $\sim300.0$ & $-860.5$(72.9) & \\	
              &       & G1,10 & $-0.2$(0.7) & 5290.0(0.0) & 150.0(0.0)
& $-34.7$(38.8) & 140.0 \\
              &       & G2,10 & $-6.5$(0.7) & 5460.0(0.0) &
147.7(13.8) & $-730.3$(38.5) & \\

              & (2,2) & D,10  & \nodata & 5231.0 & 869.0 & $-6338.3$(72.9) & \\
              &       & G1,10 & $-10.8$(0.7) & 5290.0(0.0) &
176.1(10.0) & $-2020.8$(42.0) & \\
              &       & G2,10 & $-13.0$(0.7) & 5460.0(0.0) &
189.0(8.1) & $-2607.6$(43.5) & \\
              &       & G3,10 & $-2.4$(0.7) & 5200.0(0.0) &
534.3(73.1) & $-1370.6$(73.3) & \\

              & (4,4) & D,10  & \nodata & 5363.6 & 715.9 & $-1502.6$(65.6) & \\
              &       & G1,10 & $-3.4$(0.8) & 5290.0(0.0) &
241.8(17.3) & $-879.9$(57.0) & \\
              &       & G2,10 & $-4.8$(0.8) & 5460.0(0.0) &
308.2(17.3) & $-669.0$(64.4) & \\

              & (10,9) & G2,10 & $-2.7$(0.8) & 5449.8(8.4) &
198.0(19.3) & $-568.4$(51.6) & \\

              & (3,3) & D,10  & \nodata & 5310.0 & 710.0 & $-4827.3$(212.2) &
\\
              &       & G1,10 & $-9.7$(2.3) & 5292.7(44.0) &
325.0(46.2) & $-3352.5$(193.7) & \\
              &       & G2,10 & $-6.6$(2.3) & 5437.6(31.7) &
205.4(58.5) & $-1432.5$(154.0) & \\

              & OH23827 & G2,10 & $-7.2$(2.3) & 5464.1(13.2) &
288.4(31.4) & $-2213.1$(213.1) & \\

              & (5,5) & D,10  & \nodata & 5350.4 & 610.5 & $-2211.7$(107.4) &
\\
              &       & G1,10 & $-6.2$(1.4) & 5290.0(0.0) &
221.6(18.0) & $-1453.9$(97.1) & \\
              &       & G2,10 & $-4.2$(1.4) & 5460.0(0.0) &
179.4(22.2) & $-795.2$(87.4) & \\

              & (6,6) & D,10  & \nodata & 5394.4 & 882.8 & $-2079.1$(70.5) &
\\
              &       & G1,10 & $-1.9$(0.8) & 5290.0(0.0) &
357.4(66.5) & $-728.6$(67.3) & \\
              &       & G2,10 & $-4.0$(0.8) & 5460.0(0.0) &
331.2(24.6) & $-1401.1$(64.8) & \\

              & (7,7) & D,10  & \nodata & 5396.6 & 685.2 & $-1978.1$(79.5) &
\\
              &       & G1,10 & $-2.1$(1.0) & 5290.0(0.0) &
246.2(35.7) & $-541.1$(71.5) & \\
              &       & G2,10 & $-5.0$(1.0) & 5460.0(0.0) &
248.0(14.4) & $-1330.4$(71.8) & \\

              & (8,8) & D,10  & \nodata & 5396.6 & 430.4 & $-1429.4$(38.8) &
\\
              &       & G1,10 & $-2.0$(0.6) & 5290.0(0.0) &
286.1(33.9) & $-622.8$(47.7) & \\
              &       & G2,10 & $-3.4$(0.6) & 5460.0(0.0) &
237.5(17.7) & $-857.8$(43.4) & \\

              & (9,9) & D,10  & (1.0) & \nodata & \nodata & \nodata & \\

\enddata
\end{deluxetable*} 

\addtocounter{table}{-1}
\begin{deluxetable*}{lllrrrrr} 
\tablewidth{500pt}
\tablecolumns{8}
\tablecaption{\vspace{-8.3pt} \hspace{90pt} --- {\it Continued}}
\tablehead{
\colhead{Source} & 
\colhead{Transition} & 
\colhead{Fit\tablenotemark{b}} &
\colhead{$T^*_A$} & 
\colhead{v$_{hel}$\tablenotemark{c}} & 
\colhead{FW(HM/ZI)\tablenotemark{d}} & 
\colhead{$\int$T$^*_A$dv\tablenotemark{e}} &
\colhead{S$_{cont}$}\\
&&& \colhead{(mK)} & 
\colhead{(km s$^{-1}$)} & 
\colhead{(km s$^{-1}$)} & 
\colhead{(mK km s$^{-1}$)} &
\colhead{(mJy)}
}
\startdata
NGC\,6946     & (1,1) & D,10  & \nodata & 49.2 & 265.9 & 859.6(20.5) & 33 \\
              &       & G1,10 & 3.1(0.4) &
$-8.0$(0.0)\tablenotemark{f} & 68.1(5.1) & 227.1(15.6) & \\
              &       & G2,10 & 5.5(0.4) & 64.5(3.4) &
110.8(7.4) & 646.9(19.8) & \\

              & (2,2) & D,10  & \nodata & 42.7 & 231.1 & 529.9(18.1) & \\
              &       & G1,10 & 1.1(0.4) &
$-8.0$(0.0)\tablenotemark{f} & 58.7(26.8) & 69.3(13.7) & \\
              &       & G2,10 & 3.1(0.4) & 54.5(11.8) &
123.5(16.7) & 411.4(19.9) & \\

              & (4,4) & D,10  & (0.5) & \nodata & \nodata & \nodata & \\

              & (2,1) & D,10  & (0.4) & \nodata & \nodata & \nodata & \\

NGC\,6951     & (1,1) & D,10  & (1.1) & \nodata & \nodata & \nodata &
12 \\ 
              & (2,2) & D,10  & (1.5) & \nodata & \nodata & \nodata & \\ 
              & (4,4) & D,10  & (1.3) & \nodata & \nodata & \nodata & \\ 
              & (2,1) & D,10  & (1.1) & \nodata & \nodata & \nodata & \\

\enddata
\tablenotetext{a}{~Table entries in parentheses are standard
  deviations or, in case of non-detections, RMS noise levels.}
\tablenotetext{b}{~Gn,m / D,m $\equiv$ Gaussian component number n / Direct
  measurement results with spectral resolution m\,\kms.}
\tablenotetext{c}{~Heliocentric optical velocity frame.}
\tablenotetext{d}{~Full-width half maximum (FWHM) given for gaussian
  fits; full-width zero intensity (FWZI) given for direct 
  measurements.}
\tablenotetext{e}{~Derived from direct integration of the FWZI line profile.}
\tablenotetext{f}{~NGC\,253: NH$_3$ (2,2) and (4,4) G1 fixed v$_{hel}$
  and FWHM to NH$_3$ (1,1) values; NGC\,660: NH$_3$ (2,2) and (4,4) G1
  and G2 fixed v$_{hel}$ to nominal velocity for this component (see
  Section~\ref{Comparison}), NH$_3$ (6,6) G2 and G3, and NH$_3$ (2,1)
  G4 fixed v$_{hel}$ and FWHM to nominal and/or best-fit values for
  this component (see Section~\ref{Comparison}), NH$_3$ (3,1), (3,2),
  and (4,3) G2 and (5,4) G3 fixed v$_{hel}$ and FWHM to NH$_3$ (2,1)
  values; NGC\,1365: NH$_3$ (4,4) fixed v$_{hel}$ to NH$_3$ (2,2)
  value; IC\,342: NH$_3$ (4,4) G2 fixed FWHM to NH$_3$ (2,2) value,
  NH$_3$ (5,5) fixed v$_{hel}$ to NH$_3$ (1,1) value, and NH$_3$ (6,6)
  fixed v$_{hel}$ to NH$_3$ (3,3) value; IC\,860: NH$_3$ (5,5) G2
  fixed FWHM to NH$_3$ (3,3) value; IR\,15107$+$0724: NH$_3$ (1,1) G1
  fixed FWHM to NH$_3$ (2,2) value, NH$_3$ (4,4) G1 fixed FWHM to
  NH$_3$ (1,1) value; NGC\,6946: NH$_3$ (1,1) and (2,2) G1 fixed
  v$_{hel}$ to $-8.0$\,\kms\ from H$_2$CO $2_{11}-2_{12}$ value
  \citep{Mangum2013}.} 
\end{deluxetable*} 

\section{Analysis}
\label{Analysis}

\subsection{Kinetic Temperature Derivation Using LVG Models}
\label{LVG}

To derive the kinetic temperature of the dense gas in our galaxy
sample, we use a model which incorporates the Large Velocity Gradient
(LVG) approximation \citep{Sobolev1960} to the radiative transfer.
The detailed properties of our implementation of the LVG approximation
are described in \cite{Mangum1993}.  This simplified solution to the
radiative transfer equation allows for a calculation of the global
dense gas properties in a range of environments.  As noted by
\cite{Mangum1993}, one of the major sources of uncertainty in an LVG
model prediction of the physical conditions is the uncertainty
associated with the collisional excitation rates used.  Based on the
analysis of the quantum mechanical calculations which went into the
derivation of the para-NH$_3$ to para-H$_2$ collisional excitation rates
presented by \cite{Danby1988}, \cite{Danby1987} estimate that the
calculated collisional excitation rates, which include all energy
levels up to and including J=5 for para-NH$_3$ and J=6 for
ortho-NH$_3$, are accurate to 20\% for kinetic temperatures $\lesssim
300$\,K.  We assume a similar level of accuracy for collisions
involving ortho-NH$_3$ when J$<6$.  Therefore, all physical conditions
predicted by our LVG model which use the Danby \etal\ rates are
limited to an accuracy of no better than 20\% when constrained by
measurements of NH$_3$.

As one can see from the listing of observed NH$_3$ frequencies and
their associated energies (Table~\ref{tab:frequencies}), we measure
several NH$_3$ transitions which include energy levels not included in
the \cite{Danby1988} excitation rate calculations.  In order to model
the excitation of all measured NH$_3$ transitions we have extended the
\cite{Danby1988} excitation rates using the prescription set forth in
\cite{Flower1995}.  This extrapolation to higher energy levels assumes
that the collisional de-excitation rates for NH$_3$ can be assumed to
be the same as those computed by \cite{Danby1988} for 300\,K and for
the same change in rotational quantum number J.  An average of the
existing excitation rates for a given ($\Delta$J,$\Delta$K) were used
to extrapolate to higher (J,K).  As noted by \cite{Flower1995}, this
extrapolation procedure is rather arbitrary, and we treat the results
obtained from these extrapolated excitation rates with due caution.

\begin{figure}
\centering
\includegraphics[scale=0.35,angle=-90]{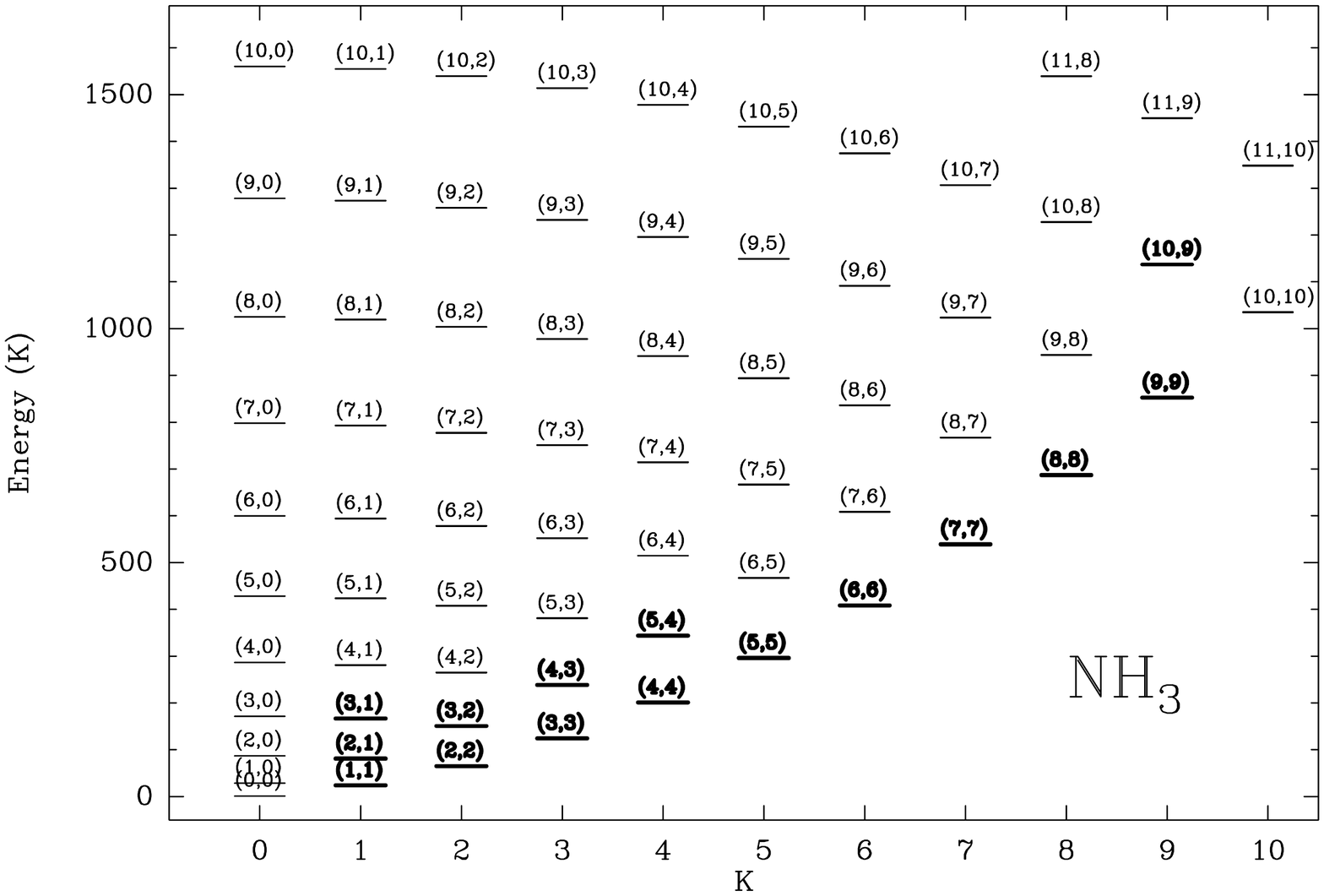}
\caption{NH$_3$ energy levels included in our LVG model analysis.
  Levels involving measured transitions are shown in \textbf{bold}.}
\label{fig:NH3Energies}
\end{figure}

After investigation of the limitations imposed by the iterative
solution speed as a function of the number of levels in our LVG model,
a uniform limit of E $\leq 1600$\,K and quantum number K $\leq 10$ was
chosen as the boundary for the extended para- and ortho-NH$_3$
excitation rates.  This resulted in a total of 84 and 43 levels for
para- and ortho-NH$_3$, respectively.  The energy levels and measured
transitions included in our LVG model analysis are shown in
Figure~\ref{fig:NH3Energies}.

\begin{figure}
\centering
\includegraphics[scale=0.40]{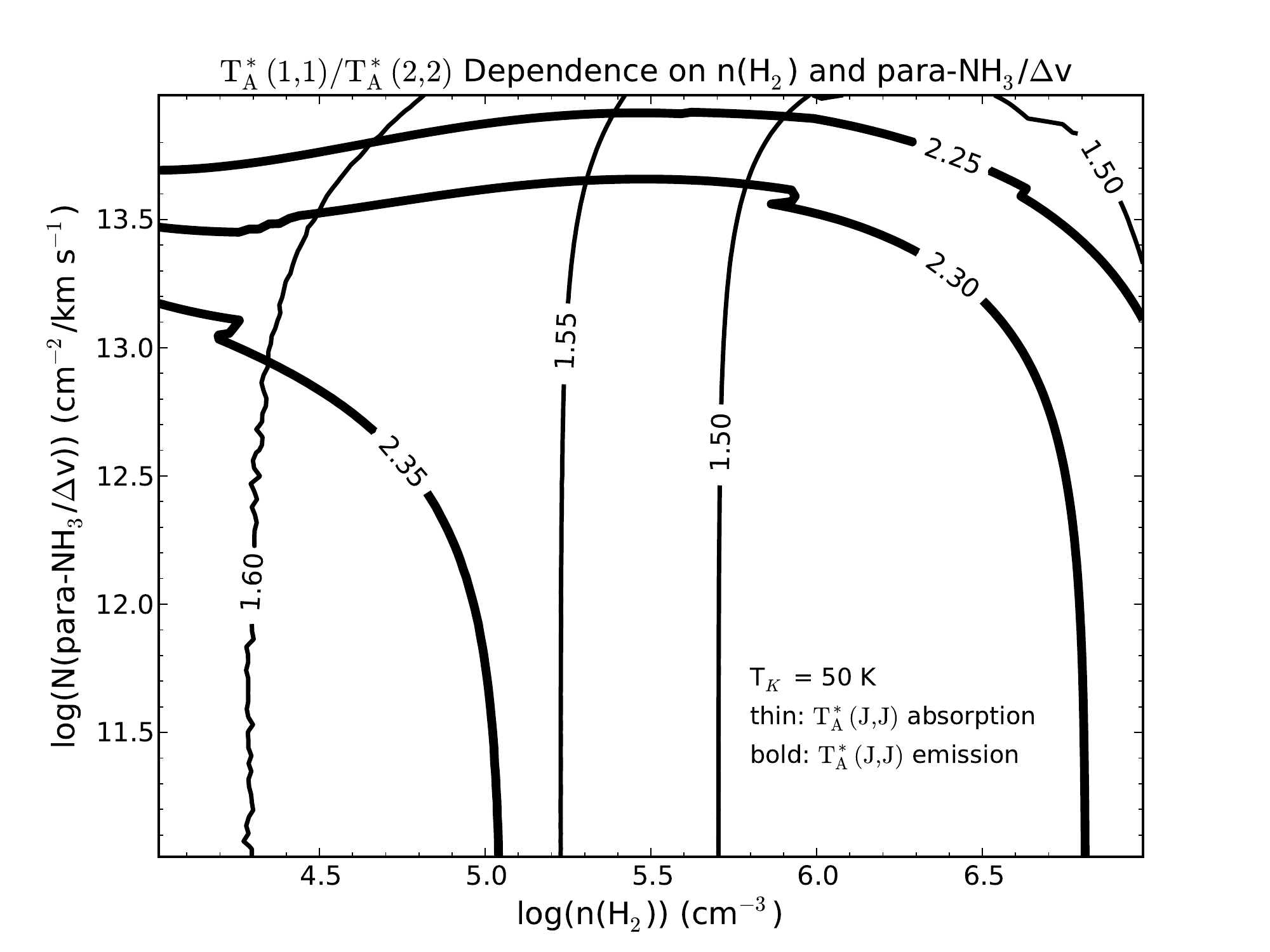}
\includegraphics[scale=0.40]{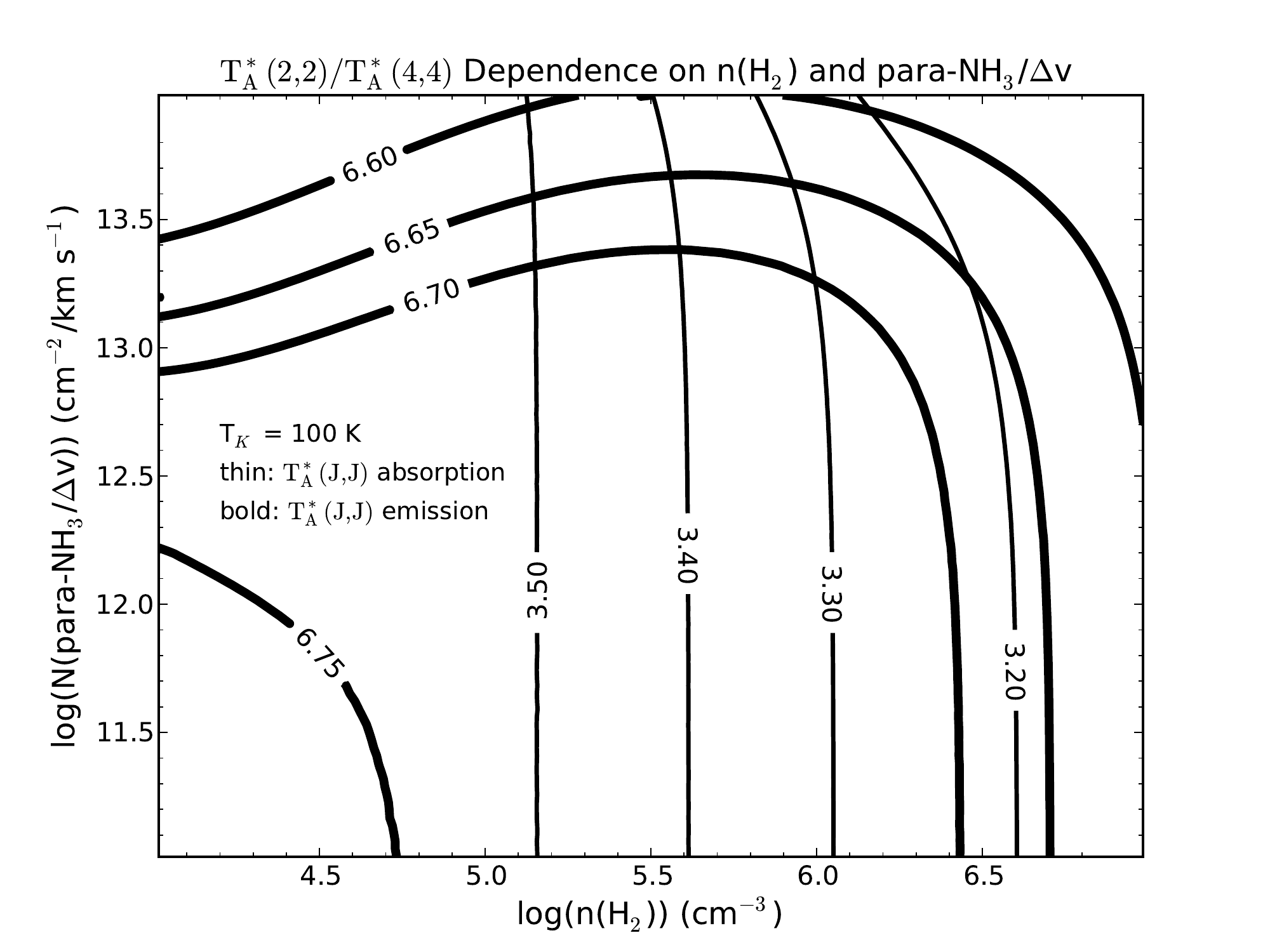}
\caption{LVG model predictions of the NH$_3$ (1,1)/(2,2) (top panel)
  and (2,2)/(4,4) (bottom panel) transition ratios, for both emission
  and absorption, as functions of spatial density and NH$_3$ column
  density.  Predicted transition ratios at representative kinetic
  temperatures, appropriate for the range of sensitivity for each
  transition ratio, are shown.}
\label{fig:LVGRatPlotGeneral}
\end{figure}

Model grids of predicted NH$_3$ transition intensities over
ranges in spatial density, para- or ortho-NH$_3$ column density per
velocity gradient, and kinetic temperature were made to accommodate
the requirements for fitting to several types of sources.  Both para-
and ortho-NH$_3$ model fits were required, in addition to model grids
which incorporated background continuum and infrared emission sources.
Table~\ref{tab:lvggrids} lists the LVG model grids and their ranges in
spatial density and NH$_3$ column density, temperature ranges,
increments, and assumed continuum temperature.  Measured NH$_3$
transition ratios were fit to these LVG model grids using two methods:
\begin{enumerate}
\item $\chi^2$ goodness-of-fit estimation of line ratios using:
\begin{equation}
\chi^2 = \sum^{i=n-1}_{i=1}{\frac{\left|\frac{\int T^*_A dv(i)}{\int
      T^*_A dv(i+1)} - \frac{\int T_{model} dv(i)}{\int T_{model}
      dv(i+1)}\right|^2}{\left[\sigma\left(\frac{\int T^*_A dv(i)}{\int
        T^*_A dv(i+1)}\right)\right]^2}}
\label{eq:chisq}
\end{equation}
\noindent{where} $n$ is the number of transitions measured in a single
galaxy.
\item NH$_3$ transition ratio best-fit using measured value
  $\pm\sigma$ comparisons to modelled ratios.  When more than two
  transition ratios are considered, the most extreme ranges in
  measurement uncertainty have been used to calculate the best-fit
  T$_K$.
\end{enumerate}
The values of $\chi^2$ which correspond to $1\sigma$, $2\sigma$, and
$3\sigma$ confidence in goodness of fit are dependent upon the number
of degrees-of-freedom (DOF) associated with a given model fit.  The
number of DOF is given by the difference between the number of NH$_3$
transitions measured and the number of free parameters considered in
the LVG model fit.  The NH$_3$ transition ratios included in
this analysis show only weak dependence on spatial density
and column density within the modeled ranges of these physical
conditions.  Specifically, over our modelled range of spatial density
and NH$_3$ column density the (1,1)/(2,2), (2,2)/(4,4), and
(5,5)/(7,7) ratios vary by $\lesssim 15$\%, with variations of
$\lesssim 5$\% over $\sim 90$\% of this modelled range.
Figure~\ref{fig:LVGRatPlotGeneral} shows examples of the LVG-modelled
NH$_3$ (1,1)/(2,2) and (2,2)/(4,4) transition ratios as functions of
n(H$_2$) and para-NH$_3$ column density per unit line width.  With a
weak dependence on spatial 
density and NH$_3$ column density, we assume DOF = number of measured
NH$_3$ transitions considered minus one.  For DOF = 1, 2, 3, and 4 the
$\chi^2$ values for $1\sigma$, $2\sigma$, and $3\sigma$ confidence
intervals are (1.0, 4.0, 9.0), (2.3, 6.2, 11.8), (3.5, 8.0, 14.2), and
(4.7, 9.7, 16.3), respectively.

%
\begin{deluxetable*}{lccccc} 
\tabletypesize{\small}
\tablewidth{0pt}
\tablecolumns{6}
\tablecaption{LVG Model Grid Parameters\label{tab:lvggrids}}
\tablehead{
\colhead{NH$_3$ Species} & 
\colhead{n(H$_2$),$\Delta\log($n(H$_2$))} &
\colhead{N$^\prime$,$\Delta\log($N$^\prime$)\tablenotemark{a}} & 
\colhead{T$_K$,$\Delta$T$_K$} & 
\colhead{T$_c$} & Used For \\
& \colhead{(cm$^{-3}$)} & \colhead{(cm$^{-2}$/km\,s$^{-1}$)} &
\colhead{(K)} & \colhead{(K)} & 
}
\startdata
para & $10^4-10^7$, (0.03,0.06) & $10^{11}-10^{14}$, (0.04,0.08) & 15--300, 3 &
T$_{cmb}$ & Emission \\
para & $10^4-10^7$, (0.03,0.06) & $10^{11}-10^{14}$, (0.04,0.08) & 10--300, 5 &
T$_{cmb}$+300 & Absorption \\
para & $10^4-10^7$, (0.03,0.06) & $10^{11}-10^{14}$, (0.04,0.08) & 15--300, 3 &
T$_{cmb}$+T$_c$\tablenotemark{b} & Absorption \\
ortho & $10^4-10^7$, (0.03,0.06) & $10^{11}-10^{14}$, (0.04,0.08) & 15--300, 3 &
T$_{cmb}$ & Emission \\
para & $10^4-10^9$, (0.05,0.10) & $10^{12}-10^{17}$, (0.05,0.10) & 15--300, 3 &
T$_{cmb}$+300 & NGC\,660 \\
ortho & $10^4-10^9$, (0.05,0.10) & $10^{12}-10^{17}$, (0.05,0.10) & 15--300, 3 &
T$_{cmb}$+300 & NGC\,660 \\
para & $10^4-10^7$, (0.03,0.06) & $10^{11}-10^{14}$, (0.04,0.08) & 15--300, 3 &
T$_{cmb}$+T$_{IR}$\tablenotemark{c} & NGC\,253, IC\,342 \\
\enddata
\tablenotetext{a}{N$^\prime$ $\equiv$ Column density per velocity
  gradient increment N(species-NH$_3$)/$\Delta v$.} 
\tablenotetext{b}{Separate model cubes with T$_c$ = 50, 100, 200, 300,
  400, and 500\,K.}
\tablenotetext{c}{T$_{IR}$ = T$_{dust}$ = 34\,K for NGC\,253 and IC\,342.}
\end{deluxetable*} 

%
\begin{deluxetable*}{lcccc} 
\tablewidth{0pt}
\tablecolumns{5}
\tablecaption{Star Forming Galaxy Derived Kinetic Temperatures\label{tab:lvg}}
\tablehead{
\colhead{Source\tablenotemark{a}} & 
\colhead{T$_{dust}$\tablenotemark{b}} & \colhead{T$_K$} &
\colhead{Notes\tablenotemark{c}} & \colhead{Fit Type\tablenotemark{d}}
\\
& \colhead{(K)} & \colhead{(K)} & &
}
\startdata
NGC\,253        & 34 & $78\pm22$/$>150$ & PF,MTC,MVC & R(5,2) \\
NGC\,253\,NE    & \ldots & $38\pm9$/$73\pm9$/$>150$ & PF,MTC & R(5,2) \\
NGC\,253\,SW    & \ldots & $78\pm22$/$>150$ & PF,MTC & R(5,2) \\
NGC\,660        & 37 & $160\pm96$\tablenotemark{e} && \ldots \\
NGC\,660\,G1    & \ldots & $149\pm72$\tablenotemark{f} & GF & $\chi^2$(2) \\
NGC\,660\,G2    & \ldots & $\gtrsim 150$\tablenotemark{f} & PF & R(2,1) \\
NGC\,660\,G3    & \ldots & $\gtrsim 100$\tablenotemark{f} & PF & R(4,1) \\
NGC\,660\,G4    & \ldots & $163\pm93$\tablenotemark{f} & GF & $\chi^2$(4,1) \\
NGC\,891        & 28 & $<30$ && \\
Maffei\,2       & 40\tablenotemark{g} & $43\pm6$ / $79\pm6$ & PF,MTC & R(2) \\
Maffei\,2\,G1   & \ldots & $40\pm3$ / $82\pm6$ & PF,MTC & R(2) \\
Maffei\,2\,G2   & \ldots & $50\pm10$ / $79\pm9$ & PF,MTC & R(2) \\
NGC\,1365       & 32 & $50\pm11$ & GF & R(2) \\
NGC\,1365\,G1   & \ldots & $110^{+50}_{-40}$ & GF & R(2) \\
NGC\,1365\,G2   & \ldots & $29\pm6$ & GF & R(2) \\
IC\,342         & 30 & $\sim 150$ & PF,MTC & $\chi^2$(4,1)\\
IC\,342\,G1     & \ldots & $24\pm7$/$115\pm17$/$>140$ & GF,MTC & R(3,1) \\
IC\,342\,G2     & \ldots & $75\pm14$/$>185$ & GF,MTC & R(4,1) \\
M~82\,SW        & 45 & $58\pm19$ & GF & R(2) \\
NGC\,3079       & 32 & $>150$ & PF & $\chi^2$(4,1) \\
NGC\,3079\,G1   & \ldots & $>100$ & PF & $\chi^2$(4,1) \\
NGC\,3079\,G2   & \ldots & $>150$ & PF & $\chi^2$(4) \\
IC\,860         & 40\tablenotemark{g} & $206\pm79$/$\gtrsim 250$ & PF,MTC &
$\chi^2$(4,1) \\ 
M\,83           & 31 & $56\pm15$ & GF & R(2) \\
IR\,15107+0724  & 40\tablenotemark{g} & $189\pm57$ & GF & R(2) \\
Arp\,220        & 44 & $234\pm52$ & GF & R(5,1) \\
Arp\,220\,G1    & \ldots & $136\pm18$ / $223\pm30$ & PF,MTC & R(4,1) \\
Arp\,220\,G2    & \ldots & $153\pm20$ / $236\pm40$ & PF,MTC & R(5,1) \\
NGC\,6946       & 30 & $47\pm8$ & GF & R(2) \\
NGC\,6946\,G1   & \ldots & $25\pm3$ & GF & R(2) \\
NGC\,6946\,G2   & \ldots & $50\pm10$ & GF & R(2) \\
\enddata
\tablenotetext{a}{Source name tags ``G1'', ``G2'', ``G3'', and ``G4''
  refer to Gaussian fit components from Table~\ref{tab:nh3results}.} 
\tablenotetext{b}{T$_{dust}$ from Table~\ref{tab:galaxies}.} 
\tablenotetext{c}{Notes: GF $\equiv$ good fit; PF $\equiv$ poor fit; MTC
  $\equiv$ multiple temperature components indicated; MVC $\equiv$
  multiple velocity components indicated.}
\tablenotetext{d}{Fit Types: R(\#,\#) $\equiv$ para- and ortho-NH$_3$
  (if available) ratio fit using \# transition ratios; $\chi^2$(\#,\#)
  $\equiv$ $\chi^2$ fit assuming \# degrees of freedom for para-NH$_3$
  and ortho-NH$_3$ (if available).} 
\tablenotetext{e}{Average and range for all four components (see
  Section~\ref{Comparison}).}
\tablenotetext{f}{Fit over high density and column density range more
  constrained.  See \S\ref{NGC660Tk}.}
\tablenotetext{g}{Assumed value.}
\end{deluxetable*} 

Table~\ref{tab:lvg} lists the derived LVG model best-fit kinetic
temperatures for the 13 galaxies where we detect at least one NH$_3$
transition and have a limit to at least one other NH$_3$ transition
(Table~\ref{tab:nh3results}).  Discussions of these LVG model kinetic
temperature fits, along with a comparison to previous NH$_3$
measurements of individual galaxies, are given in \S\ref{Comparison}.

\pagebreak
\subsection{LVG Models with Background Continuum and Infrared Emission}
\label{LVGContIR}

\subsubsection{Background Continuum Emission}
\label{BackCont}

For the five galaxies in our sample which exhibit NH$_3$ absorption a
background continuum source must be added to our LVG model analysis.
Even though we have estimates of the single-dish background continuum
flux from our extragalactic NH$_3$ measurements
(Table~\ref{tab:nh3results}), in general we do not have a measurement
of the 24\,GHz continuum source size.  We can estimate the likely
continuum source size from other centimeter-wavelength continuum
measurements.  In \S\ref{Comparison} we list the relevant
centimeter-wavelength continuum measurements for these five galaxies,
all of which imply continuum source sizes of $\theta_s \lesssim
0.5$\,arcsec.  For 24\,GHz continuum fluxes
(Table~\ref{tab:nh3results}) which range from 15\,mJy
(IR\,15107$+$0724) to 140\,mJy (Arp\,220) within our $\theta_B \simeq
30$\,arcsec beam, the continuum brightness temperatures from these
five galaxies are estimated to be in the range 100 to 1000\,K.

To further guide the allowable range of continuum brightness
temperatures that we must consider, we note that four of the five
NH$_3$ absorption galaxies are detected in the high-excitation NH$_3$
(7,7), (8,8), and/or (9,9) transitions.  Since for thermalized energy
levels an absorption line is produced for $T_c \gtrsim T_{ex} \simeq
T_K$, significant (7,7) \textit{absorption} is produced only for $T_c
\gtrsim 200$\,K.  For IR\,15107$+$0724, where the highest excitation
transition detected is the NH$_3$ (4,4) transition, significant
absorption is produced for $T_c \gtrsim 100$\,K.

To derive the kinetic temperature within these five NH$_3$ absorption
galaxies we have constructed LVG models over similar ranges in
(n(H$_2$),N(para-NH$_3$)/$\Delta v$,T$_K$) as used to constrain our
extragalactic NH$_3$ emission measurements and added
blackbody background continuum sources with brightness
temperatures of T$_c$ = 50, 100, 200, 300, 400, and 500\,K
(Table~\ref{tab:lvggrids}).  With the lower-limits to $T_c$ 
noted above in mind, we investigated the behavior of the
higher-excitation absorption transition ratios within these NH$_3$ LVG 
models.  We find that there is a minimal ($\lesssim 10$\%) dependence
of the derived kinetic temperature on the exact continuum brightness
temperature chosen.  Since the limitations of our NH$_3$ excitation 
rates restrict us to studies of gas with T$_K \lesssim 300$\,K, we have
adopted a uniform T$_c = 300$\,K for all five of our NH$_3$ absorption
galaxies.  The kinetic temperatures listed in Table~\ref{tab:lvg} are
derived from fits to the LVG model cubes with T$_c = 300$\,K.  Note,
though, as noted above, that fits to LVG model cubes with values of
T$_c$ ranging from 100 to 500\,K do not produce appreciably different
kinetic temperatures.

\subsubsection{Embedded Infrared Emission}
\label{EmbedIR}

In NGC\,253 and IC\,342 we detect NH$_3$ (2,1) emission.  As described
in \S\ref{NH3Probe}, \cite{Morris1973} showed that the level
populations of the NH$_3$ (2,1) transition become inverted when
n(H$_2$) $<10^6$\,cm$^{-3}$ and T$_K = 50$\,K (or n(H$_2$)
$<10^5$\,cm$^{-3}$ and T$_K = 100$\,K, as shown in
Figure~\ref{fig:NH3Plots}) in the presence of a blackbody 
infrared radiation field.  This condition results in detectable
non-metastable NH$_3$(2,1) emission under a wide range of physical
conditions that our LVG models reproduce (see
Figure~\ref{fig:NH3Plots}).  Higher-lying non-metastable (J$\not=$K)
transitions also become inverted, but require higher infrared
radiation fields and/or higher kinetic temperatures to produce the
inverted populations.

One should also be concerned about the effects of the insertion of an
embedded infrared source on the populations of the energy levels
which comprise the NH$_3$ (1,1) and (2,2) transitions.  Comparing LVG
model predictions of the NH$_3$ (1,1) and (2,2) radiation temperatures
with an undiluted (\ie\ completely coupled to the source of
  molecular emission) 34\,K blackbody infrared source
(indicated with a superscript IR below) with those 
models which do not have infrared emission (see
Table~\ref{tab:lvggrids}), we find that, for n(H$_2$) = $10^4$ to
$10^7$\,cm$^{-3}$: 
\begin{itemize}
\item $\frac{T_R(1,1)}{^{IR}T_R(1,1)}$ = 0.65 to 1.25,
  increasing monotonically over the density range at all modelled
  T$_K$.
\item $\frac{T_R(2,2)}{^{IR}T_R(2,2)}$ = 0.70 to 1.10,
  increasing monotonically over the density range for T$_K$ $\gtrsim
  50$\,K
\item $\frac{T_R(2,2)}{^{IR}T_R(2,2)}$ = 0.60 to 0.90,
  with minima at the extremes of the modelled density range and a
  maximum near n(H$_2$) = $10^5$\,cm$^{-3}$, all for T$_K$ = 15 to
  50\,K
\end{itemize}
\noindent{Based} on our LVG modelling, it appears that the
kinetic temperature sensitive $\frac{T_R(1,1)}{T_R(2,2)}$ ratio, for a
given (n,N,T$_K$), is decreased by $\lesssim 30$\% over the range of
physical conditions considered when a moderate embedded infrared
source is added.

Fundamentally, this behavior is due to the influence of the
non-metastable excitation state on the metastable transitions.  When
the non-metastable transitions are driven into maser emission by the
insertion of an infrared source, these non-metastable transitions
occupy a larger fraction of the population distribution within NH$_3$.
This results in a lower intensity for the metastable transitions at
the bottom of the K-ladders of NH$_3$ which has a stronger effect on
lower-energy transitions.  One can see this effect on the
$\frac{T_R(1,1)}{T_R(2,2)}$ and $\frac{T_R(2,2)}{T_R(4,4)}$ transition
ratios from the ``cool component'' of NGC\,253NE and NGC\,253SW shown
in Figures~\ref{fig:NGC253NETkFit} and \ref{fig:NGC253SWTkFit},
respectively.  Effectively this imparts a slight spatial density
dependence to the transition ratio.  Furthermore, the density at which
a transition ratio (\ie\ $\frac{T_R(1,1)}{T_R(2,2)}$,
$\frac{T_R(2,2)}{T_R(4,4)}$, $\frac{T_R(5,5)}{T_R(7,7)}$, \etc),
begins to decrease is driven to higher densities as T$_{IR}$ is
increased.  Quantitatively, for T$_{IR} = 100$ and 200\,K, the kinetic
temperature derived for a given NH$_3$ metastable transition ratio is
decreased by $\sim 15$ and $\sim 30$\% while the uncertainty
associated with this kinetic temperature measurement is increased by a
factor of two and three, respectively.  Recall that our conservative
error estimation incorporates any spatial density dependence to a
kinetic temperature sensitive transition ratio.

\subsection{Extragalactic Kinetic Temperature Fits and Comparison to Previous Ammonia Measurements}
\label{Comparison}

In the following we discuss our LVG model fits to our NH$_3$
measurements of the kinetic temperatures in our galaxy sample.  We
also compare our current extragalactic NH$_3$ detections and kinetic
temperature measurements to existing measurements.  Detailed summaries
of existing dense-gas molecular line measurements toward these
galaxies can be found in \citet{Mangum2008} and \citet{Mangum2013}.

\medskip\noindent
{\bf NGC\,253:}  NH$_3$ measurements of the (1,1), (2,2), (3,3),
(4,4), and (6,6) transitions toward this nearby starburst galaxy have
been reported by several authors.  Most of these measurements are able
to spectrally, and in the case of imaging measurements spatially
resolve the NH$_3$ emission toward NGC\,253 into two distinct velocity
components: a north-east (NE) component at $v \simeq 180$\,km\,s$^{-1}$ and
a south-west (SW) component at $v \simeq 300$\,km\,s$^{-1}$.  

The (1,1), (2,2), and (3,3) transitions have been detected by
\cite{Takano2002} with spatial and spectral resolution $\theta_b$ =
71$^{\prime\prime}$ and $\Delta v \simeq 20$\,km\,s$^{-1}$,
respectively.  The the 
(4,4) transition was searched for but not detected, while the (2,2)
transition was detected in only the 300\,km\,s$^{-1}$ component.
Using only the (1,1), (2,2), and a limit to the (4,4) transition
\cite{Takano2002} derive T$_K \lesssim 17$ and $37^{+9}_{-6}$\,K for the
180 and 300\,km\,s$^{-1}$ components, respectively.

\cite{Mauersberger2003} measured the (1,1), (2,2), (3,3), (4,4), and
(6,6) transitions at spatial and spectral resolution of $\theta_b$ =
38$^{\prime\prime}$ and $\Delta v \simeq 15$\,km\,s$^{-1}$,
respectively.  Using a rotation diagram analysis
\cite{Mauersberger2003} derived T$_K = 142\pm14$ and $100\pm3$\,K for
the 180 and 300\,km\,s$^{-1}$ components, respectively.  These
relatively high temperatures are dominated by the emission from the
(3,3) and (6,6) transitions, which has been shown to produce maser
emission under a wide range of physical conditions in star formation
regions \citep[see][]{Mangum1993}.  If the (3,3) and (6,6)
transitions are excluded from their fit to T$_K$, the rotation
temperatures derived are $\sim 60$\,K for both velocity components.

\cite{Ott2005} imaged the (1,1), (2,2), (3,3), and (6,6) transitions
at spatial and spectral resolution of $\theta_b$ =
$30^{\prime\prime}\times5^{\prime\prime}$, position angle
$-12^\circ$, and $\Delta v \simeq$ 
12\,km\,s$^{-1}$, respectively.  Six dense gas condensations were
measured within a $\sim 30^{\prime\prime}$ region around the nucleus
which generally fall into two groups: a NE group with $v \simeq
180$\,km\,s$^{-1}$ and a SW group with $v \simeq 300$\,km\,s$^{-1}$.
These measurements associated for the first time the 180 and
300\,km\,s$^{-1}$ components noted previously with dense gas structures
within NGC\,253.  \cite{Ott2005} also note NH$_3$ \textit{absorption}
in the (1,1), (2,2), and (6,6) transitions toward the central compact
continuum source in NGC\,253.  Using an assumed ortho-to-para NH$_3$
ratio of unity, \cite{Ott2005} derive T$_K$ $\simeq 140$ and 200\,K
toward the spatial 
components in the NE and SW of NGC\,253, respectively.  As was the
case with the kinetic temperatures derived by \cite{Mauersberger2003}
the high kinetic temperatures derived are largely dependent upon the
intensities of the (3,3) and (6,6) transitions.  Since
\cite{Ott2005} note that the (3,3) transition is observed in emission
toward the central compact continuum source position which produces
NH$_3$ (1,1), (2,2), and (6,6) absorption, maser amplification of the
(3,3) transition may be occurring.  As noted above, both the (3,3) and
(6,6) transitions are known to produce maser emission in star
formation regions in our Galaxy \citep[see][]{Mangum1993}.  Using
separate measurements of the NH$_3$ rotation temperature from the
(1,1)/(2,2) (T$_{12}$) and (3,3)/(6,6) (T$_{36}$) transition ratios,
\cite{Ott2005} derive NH$_3$ rotation temperatures of T$_{12} \simeq$
37 and 51\,K and T$_{36} \simeq$ 108 and 127\,K, for the NE and SW
components, respectively.

In an imaging measurement similar to that reported by \cite{Ott2005},
\cite{Takano2005} imaged the (1,1), (2,2), and (3,3) transitions at
spatial and spectral resolution of $\sim 4\times3^{\prime\prime}$ and
$\Delta v \simeq 20$\,km\,s$^{-1}$, respectively.  As with past NH$_3$
measurements the NE (180\,km\,s$^{-1}$) and SW (300\,km\,s$^{-1}$)
spatial/spectral components were detected, along with NH$_3$ (1,1)
\textit{absorption} toward the central continuum source.
Using only their (1,1) and (2,2) measurements \cite{Takano2005} derive
T$_K \simeq 15$ to 28\,K and $\simeq 20$ to 30\,K toward the NE and SW
components, respectively.  The low kinetic temperatures derived likely
reflect the low-excitation transitions employed in this study.

The $\sim 2.^{\prime\prime}5$ $\Delta v = 20$\,\kms\ images of the
NH$_3$ (3,3) emission from NGC\,253 reported by \cite{Lebron2011}
arise from six GMC components located within a
$40^{\prime\prime}\times 5^{\prime\prime}$ structure with position
angle $\sim 45^\circ$.  The emission distribution measured is similar
to previous high-resolution NH$_3$ studies of NGC\,253.

Our $30^{\prime\prime}$ spatial resolution measurements of the
para-NH$_3$ (1,1), (2,2), and (4,4) 
transitions and the ortho-NH$_3$ (3,3) and (6,6) transitions 
toward NGC\,253 are in good agreement with many of the aspects of
these previous studies.  The addition of the (5,5), (7,7), (8,8), and
(9,9) transitions allows us to better characterize the high
temperature environment in this starburst galaxy.  The HC$_3$N J=$3-2$
spectrum (Figure~\ref{fig:NGC253NH3Spec}) was measured \textit{gratis}
as part of our NH$_3$ (7,7), (8,8), and (9,9) observing
configuration.  For both NH$_3$ and HC$_3$N we measure velocity
components at $\sim180$ and $\sim300$\,km\,s$^{-1}$, consistent with
the NE and SW components noted previously.  Employing our LVG-based
kinetic temperature analysis (\S\ref{LVG}) we obtain a poor fit to the
kinetic temperature if an overall fit to both velocity components is
made (see Table~\ref{tab:lvg}).  T$_K$ derived for each of the  
two velocity components produces fits which strongly suggest
contributions due to multiple temperature components.  Summarizing our
T$_K$ fits made toward NGC\,253:
\begin{itemize}
\item \textit{NE (180\,km\,s$^{-1}$) Component:}  This velocity component
  yielded a three-temperature fit: T$_K = 38\pm9$\,K / $73\pm9$\,K /
  $>150$\,K using the $\frac{(1,1)}{(2,2)}$ (low-temperature
  component), $\frac{(2,2)}{(4,4)}$ (warm-temperature component), 
  and combination of the para-NH$_3$ $\frac{(4,4)}{(5,5)}$,
  $\frac{(5,5)}{(7,7)}$, $\frac{(7,7)}{(8,8)}$, and ortho-NH$_3$
  $\frac{(3,3)}{(6,6)}$, and $\frac{(6,6)}{(9,9)}$ (high-temperature
  component) transition ratios (Figure~\ref{fig:NGC253NETkFit}).

\begin{figure}
\centering
\includegraphics[scale=0.45]{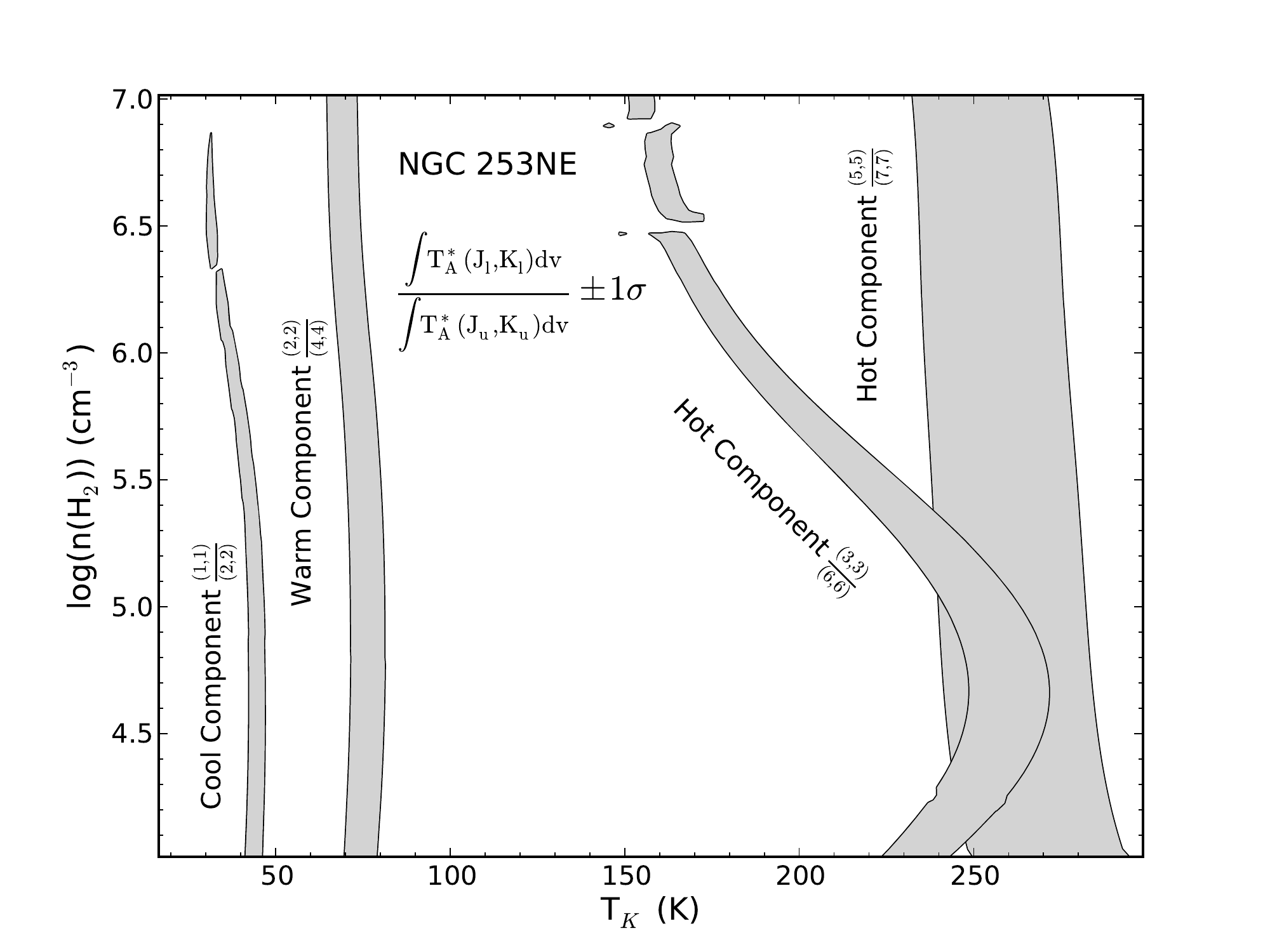}
\caption{LVG model kinetic temperature fit to the $v_{opt} \simeq
  180$\,\kms\ (NE) component of NGC\,253.  Shown as grey contours are
  the best-fit ratios at a representative N(species-NH$_3$)/$\Delta v =
  10^{13.5}$\,cm$^{-2}$/\kms\ within one standard deviation for this
  spectral component using the (1,1)/(2,2), (2,2)/(4,4), (3,3)/(6,6),
  and (5,5)/(7,7) ratios which characterize the cool, warm, hot, and
  hot kinetic temperature components, respectively.  The broken
  contours at n(H$_2$) $\gtrsim 10^{6.5}$\,cm$^{-3}$ for the
  ortho-NH$_3$ (3,3)/(6,6) ratio are due to the onset of
  maser amplification in the (3,3) and (6,6) transitions.}
\label{fig:NGC253NETkFit}
\end{figure}

\item \textit{SW (300\,km\,s$^{-1}$) Component:}  This velocity
  component yielded a two-temperature fit: T$_K = 78\pm22$\,K /
  $>150$\,K using the combined $\frac{(1,1)}{(2,2)}$ and
  $\frac{(2,2)}{(4,4)}$ transition ratios (low-temperature component)
  and the combined para-NH$_3$ $\frac{(4,4)}{(5,5)}$,
  $\frac{(5,5)}{(7,7)}$, $\frac{(7,7)}{(8,8)}$, and ortho-NH$_3$
  $\frac{(3,3)}{(6,6)}$, and $\frac{(6,6)}{(9,9)}$ transition ratios
  (high-temperature component; Figure~\ref{fig:NGC253SWTkFit}).

\begin{figure}
\centering
\includegraphics[scale=0.45]{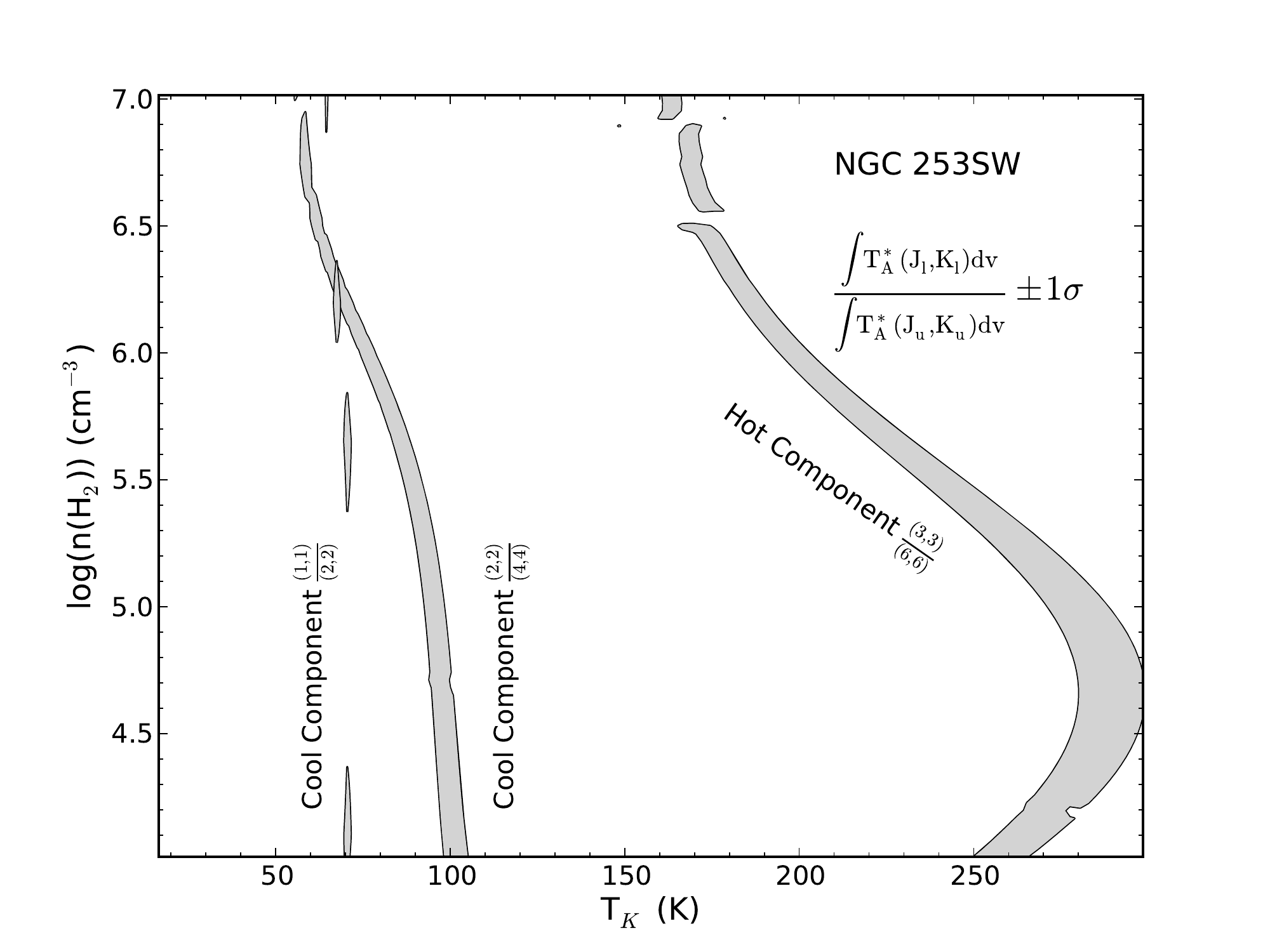}
\caption{LVG model kinetic temperature fit to the $v_{opt} \simeq
  300$\,\kms\ (SW) component of NGC\,253.  Shown as grey contours are
  the best-fit ratios at a representative N(species-NH$_3$)/$\Delta v
  = 10^{13.5}$\,cm$^{-2}$/\kms\ within one standard deviation for this
  spectral component using the (1,1)/(2,2), (2,2)/(4,4), and
  (3,3)/(6,6) ratios which characterize the cool, cool, and hot
  kinetic temperature components, respectively.  Transition
  measurement ratios involving the higher-lying (5,5), (7,7), (8,8),
  and (9,9) transitions yield T$_K \gtrsim 300$\,K.  The broken
  contours at n(H$_2$) $\gtrsim 10^{6.5}$\,cm$^{-3}$ for the
  ortho-NH$_3$ (3,3)/(6,6) ratio are due to the onset of
  maser amplification in the (3,3) and (6,6) transitions.  Broken
  contours for the para-NH$_3$ (1,1)/(2,2) cool component fit are due
  to the best-fit range being comparable to our LVG model cube kinetic
  temperature increment.}
\label{fig:NGC253SWTkFit}
\end{figure}

\item \textit{Whole Galaxy:} Due to the disparate kinetic temperatures
  derived for the NE and SW components, the kinetic temperature fit to
  the total galaxy NH$_3$ emission is poor.  Taking an average of the
  kinetic temperature fits to the NE and SW components:
  \begin{itemize}
  \item T$_K = 78\pm22$\,K / $>150$\,K from levels lower and higher
    than (4,4), respectively.
  \end{itemize}
\end{itemize}

\medskip\noindent
{\bf NGC\,660:} Our detections of the NH$_3$ (1,1) through (7,7)
transitions in absorption are the first reported NH$_3$ measurements
toward this polar ring galaxy.  By far the most diverse and
complicated NH$_3$ measurements in our galaxy sample, we present a
detailed analysis of these remarkable spectra in
Section~\ref{NGC660NH3}.

\medskip\noindent
{\bf NGC\,891:}  This is a new detection of the NH$_3$ (1,1)
transition toward this galaxy.  With only a limit to the (2,2) and
(4,4) intensities we derive only an upper limit to the kinetic
temperature.  High spatial resolution measurements of dense gas
tracers in NGC\,891 \citep[see summary in][]{Mangum2013} suggest that
the molecular emission is mainly confined to a nuclear condensation
that is $\sim 20^{\prime\prime}\times10^{\prime\prime}$ in extent.

\medskip\noindent
{\bf Maffei\,2:} \cite{Takano2000} measured the NH$_3$ (1,1), (2,2), (3,3),
and (4,4) transitions (did not detect the (4,4) transition) at
$\theta_b \simeq 70^{\prime\prime}$ and $\Delta v \simeq
10$\,km\,s$^{-1}$.  \cite{Henkel2000} measured the (1,1), (2,2),
(3,3), and (4,4) transitions, while \cite{Mauersberger2003} measured
the (6,6) transition toward this galaxy, all at $\theta_b \simeq
40^{\prime\prime}$ and $\Delta v \simeq 15$\,km\,s$^{-1}$.  The
NH$_3$ emission in all of these measurements is comprised of two
velocity components at $\sim -80$ and $\sim +6$\,km\,s$^{-1}$.
\cite{Takano2000} measure rotation temperatures T$_{rot} \simeq 30$\,K
over all emission velocities.  \cite{Henkel2000} derive T$_{rot} =
85\pm5$\,K for the $-80$\,\kms\ component and T$_{rot} = 55\pm10$\,K
for the $+6$\,\kms\ component using only the (1,1), (2,2), and (3,3)
transitions (deconvolution of the (4,4) transition into two velocity
components was considered tentative by \cite{Henkel2000}).  Combining
their NH$_3$ (6,6) measurement to the \cite{Henkel2000} measurements, 
\cite{Mauersberger2003} derive T$_{rot} = 48\pm15$\,K using NH$_3$
(1,1) and (2,2) and T$_{rot} = 132\pm12$\,K using NH$_3$ (3,3) and
(6,6) over both velocity components.

The $\theta_b \sim 3.^{\prime\prime}5$ $\Delta v = 20$\,\kms\ images of the
NH$_3$ (1,1) and (2,2) emission from Maffei\,2 reported by
\cite{Lebron2011} arise from four main GMC components grouped into
three velocity ranges near $-80$, 0, and $+20$\,\kms.  The overall
NH$_3$ emission is located within a $\sim 20^{\prime\prime}\times
5^{\prime\prime}$ structure with position angle $\sim 30^\circ$.  In
addition to the emission detected from all four GMCs, NH$_3$ (2,2)
absorption from ``peak B'' ($v_{hel} \simeq -85$\,\kms) against the
centimeter continuum emission peak in Maffei\,2 was noted. Derived
T$_{rot}$ for the four GMCs imaged range from 25--50\,K.

Our measurements of the NH$_3$ (1,1), (2,2), and (4,4) transitions
toward Maffei\,2 reflect the same two-velocity component structure
previously noted.  Using our NH$_3$ (1,1), (2,2), and (4,4)
measurements, both the $-80$\,km\,s$^{-1}$ (Maffei2\,G1) and
$+20$\,km\,s$^{-1}$ (Maffei2\,G2) velocity components deconvolve into
low-temperature (T$_K \simeq 40$\,K) and high-temperature (T$_K \simeq
80$\,K) components.  The kinetic temperature fits to both velocity
components is poor due to this multiple-temperature component
structure. 

\begin{figure}
\centering
\includegraphics[scale=0.45]{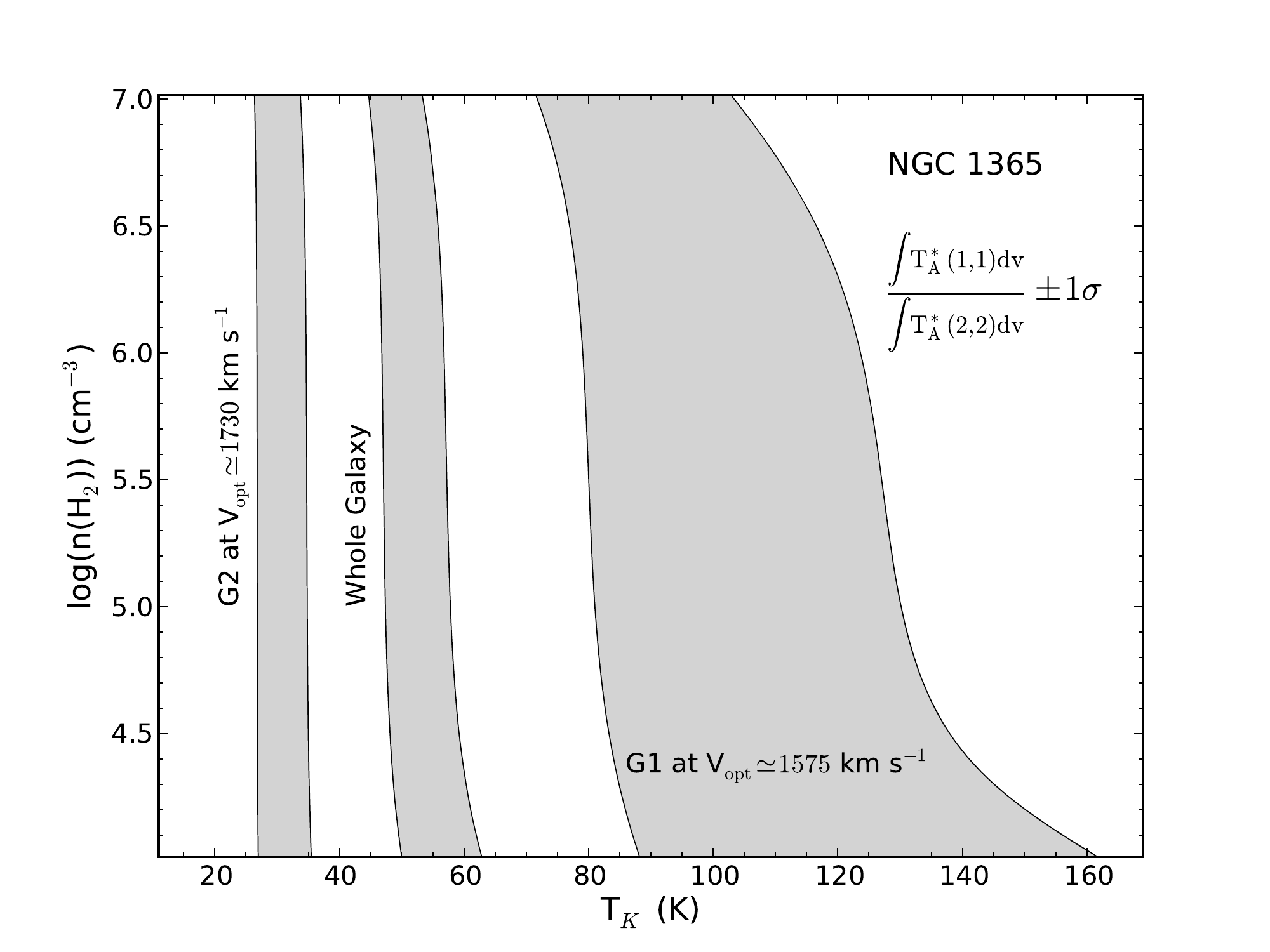}
\caption{LVG model kinetic temperature fit to the individual velocity
  components in NGC\,1365.  Shown as grey contours are the best-fit
  ratios within one standard deviation for the two velocity components
  of NGC\,1365 and for the galaxy as a whole.}
\label{fig:NGC1365TkFit}
\end{figure}

\medskip\noindent
{\bf NGC\,1365:} The NH$_3$ emission from this galaxy appears to
emanate from two velocity components; at $v_{hel} \sim 1575$ and $\sim
1730$\,km\,s$^{-1}$.  The H$_2$CO $1_{10}-1_{11}$ absorption spectrum
reported by \cite{Mangum2013} is suggestive of this same two-velocity
component structure.  The quality of the kinetic temperature fit for this
galaxy is good, well constrained by the NH$_3$ (1,1) and (2,2)
transitions detected towards both velocity components.  The velocity
component near $v_{hel} \simeq 1575$\,\kms\ (NGC\,1365\,G1) appears to
be significantly warmer (T$_K \simeq 100$\,K) than the velocity
component near $v_{hel} \simeq 1730$\,\kms\ (NGC\,1365\,G2; T$_K
\simeq 30$\,K).  Figure~\ref{fig:NGC1365TkFit} shows the LVG model fit
to the kinetic temperature in this galaxy.  We are not aware of any
previous measurements of NH$_3$ toward this galaxy.

\medskip\noindent
{\bf IC\,342:} The NH$_3$ emission from this galaxy, the first
extragalactic object ever detected in NH$_3$ \citep{Martin1979},
appears to originate in two velocity components; at $v_{hel} \simeq
25$ (G1) and $\simeq 50$\,\kms\ (G2).  The NH$_3$ (1,1) through (9,9)
transition measurements reported in \cite{Mauersberger2003} and the
H$_2$CO $1_{10}-1_{11}$ and $2_{11}-2_{12}$ absorption spectra
reported by \cite{Mangum2008,Mangum2013} originate from a single
velocity component with $v_{hel} \simeq 25$\,\kms.
\cite{Mauersberger2003} derive NH$_3$ rotational temperatures of $\sim
50$\,K (from their (1,1) through (3,3) measurements) and $\sim 400$\,K
(from their (5,5) through (9,9) measurements), suggesting
contributions from warm and hot dense gas.

The $\theta_b \sim 3.^{\prime\prime}5$ $\Delta v = 20$\,\kms\ images of the
NH$_3$ (1,1) and (2,2) emission from IC\,342 reported by
\cite{Lebron2011} and the $\theta_b \sim 6^{\prime\prime}$ $\Delta v =
9$\,\kms\ images of the NH$_3$ (6,6) emission from IC\,342 reported by
\cite{MonteroCastano2006} both identified four of the five giant
molecular cloud (GMC) complexes previously noted by \cite{Downes1992}
and \cite{Meier2001}.  In their $\sim 2^{\prime\prime}$ $\Delta v =
10$\,\kms\ images of the HC$_3$N J=$5-4$ and J=$16-15$ emission from
IC\,342 \cite{Meier2011} identified all five of these GMC complexes.  The
central velocities of these GMC complexes 
(identified as GMC A through E) fall into three groups; a complex with
$v_{hel} \simeq 12$\,\kms\ (GMC E, comprising the SW spiral arm of
IC\,342), two complexes with $v_{hel} \simeq 25$\,\kms\ (GMC A and B,
comprising the nuclear star forming region and central bar of
IC\,342), and two complexes with $v_{hel} \simeq 50$\,\kms\ (GMC C and
D, comprising the NE spiral arm of IC\,342).  The NH$_3$ velocity
component G1 ($v_{hel} \simeq 
25$\,\kms) corresponds to the GMC complexes A and B, while G2
($v_{hel} \simeq 50$\,\kms) corresponds to the GMC complexes C and D.
\cite{Meier2011} and \cite{MonteroCastano2006} note that the HC$_3$N
J=$16-15$ and NH$_3$ (6,6) transitions, respectively, are most intense
toward GMC complex C, which corresponds to the NH$_3$ velocity
component G2 near $v_{hel} \simeq 50$\,\kms.

\cite{Lebron2011} derive T$_{rot} \simeq$ 25--35\,K from their NH$_3$
(1,1) and (2,2) images, limited by the low excitation of these two
NH$_3$ transitions.  The high-excitation NH$_3$ (6,6) (E$_u = 408$\,K)
measurements reported by \cite{MonteroCastano2006} suggest higher
kinetic temperatures within both of the velocity components within
this galaxy.  Using their images of the HC$_3$N J=$5-4$, $10-9$
\citep{Meier2005}, and $16-15$ emission, \citet{Meier2011} constrain
LVG models of the spatial density and kinetic temperature toward the
GMC complexes in this galaxy.  They derive kinetic temperatures in the
range 30--70\,K and spatial densities in 
the range $10^3$--10$^5$\,cm$^{-3}$.  The lower spatial densities in
these LVG models correspond to higher kinetic temperatures, and
vice-versa, which is a reflection of the degeneracy within LVG models
which attempt to derive both kinetic temperature and spatial density
from molecular transitions with coupled sensitivity to both physical
parameters.

Our NH$_3$-constrained derivations of the kinetic temperatures from
the two velocity components G1 and G2 point to the existence of kinetic
temperature gradients or multiple temperature components.  The
$v_{hel} \simeq 25$\,\kms\ (G1, or nuclear star forming region)
component clearly possesses multiple kinetic temperature components.
Individual NH$_3$ transition ratios result in the following LVG model
derived kinetic temperatures for this velocity component:
\begin{itemize}
\item T$_K = 24\pm7$\,K using (1,1)/(2,2)
\item T$_K = 115\pm17$\,K using (2,2)/(4,4)
\item T$_K > 188$\,K using (4,4)/(5,5)
\item T$_K > 141$\,K using (3,3)/(6,6)
\end{itemize}
\noindent{suggesting} that there are at least three temperature
components comprising the G1 velocity component of IC\,342.  The
$v_{hel} \simeq 50$\,\kms\ (G2, or NE spiral arm)
component also appears to possess 
multiple kinetic temperature components.  Individual NH$_3$ transition
ratios result in the following LVG model derived kinetic temperatures
for this velocity component:
\begin{itemize}
\item T$_K > 90$\,K using (1,1)/(2,2)
\item T$_K = 75\pm14$\,K using (2,2)/(4,4)
\item T$_K > 300$\,K using (4,4)/(5,5)
\item T$_K > 186$\,K using (3,3)/(6,6)
\end{itemize}
\noindent{suggesting} that there are at least two temperature
components comprising the G2 velocity component of IC\,342.  Note also
that the G2 velocity component is a source of weak NH$_3$ (2,1) emission,
suggesting intense infrared emission associated with this component.
This fact, coupled with the lack of low-temperature gas in the
G2 component suggests a more energetic star formation event in the
NE spiral arm region of IC\,342 than that found in the nuclear
region of this galaxy.

\begin{figure}
\centering
\includegraphics[scale=0.45]{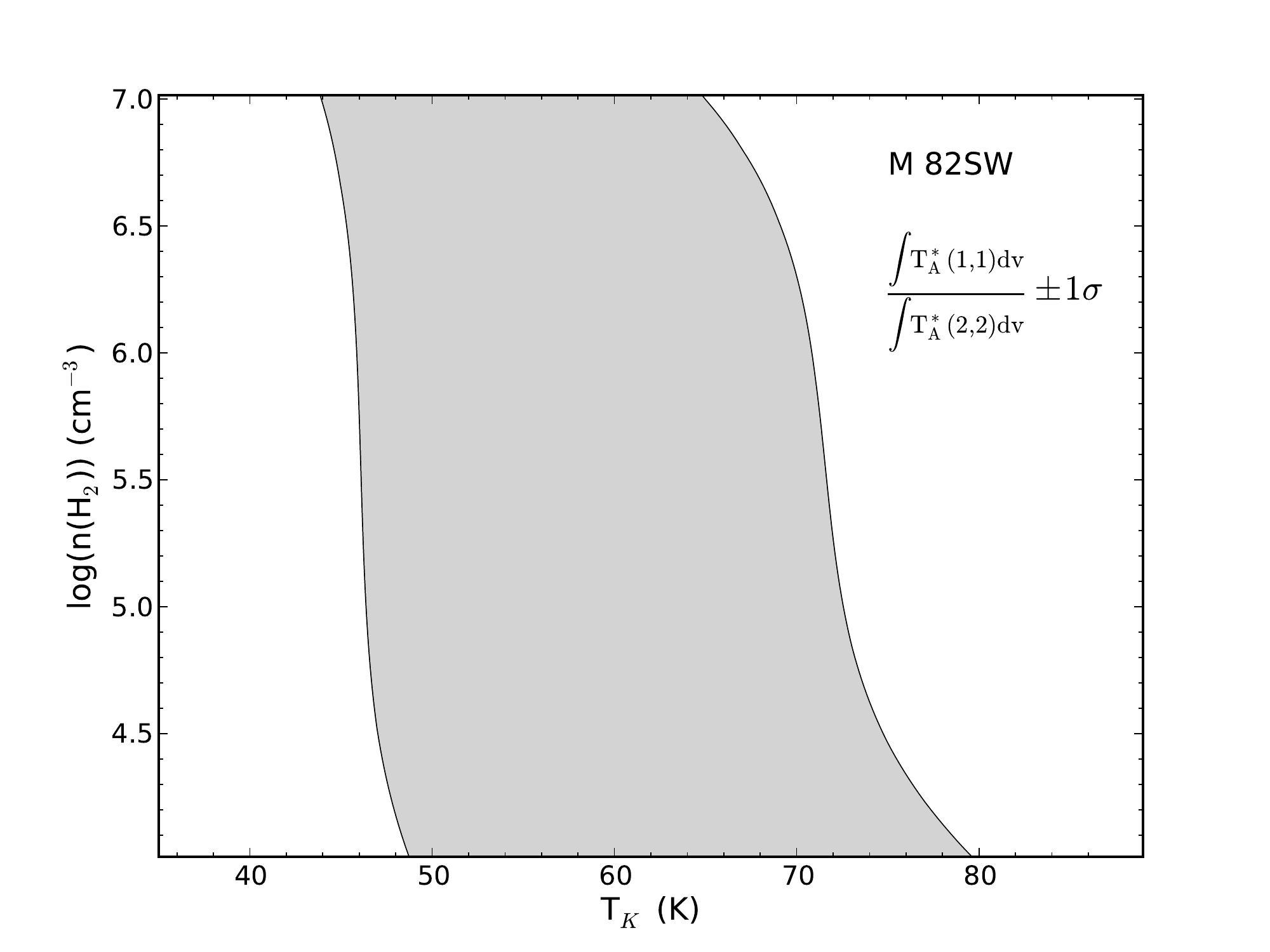}
\caption{LVG model kinetic temperature fit to M\,82SW.  Shown as grey
  contours is the best-fit ratio within one standard deviation for the
  NH$_3$ (1,1) and (2,2) emission toward M\,82SW.}
\label{fig:M82SWTkFit}
\end{figure}

\medskip\noindent
{\bf M\,82SW:} In addition to the central position in M\,82 (a
non-detection), we searched for NH$_3$ emission at a position offset
of ($-12$,$-4$) 
arcsec from the nominal M\,82 position known to be a source of NH$_3$
emission \citep[][ identified as the ``southwestern molecular
  lobe'']{Weiss2001}. The $\theta_b \simeq 40^{\prime\prime}$ and
$\Delta v \simeq$ 16\,\kms\ measurements of the (1,1), (2,2),
(3,3), and (4,4) transitions by \cite{Weiss2001} resulted in a kinetic
temperature of $\simeq 60$\,K based on a rotation diagram analysis of
their (1,1), (2,2), and (3,3) spectra (they did not detect the (4,4)
transition).  The LVG model fit to our NH$_3$ (1,1) and (2,2) spectra
result in a similar kinetic temperature (Figure~\ref{fig:M82SWTkFit}).
High spatial resolution measurements of dense gas tracers in
M\,82 \citep[see summary in][]{Mangum2013} suggest that the
molecular emission extends over a
$40^{\prime\prime}\times15^{\prime\prime}$ disk-like structure.

\medskip\noindent
{\bf NGC\,3079:}  The NH$_3$ absorption toward NGC\,3079 possesses the
same two-component spectral structure as that observed in the H$_2$CO 
$2_{11}-2_{12}$ measurements reported in \citet{Mangum2013}.  The
H$_2$CO $1_{10}-1_{11}$ measurements reported in
\citet{Mangum2008,Mangum2013} also have the same velocity structure as
that observed in our NH$_3$ measurements, with the exception that
there is a tentative, weak H$_2$CO $1_{10}-1_{11}$ absorption component at
$v_{hel} \simeq 1500$\,\kms\ that is just outside the bandwidth of our NH$_3$
measurements.  We are not aware of any previous NH$_3$ measurements
toward this galaxy.  The background continuum source in NGC\,3079 is
estimated to be $\lesssim 0.5^{\prime\prime}$ in extent
\citep[see][]{Duric1983,Kukula1995}.

Unexplained is the non-detection of NH$_3$ (3,3) toward this galaxy,
even though we detect the higher-excitation (6,6) and (9,9)
transitions.  Attempts were made to detect NH$_3$ (3,3) using four
different correlator configurations spanning four observing runs
during the time period 2011/10/30 through 2011/11/04, one
of which was tuned exclusively to the (3,3) transition.  During two of
the four observing runs NH$_3$ (3,3) was detected in three other
targets; the Galactic molecular cloud W3(OH) and starburst galaxies
NGC\,253 and NGC\,660.  These checks of the observing setup involving
NH$_3$ (3,3) appear to eliminate faulty observing configuration as the
reason for the non-detection of NH$_3$ (3,3) in NGC\,3079.

With the exception of ad-hoc and improbable explanations which involve
inconsistent source size distributions between NH$_3$ (3,3) and all
other para- and ortho-NH$_3$ transitions, our non-detection of NH$_3$
(3,3) toward NGC\,3079 is difficult to explain.  One possibility is
the existence of a separate physical component within NGC\,3079 which
has physical conditions which result in NH$_3$ (3,3) maser emission
which cancels any NH$_3$ (3,3) absorption from the physical structures
which also produce absorption in the other NH$_3$ transitions.  To
study this possibility we have investigated the optical depth values
produced by our LVG models of the NH$_3$ (3,3), (6,6), and (9,9)
transitions.

\begin{figure}
\centering
\includegraphics[scale=0.45]{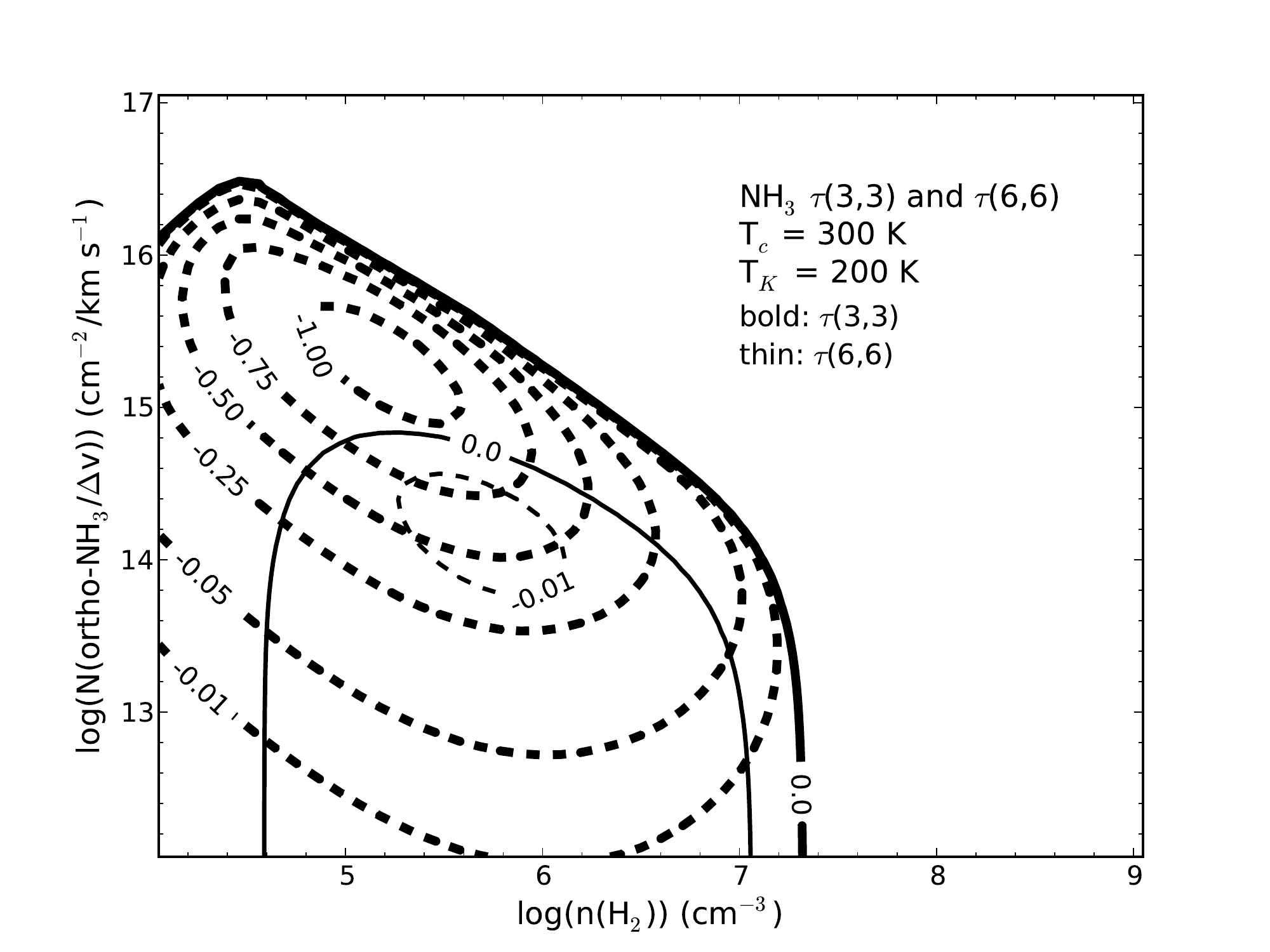}
\caption{LVG model predictions of NH$_3$ (3,3) and (6,6) maser ($\tau < 0$)
  emission for an assumed T$_c$ = 300\,K and T$_K$ = 200\,K.}
\label{fig:ONH3TauMaser}
\end{figure}

As there is only a minimal dependence of the absolute NH$_3$ (3,3),
(6,6), and (9,9) optical depths on the model input continuum
background temperature T$_c$, we have restricted our optical depth
comparison as a function of spatial density, NH$_3$ column density,
and kinetic temperature to T$_c = 300$\,K.  For this assumed
background continuum temperature, our LVG models predict:
\begin{itemize}
\item $\tau$(3,3) $\leq 0$ for T$_K \gtrsim 25$\,K and a range of
  n(H$_2$) and N(ortho-NH$_3$)/$\Delta$v which peaks at $\tau$(3,3)
  $\simeq -1.0$ near n(H$_2$) $\simeq 10^{5.25}$\,cm$^{-3}$ and
  N(ortho-NH$_3$) $\simeq 10^{15.5}$\,cm$^{-2}$/\kms\ and widens in
  (n,N) as T$_K$ increases.
\item $\tau$(6,6) $\leq 0$ for T$_K \gtrsim 100$\,K and a range of
  n(H$_2$) and N(ortho-NH$_3$)/$\Delta$v which peaks at $\tau$(6,6)
  $\simeq -0.01$ near n(H$_2$) $\simeq 10^{5.75}$\,cm$^{-3}$ and
  N(ortho-NH$_3$) $\simeq 10^{14.25}$\,cm$^{-2}$/\kms\ and, like
  $\tau$(3,3), widens in (n,N) as T$_K$ increases.
\item $\tau$(9,9) is never less than 0 over the range of physical
  conditions modelled.
\end{itemize}
Figure~\ref{fig:ONH3TauMaser} shows the LVG modelled optical depth
when $\tau\leq 0$ for the NH$_3$ (3,3) and (6,6) transitions at a
representative T$_K$ of 200\,K (appropriate for NGC\,3079).  As the
LVG model predictions suggest that it is relatively easy to drive the
NH$_3$ (3,3) transition into maser emission, a model which produces
cancelling NH$_3$ (3,3) emission and absorption from two spectral
components, with the same velocity and line width, 
along the line-of-sight seems plausible.  Furthermore, this scenario is
lent some credence by the fact that the H$_2$CO $1_{10}-1_{11}$
transition profile presented in \citet{Mangum2013} suggests both
emission and absorption contributions.  Low spatial densities
(n(H$_2$) $\lesssim 10^{5.5}$\,cm$^{-3}$) result in H$_2$CO
$1_{10}-1_{11}$ absorption, while higher spatial densities (n(H$_2$)
$\gtrsim 10^{5.5}$\,cm$^{-3}$) result in H$_2$CO $1_{10}-1_{11}$ emission.

Our NH$_3$ (1,1) through (9,9) spectra provide only lower limits of
100 and 150\,K to the kinetic temperature in both of the v$_{opt}
\simeq 1020$\,\kms\ (NGC\,3079G1) and 1115\,\kms\ (NGC\,3079G2)
velocity components.  The highest-excitation para-NH$_3$ ((5,5)/(7,7))
and ortho-NH$_3$ ((6,6)/(9,9)) ratios imply T$_K \gtrsim 230$\,K for
both velocity components.

\medskip\noindent
{\bf IC\,860:}  The NH$_3$ absorption in this galaxy, which is to our
knowledge the first reported, corresponds
spectrally to the H$_2$CO $1_{10}-1_{11}$ emission and $2_{11}-2_{12}$
absorption components \citep{Mangum2013}.  The background continuum
source size is estimated to be $\lesssim 0.5^{\prime\prime}$
\citep[see][ and references therein]{Condon1990}.  The LVG-derived
kinetic temperature fit to our ortho- and para-NH$_3$ (1,1) through
(7,7) measurements is poorly constrained due to the influence of
multiple high-temperature components.  The (5,5)/(7,7) ratio
implies a kinetic temperature $\gtrsim 250$\,K, while all other line
ratios are best described by T$_K$ in the range $206\pm79$\,K.

\begin{figure}
\centering
\includegraphics[scale=0.45]{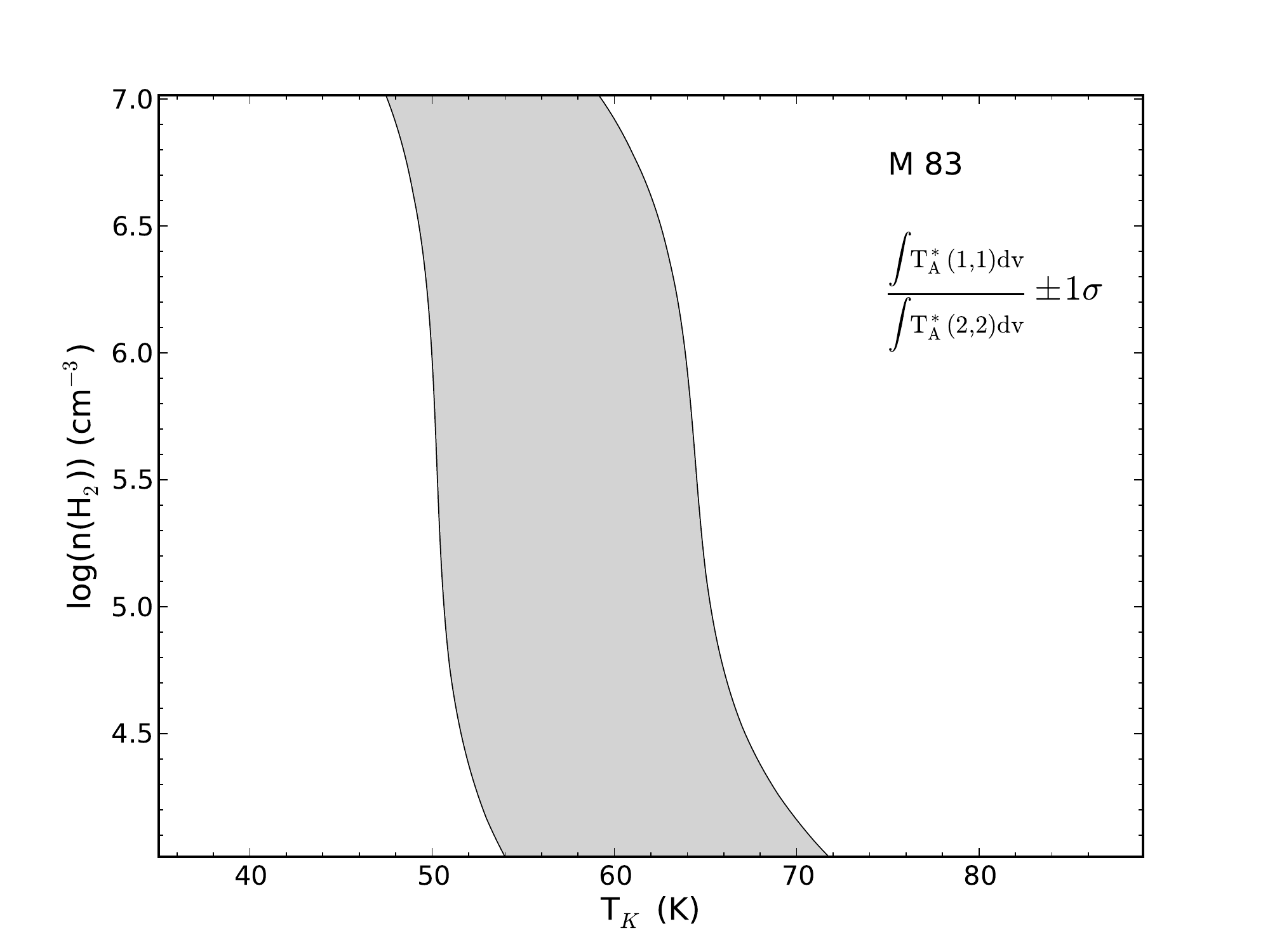}
\caption{LVG model kinetic temperature fit to M\,83.  Shown as grey
  contours is the best-fit ratio within one standard deviation for the
  NH$_3$ (1,1) and (2,2) emission toward M\,83.}
\label{fig:M83TkFit}
\end{figure}

\medskip\noindent
{\bf M\,83:}  The quality of the kinetic temperature fit for this
galaxy is good, well constrained by the NH$_3$ (1,1) and (2,2)
transitions detected (Figure~\ref{fig:M83TkFit}).  We are not aware of
any previous measurements of NH$_3$ toward this galaxy.  High
spatial resolution measurements of dense gas tracers in M\,83
\citep[see summary in][]{Mangum2013} suggest that the molecular
emission extends over a $45^{\prime\prime}\times15^{\prime\prime}$
structure.

\medskip\noindent
{\bf IR\,15107+0724:} The NH$_3$ (1,1), (2,2), and (4,4) absorption
spectra measured toward this galaxy, representing new
discoveries of NH$_3$, have the same velocity
structure measured in the H$_2$CO $1_{10}-1_{11}$ and $2_{11}-2_{12}$
transitions reported in \citet{Mangum2013}.  Due to the relative
weakness of the NH$_3$ (4,4) absorption ($\sim 4.4\sigma$ in
integrated intensity), the kinetic temperature derived is based on the
NH$_3$ (1,1) and (2,2) measurements toward this galaxy.  The
background continuum source size is estimated to be $\lesssim
1^{\prime\prime}$ \citep[see][ and references therein]{Condon1990}.

\medskip\noindent
{\bf Arp\,220/IC\,4553:} \citet{Mangum2008,Mangum2013} report
detections of the H$_2$CO $1_{10}-1_{11}$ transition in emission and
the $2_{11}-2_{12}$ transition in absorption.  These H$_2$CO
measurements exhibit a primarily two-velocity component structure with
v$_{hel} \sim 5350$ and 5500\,\kms, with a tentative detection of a
third component near v$_{hel}\sim 5600$\,\kms.  This primarily
two-component velocity structure has been noted in previous studies of
the high-density emission from Arp\,220 \citep[HCN, HCO$^+$,
HNC; see ][]{Greve2009},
whereby the $\sim 5350$\,\kms\ component has been associated with the
western nucleus of the merger, while the eastern nucleus produces the
$\sim 5500$\,\kms\ velocity component.  The \citet{Greve2009} analysis
of these two velocity components suggests that the western nucleus
possesses a lower spatial density than the eastern nucleus, but our
H$_2$CO measurements do not necessarily support this interpretation
\citep{Mangum2008,Mangum2013}.  The background continuum source size
is estimated to be $\lesssim 0.6^{\prime\prime}$ \citep[see][ and references
  therein]{Ott2011}.

Our NH$_3$ measurements replicate the two-velocity component structure
noted in previous studies (see Table~\ref{tab:nh3results} 
and Figure~\ref{fig:Arp220NH3Spec}).  The NH$_3$ (1,1) through (8,8)
metastable (J=K) transitions all exhibit v$_{hel} \sim
5350$ and 5500\,\kms\ components.  The non-metastable NH$_3$ (10,9)
(in the same spectrum as the NH$_3$ (4,4) transition; see
Figure~\ref{fig:Arp220NH3Spec}) and OH $^2\Pi_{3/2}$\,J=9/2\,F=$5-5$
transition (in the same spectrum as NH$_3$ (3,3); see
Figure~\ref{fig:Arp220NH3Spec}) only appear to have v$_{hel} \sim
5500$\,\kms\ components.  A broad ($\Delta v \simeq 530$\,\kms)
low-velocity component with v$_{hel} \simeq 5200$\,\kms\ is apparent
in our NH$_3$ (1,1)/(2,2) spectrum.  This component might be a broad
blue-shifted wing of a dense molecular outflow.

To investigate the possible sources of this broad low-velocity
spectral component in Arp\,220, we consider two possibilities:
\begin{itemize}
\item \textit{There is an additional molecular or atomic species that
  produces this spectral feature.}  A search of molecular and atomic
transition frequency lists indicates that only the H109$\epsilon$
recombination line at 23743.83\,MHz, which would correspond to
emission from the G2 (v$_{hel} \simeq 5500$\,\kms) component in
Arp\,220, could produce the additional broad, low-velocity component
we measure.  Radio recombination transitions have been detected toward
Arp\,220 \citep{Mangum2013,Anantharamaiah2000}, but only from the
lower-excitation H-$\alpha$ transitions.  Furthermore, these H-$\alpha$
recombination transitions have FWHM line widths $\sim 200$\,\kms, much
narrower than the FWHM $\sim 530$\,\kms\ H$_2$CO component we measure.  
\item \textit{The (1,1) transition is anomalously weak, or that
  the (2,2) transition is anomalously broad and intense.}
  \cite{Ott2011} proposed that the lower intensity of the NH$_3$ (1,1)
  transition was due to an undetected cold (T$_K \lesssim 20$\,K)
  NH$_3$ emission component that is not spatially aligned with the
  background continuum source in Arp\,220, but is spectrally
  overlapped with the main NH$_3$ (2,2) components.  An alternative
  explanation is that the NH$_3$ (2,2) transition is anomalously
  broadened and intensified by a third velocity component.  This
  scenario seems to be supported by the evidence for a broad velocity
  component shifted by approximately 200\,\kms\ relative to the
  nominal v$_{hel} = 5290$ and 5460\,\kms\ velocity components in our
  NH$_3$ (1,1) and (2,2) spectrum (Figure~\ref{fig:Arp220NH3Spec}).
  Furthermore, \textit{Herschel}/PACS spectroscopic measurements of OH
  (163\,$\mu$m), [O I] (145\,$\mu$m), and H$_2$O (82--90\,$\mu$m)
  toward Arp\,220 point to a blueshifted 550--1000\,\kms\ wide
  velocity component offset by $\sim 400$\,\kms\ from the systemic
  velocity of this merger system \citep{GonzalezAlfonso2012}.  These
  \textit{Herschel}/PACS observations appear to be consistent with an
  interpretation whereby our GBT NH$_3$ (1,1) and (2,2) measurements
  possess a blueshifted wide-linewidth component which can be
  interpreted as evidence for a dense outflow in Arp\,220.
\end{itemize}

\begin{figure}
\centering
\includegraphics[scale=0.45]{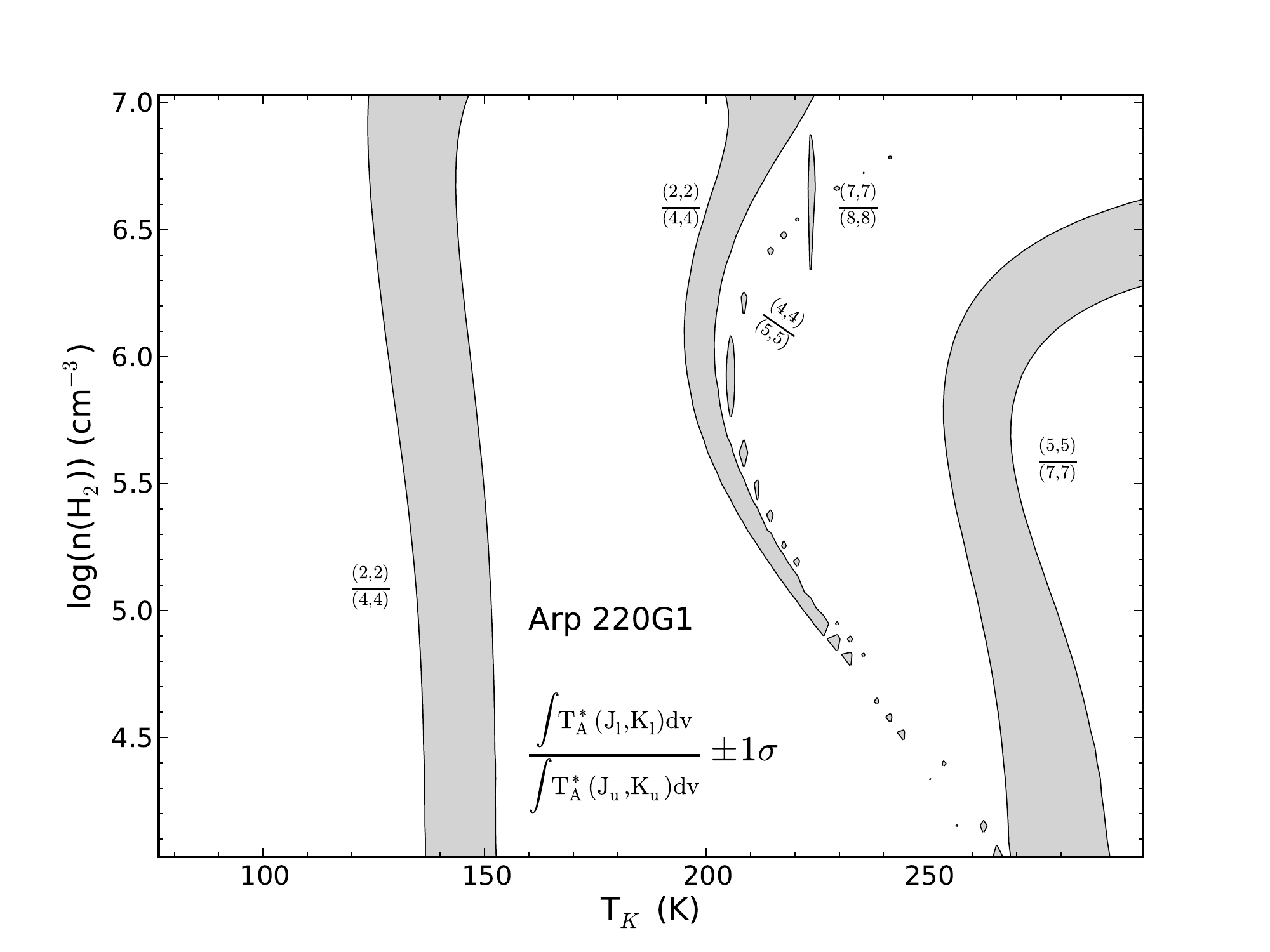}
\caption{LVG model kinetic temperature fit to the G1 (v$_{opt} \simeq
  5350$\,\kms) component of Arp\,220.  Shown as grey contours are the
  best-fit ratios within one standard deviation for the (2,2) through (8,8)
  transitions.}
\label{fig:Arp220G1TkFit}
\end{figure}

A global single-temperature fit to either velocity component or the
total Arp\,220 NH$_3$ emission fails due to the apparent existence of
kinetic temperature structure.  Para-NH$_3$ transition ratio fits
to the v$_{hel} \sim 5350$\,\kms\ (Arp\,220G1) component result in an
apparent two-temperature structure: T$_K$ = $136\pm18$\,K derived from
NH$_3$ (2,2) and (4,4), T$_K = 223\pm30$\,K derived from all five
para-NH$_3$ transitions (Figure~\ref{fig:Arp220G1TkFit}).  For the
v$_{hel} \sim 5500$\,\kms\ (Arp\,220G2) component a two-temperature
structure is 
also observed: T$_K = 153\pm20$\,K, derived from NH$_3$ (1,1) and
(2,2), T$_K = 236\pm40$\,K, derived from NH$_3$ (4,4) through (8,8).
Fits to the ortho-NH$_3$ (3,3) and (6,6) ratio proved to be very
complex, yielding a good fit only for T$_K > 90$\,K and $156\pm15$\,K
for Arp\,220G1 and G2, respectively, over n(H$_2$) $>
10^{6.7}$\,cm$^{-3}$ and N $< 10^{16.75}$\,cm$^{-2}$/\kms\ for G1 and
n(H$_2$) $\simeq 10^{6.4}$\,cm$^{-3}$ and N $\gtrsim
10^{16.9}$\,cm$^{-2}$/\kms\ for G2.  As noted in the discussion about
the non-detection of NH$_3$ (3,3) absorption toward NGC\,3079, the
ortho-NH$_3$ (3,3) and (6,6) transitions mase at a wide range of
spatial densities, NH$_3$ column densities, and kinetic temperatures.
As Arp\,220 is known to be a spatially complex source, including
spatial densities which are apparently high enough to produce H$_2$CO
$1_{10}-1_{11}$ emission \citep{Mangum2013}, it seems plausible to
associate the complex and nearly uninterpretable NH$_3$ (3,3) and
(6,6) absorption from this merger as being due to spatial and kinetic
structure.

The kinetic temperature derived from our GBT NH$_3$ (1,1) through (8,8)
measurements is consistent, though somewhat larger, than the rotation
diagram analysis of the kinetic temperature of $186\pm55$\,K derived
from ATCA measurements of the (2,2) through (6,6) transitions of para-
and ortho-NH$_3$ presented by \citet{Ott2011}.  Clearly the inclusion
of higher excitation NH$_3$ transitions in the present analysis has
driven the derived kinetic temperature higher.

\begin{figure}
\centering
\includegraphics[scale=0.45]{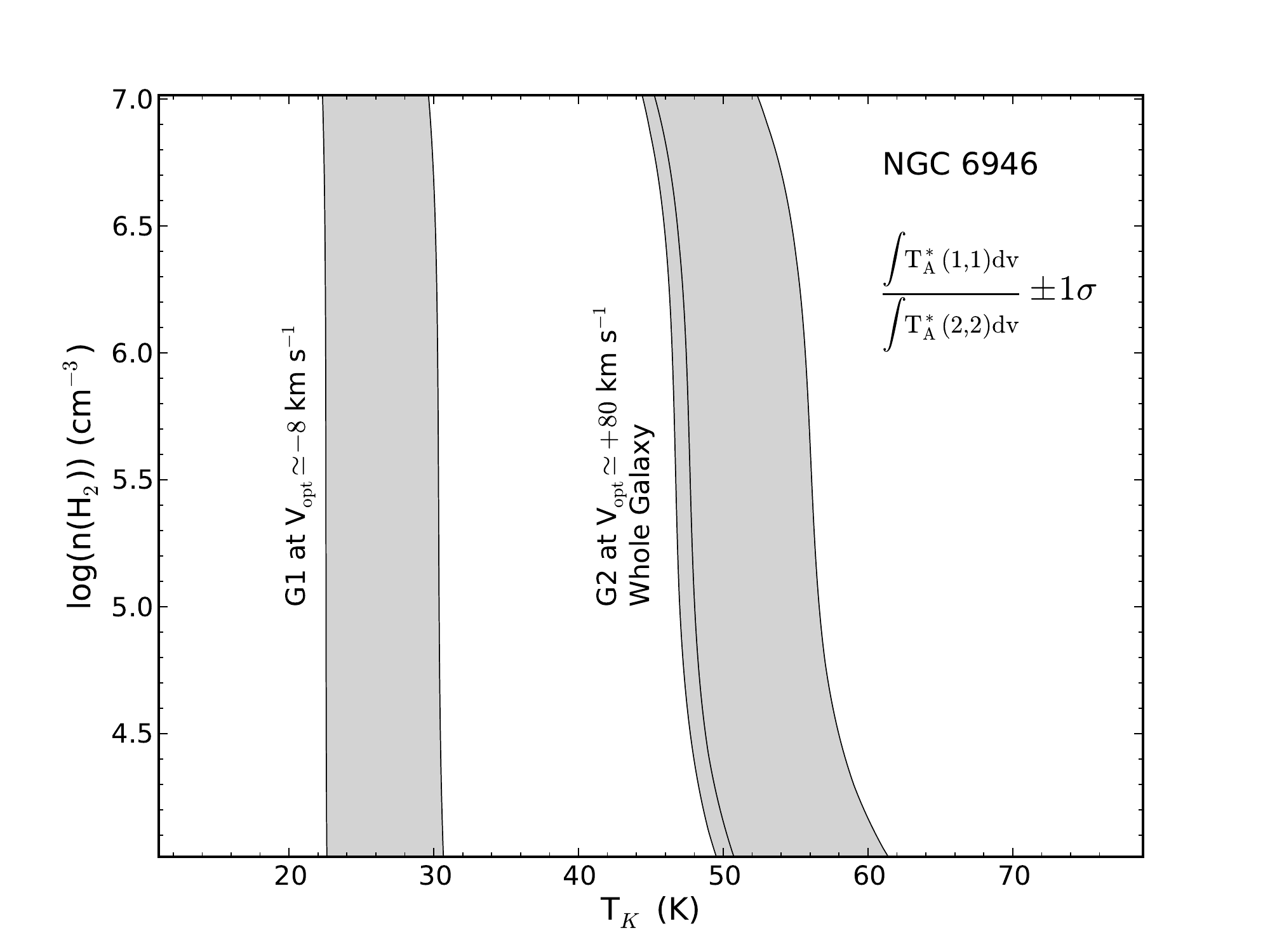}
\caption{LVG model kinetic temperature fit to the individual velocity
  components in NGC\,6946.  Shown as grey contours are the best-fit
  ratio within one standard deviation for the two velocity components
  of NGC\,6946 and for the galaxy as a whole.}
\label{fig:NGC6946TkFit}
\end{figure}

\medskip\noindent
{\bf NGC\,6946:}  \cite{Mangum2008,Mangum2013} report detections of
the H$_2$CO $1_{10}-1_{11}$ and $2_{11}-2_{12}$ transitions in
absorption toward this galaxy, where a two-velocity component
structure was noted with $v_{hel} \simeq -8$ and $\simeq +80$\,\kms.
Our measurements of the NH$_3$ (1,1) and (2,2) transitions are
consistent with this two-velocity component structure.  The quality of
the kinetic 
temperature fits to both velocity components for this galaxy are good,
well constrained by the NH$_3$ (1,1) and (2,2) transitions detected,
suggesting that the $+80$\,\kms\ component (NGC\,6946\,G2) is
significantly warmer than the $-8$\,\kms\ component (NGC\,6946\,G1;
see Figure~\ref{fig:NGC6946TkFit}).  We are not aware of any previous
measurements of NH$_3$ toward this galaxy.  High spatial
  resolution measurements of dense gas tracers in NGC\,6946 \citep[see
    summary in][]{Mangum2013} suggest that the molecular emission
  emanates from two compact components: a compact nuclear source of
  size $\sim 2^{\prime\prime}$ and a ``nuclear spiral'' which is $\sim
  5^{\prime\prime}\times10^{\prime\prime}$ in extent.

\medskip\noindent
{\bf Non-Detected Galaxies:} Ten galaxies were not detected in NH$_3$
to single-channel ($\Delta v \simeq 5-10$\,km\,s$^{-1}$) RMS levels
of $\sim 0.8$ to 2\,mK (see 
Table~\ref{tab:nh3results}).  Six of these ten galaxies: NGC\,598,
IR\,01418+1651, NGC\,3690, NGC\,4214, NGC\,4418, and
NGC\,6951, are not known to be dense gas emission sources.  Two
galaxies: NGC\,598 
and IR\,01418+1651, were also not detected in the H$_2$CO survey of
\cite{Mangum2008,Mangum2013}.  Mrk\,231 is known to be a
  source of dense gas emission \citep[\eg ][]{Aalto2012}, but was not
  included in this H$_2$CO survey.  The remaining three galaxies:
NGC\,1144, NGC\,2146, and NGC\,3628, are all detected in H$_2$CO
$1_{10}-1_{11}$ absorption, and with the exception of NGC\,1144 also
detected in $2_{11}-2_{12}$ absorption \citep{Mangum2013}.  All three
galaxies are also HCN emission sources \citep{Gao2004a}.  The measured
H$_2$CO $1_{10}-1_{11}$ intensities of NGC\,1144, NGC\,2146, and NGC\,3628 are not
particularly large (peak T$^*_A \simeq -1$ to $-8$\,mK), implying
H$_2$CO column densities N(ortho-H$_2$CO)/$\Delta$v $\simeq
10^{11}$\,cm$^{-2}$\,km\,s$^{-1}$, which are on the low end of the
range of H$_2$CO-detected galaxies \citep{Mangum2013}.  Although we
might have hoped to detect weak NH$_3$ emission from these three
galaxies, the fact that we did not detect NH$_3$ in these galaxies is
not a surprise.

\section{Detection of OH $^2\Pi_{3/2}$\,J=9/2 Absorption Toward
  Arp\,220}
\label{Arp220OH}

\citet{Ott2011} reported a tentative detection of the F=$4-4$ and
F=$5-5$ magnetic hyperfine structure transitions from the
$^2\Pi_{3/2}$\,J=9/2 lambda doublet state of OH toward Arp\,220.
The \cite{Ott2011} detection of this OH doublet, which lies at an
energy above ground of $\sim 512$\,K, was considered tentative as they
detected an absorption line at approximately the correct frequency in
their Australia Telescope Compact Array (ATCA) spectra, but not in
their GBT spectra measuring the same frequency.  Our measurements of
the NH$_3$ (3,3) transition toward Arp\,220 include the frequency
range occupied by the OH $^2\Pi_{3/2}$\,J=9/2 doublet
(Figure~\ref{fig:Arp220NH3Spec}).  The absorption component located
$\sim +650$\,\kms\ from the NH$_3$ (3,3) absorption line is best-fit
to a v$_{opt} = 5464.1\pm13.2$\,\kms\ (Table~\ref{tab:nh3results}) at
the rest frequency of the OH $^2\Pi_{3/2}$\,J=9/2\,F=$5-5$ transition
(23826.6211\,MHz; see Table~\ref{tab:frequencies}).  The F=$4-4$
component would be located $\sim +150$\,\kms\ ($\sim 10$\,MHz) from
the F=$5-5$ transition.  Examination of the NH$_3$ (3,3) spectrum in
Figure~\ref{fig:Arp220NH3Spec} suggests that the F=$4-4$ component may
also have been detected, but that the purported OH
$^2\Pi_{3/2}$\,J=9/2 doublet absorption is dominated by the F=$5-5$
transition.  Note that absorption in the far infrared rotational OH
transitions from both the $^2\Pi_{3/2}$ and $^2\Pi_{1/2}$ states with
upper state energies as high as $\sim 900$\,K have been measured 
toward Arp\,220 \citep{GonzalezAlfonso2012}.  A significant
contributor to this high-excitation OH absorption appears to originate
from a dense outflow component in Arp\,220.  Other hyperfine
transitions from within rotationally excited OH doublet states within 
both the $^2\Pi_{3/2}$ and the $^2\Pi_{1/2}$ ladders have recently been
detected toward Arp\,220 by \citet{Salter2008} with the Arecibo 305
meter telescope.

\section{The Remarkable NH$_3$ Absorption Toward NGC\,660}
\label{NGC660NH3}

NGC\,660, perhaps best known for its exquisite crossing dust lanes, is
a nearly edge-on SB(s)a LINER galaxy with a polar ring. The radio
continuum images presented by \cite{Condon1980} and
\cite{Condon1982,Condon1990} show bright, compact emission from the
nucleus of NGC\,660.  The nuclear region harbors a continuum source
which has been interpreted as a ring of high density and compact star 
formation regions \citep{Carral1990}.  There is no observed central,
compact core to indicate activity from an AGN and no X-ray counterpart
to the nucleus \citep{Dudik2005}. The relatively high infrared
luminosity ($10^{10.3} L_{\odot}$), optical LINER spectrum, and
indications that star formation processes dominate the contribution to
the radio flux establish NGC\,660 as a moderate starburst.  H$_2$CO
\citep{Mangum2008,Mangum2013}, NH$_3$ (this work), OH, and HI
\citep{Baan1992} have all been observed in absorption against this
compact continuum source.

An extensive study of NGC\,660 has been carried out by
\cite{vanDriel1995}.  The disk is seen almost edge-on with inclination
$i \sim 70^{\circ}$ at a position angle of $\mbox{PA} \approx
45^{\circ}$. The polar ring is inclined on average
$i \approx 55^{\circ}$ at $\mbox{PA} \approx 170^{\circ}$.  A two
component dust model with $T_{d} \approx 19$ K and $T_{d} \approx 50$
K has been fit to the millimeter/submillimeter dust emission from
NGC\,660 \citep{Chini1993}, while a single temperature dust model
yields $T_{d} \approx 37$ K. 

High spatial resolution imaging of the HI and OH absorption from the
nuclear region of NGC\,660 have been carried out by \cite{Baan1992a}.
H$\alpha$ images show active star formation and HII regions throughout
the polar ring \citep{vanDriel1995}.  The rotation curve, mass of the
polar ring ($M_{ring} \sim 15.5\times10^9$\,\msun), and mass of the
disk ($M_{disk} \sim 19.4\times10^9$\,\msun ) have been derived 
from HI observations \citep{vanDriel1995}.  An outflow has been
observed normal to the disk axis \citep{Heckman2000}.

Radio continuum images of the nuclear region of NGC\,660 show compact
($\theta \lesssim 6^{\prime\prime}$) morphological components which
can be interpreted as an edge-on ring or linear jet in the
northeast--southwest direction.  Without a clearly identifiable radio
core, indicative of an AGN, large complexes of high-density star
formation regions are proposed as the source of the compact radio
emission \citep{Carral1990}. The optical nucleus is located $\sim
3^{\prime\prime}$ south of the strongest 15\,GHz continuum peak
\citep{Filho2002}.

To further understand the radio continuum emission source we have
compiled measurements of the radio continuum flux from the literature
to calculate the spectral index in the frequency range 408\,MHz to
5\,GHz measured with similar spatial resolution
(Table~\ref{tab:specindex}). The best fit spectral index is $\alpha =
0.64^{+0.06}_{-0.07}$.  \cite{Condon1982} derived a spectral index of
$\alpha = 0.6$ using 1.4 
and 4.85\,GHz VLA measurements, while \cite{vanDriel1995} derived a mean
radio spectral index of $\alpha=0.57$.  The observed slope of
the radio spectral index is indicative of non-thermal synchrotron
radiation from optically thin supernova remnants \citep[SNR; $\alpha \simeq
0.7$][]{Filho2002}.  However, the fact that the spectral index is a
bit shallower than pure synchrotron suggests that there
is a non-negligible contribution from thermal free-free
emission. This scenario is supported by H92$\alpha$ radio
recombination line emission measurements from NGC\,660, indicative of
high density ionized gas \citep{Phookun1998}. 

%
\begin{deluxetable}{lccl}
\tabletypesize{\small}
\tablewidth{0pt}
\tablecolumns{4}
\tablecaption{Continuum Measurements of NGC\,660\label{tab:specindex}}
\tablehead{
\colhead{$\nu$} & \colhead{S$_\nu$} & \colhead{$\theta_b$} &
\colhead{Reference} \\
\colhead{(MHz)} & \colhead{(mJy)} &&
}
\startdata
408  & 830$\pm$70 & $2^\prime 30^{\prime\prime}$ & \cite{Large1981} \\
2400 & 255$\pm$13 & $2^\prime 42^{\prime\prime}$ & \cite{Dressel1978} \\
4850 & 187$\pm$26 & $3^\prime 30^{\prime\prime}$ & \cite{Gregory1991} \\
4850 & 184$\pm$28 & $3^\prime 30^{\prime\prime}$ & \cite{Becker1991} \\
5000 & 156$\pm$18 & $2^\prime 42^{\prime\prime}$ & \cite{Sramek1975} \\
\enddata
\end{deluxetable}

The NH$_3$ spectra from NGC\,660 possess a high degree of structure
(see Figure~\ref{fig:NGC660NH3spec}).  Furthermore, although the
NH$_3$ and H$_2$CO absorption features overlap in velocity, they differ
in velocity width by almost an order of magnitude; FWZI(NH$_3$)
$\simeq 150$ to 325\,\kms\ (Table~\ref{tab:nh3results}) and FWZI(H$_2$CO)
$\simeq 1000$\,\kms\ (Table~2 of \citet{Mangum2013}).  These facts
suggest that the NH$_3$ and H$_2$CO transitions probe different
spectral components of the galaxy.

Spatially resolved measurements of the nuclear region of NGC\,660 make
it possible to connect the NH$_3$ spectral components to spatial
components in the galaxy.  \cite{Baan1992a} have imaged the HI and OH
absorption toward the nucleus of NGC\,660 at $\theta_b =
1.^{\prime\prime}47$ using the VLA in its A-configuration.  These
observations have been supplemented with single dish observations from
the Arecibo Observatory.  \cite{Baan1992a} identify 12 HI absorption
features and 7 distinct OH absorption features.  The HI and OH
absorption features have FWHM $\approx 90$ km\,s$^{-1}$, similar to the
FWHM $\sim 30 - 80$ km\,s$^{-1}$ of our NH$_3$ observations.  The
approximately 30 km s$^{-1}$ discrepancy between the HI/OH and NH$_3$
line widths can be attributed to HI and OH tracing a less dense phase
of the Interstellar Medium (ISM) than NH$_3$.

\cite{Baan1992a} place the HI features with central velocities at 675,
687, 761, 862, 965, 974, 1002, and 1018\,\kms\ in the disk; 768,
851, and 870\,\kms\ in the polar ring\footnote{\cite{Baan1992a} refer
  to the part of the polar ring component which passes in front of the
  nucleus as the ``warp''.}; and a feature at 720\,\kms\
in an ``anomalous ridge''. \cite{Baan1992a} also place the OH features
with central velocities at 675, 765, 923, 992, and 1008\,\kms\ in the
disk; one at 861\,\kms\ in the polar ring; and one at 750\,\kms\ in the
anomalous ridge.  The NH$_3$ absorption features presented in this work have
velocity centroids between $765 - 769$\,\kms\ (G1), $785 -
800$\,\kms\ (G2), $828 - 839$\,\kms\ (G3), and $979 -
989$\,\kms\ (G4). Comparing the central velocities of our NH$_3$
measurements to those of \cite{Baan1992a}, we place G1 and G4 in the
disk and find the location of G2 to be indeterminate.

For the NH$_3$ G3 component, the HI observations are ambiguous with a
disk component at 862\,\kms\ and polar ring components at
851\,\kms\ and 870\,\kms. The OH observations suggest clear placement
in the polar ring for G3, but keep in mind that the OH measurements of
the polar ring are actually measuring the part of this component that
is projected against the nucleus of the galaxy. \cite{Combes1992} have
detected CO $1-0$ emission toward two of the HI maxima
\citep{vanDriel1995} identified with the polar ring component in
NGC\,660 at $v_{opt} = 723 - 726$ and $980 - 945$\,\kms, velocities
inconsistent with our G3 component at $828 - 839$\,\kms.  A more
likely association of the G3 NH$_3$ velocity component and the narrow
line K-doublet H$_2$CO feature at 840\,\kms\ \citep{Mangum2013} would
place these spectral components in the nuclear region, near the
nuclear starburst ring noted by \cite{Carral1990}.

The HI absorption measurements point to a velocity width of $\sim
60$\,\kms\ in the polar ring, much smaller than the galactic disk
component, with emission line width of $\sim 350$\,\kms. Although HI
is not necessarily associated with the dense gas traced by NH$_3$ and
H$_2$CO, the order of magnitude difference in line width between the
narrow absorption features and broad galactic emission support the
interpretation of the dense NH$_3$ components as inhabitants of the
nuclear region who are superimposed directly in front of a bright,
compact continuum source.

\subsection{Kinetic Temperature Within the Spatial/Spectral
  Components of NGC\,660}
\label{NGC660Tk}

In Table~\ref{tab:lvg} we summarize the results from our LVG model fits
to the NH$_3$ measurements of the four velocity components which
comprise NGC\,660.  As summarized in \S\ref{NGC660NH3}, components G1
and G4 are associated with the disk component in NGC\,660, G3 is
associated with the starburst ring, while the location of component G2 is
indeterminant.  The introduction of a background continuum source to
our LVG analysis of the NH$_3$ absorption toward this galaxy, coupled
with the detection of five non-metastable (J$\neq$K) transitions, has
resulted, in many cases, to a more complex model fit.

\subsubsection{Spectral Component G1}
\label{NGC660G1}

The spectral component with the lowest velocity (v$_{hel} =
767\pm2$\,km\,s$^{-1}$), the G1 component is only detected in the
low-lying NH$_3$ (1,1), (2,2), and (4,4) transitions.  The lower-limit
kinetic temperature of T$_K \gtrsim 80$\,K is applicable over all
spatial and column densities investigated for T$_c \lesssim 300$\,K.
Over more restricted ranges in spatial density and NH$_3$ column
density of $\gtrsim 10^{6.5}$\,cm$^{-3}$ and $\lesssim
10^{16.5}$\,cm$^{-2}$/km\,s$^{-1}$, though, the kinetic temperature is
better constrained: T$_K = 149\pm72$\,K
(Figure~\ref{fig:NGC660G1G4TkFit}).  As the detection of this 
component in only low-lying NH$_3$ transitions suggests a relatively
low kinetic temperature, we believe that the kinetic temperature
derived from this more restricted range in (n,N) better represents the
true kinetic temperature of this spectral component.

\subsubsection{Spectral Component G2}
\label{NGC660G2}

Detected in a total of nine NH$_3$ transitions (6 para, 3 ortho, 4
non-metastable), best fit cumulative and individual ratio solutions
are found for T$_c \simeq 300$\,K.  Apparent contributions due to
multiple temperature components makes the cumulative fits using all
para-NH$_3$ (both metastable and non-metastable) transitions poor.
Individual ratio fits to our (2,2)/(4,4) and (4,4)/(5,5) para-NH$_3$
measurements yields T$_K \gtrsim 40$\,K and $\gtrsim 220$\,K,
respectively. 

As the non-metastable transitions can be affected by infrared
excitation, we have performed cumulative LVG model fits constrained by
our para-NH$_3$ metastable (J=K) transitions.  These fits produce a
lower-limit T$_K \gtrsim 260$\,K over a limited range in (n,N) of
$\lesssim 10^{5.0}$\,cm$^{-3}$ and $\gtrsim
10^{16.5}$\,cm$^{-2}$/km\,s$^{-1}$, respectively.  Finally, LVG model
fits to our metastable ortho-NH$_3$ (3,3)/(6,6) measurements yield a
rather poorly-constrained T$_K = 173\pm106$\,K for a limited range in
(n,N) of $\gtrsim 10^{6.0}$\,cm$^{-3}$ and $\gtrsim
10^{16.5}$\,cm$^{-2}$/km\,s$^{-1}$, respectively.  These LVG model
fits, taken with the detection of the high-excitation (5,5) and (6,6)
transitions, suggest that a good lower-limit to the kinetic
temperature of the G2 component is 150\,K.

\subsubsection{Spectral Component G3}
\label{NGC660G3}

As with G1 and G2, cumulative best fit solutions assuming T$_c \simeq
300$\,K, are all rather poor.  As was done with the G2 spectral
component, using only metastable para-NH$_3$ transitions in our LVG
model fit resulted in only slight improvement, suggesting that T$_K
\gtrsim 100$\,K.  Looking at fits to individual ratios:
\begin{itemize}
\item (1,1)/(2,2): T$_K = 154\pm96$\,K
\item (2,2)/(4,4): T$_K \gtrsim 68$\,K
\item (3,3)/(6,6): T$_K = 179\pm121$\,K
\item (4,4)/(5,5): T$_K \gtrsim 200$\,K
\item (5,5)/(7,7): T$_K \gtrsim 284$\,K for N $\lesssim
  10^{14.5}$\,cm$^{-2}$/km\,s$^{-1}$ and T$_K \gtrsim 88$\,K for (n,N) 
  $\gtrsim$ ($10^{6.5}$\,cm$^{-3}$, $10^{14.5}$\,cm$^{-2}$/km\,s$^{-1}$).
\end{itemize}

As noted with the analysis of our ortho-NH$_3$ measurements of
NGC\,3079, one should be cautious with fits to the (3,3) and (6,6)
transitions as slight differences in the spatial and column densities
and kinetic temperatures at which the (3,3) and (6,6) transitions
produce emission and absorption corrupt the uniqueness of their
kinetic temperature sensitivity.  This results in a  ``double-value''
to the best-fit kinetic temperature for n(H$_2$) $>
10^{4.5}$\,cm$^{-3}$.  The kinetic temperature range indicated includes
the limits of this double-value fit.  The kinetic temperature for the
G3 spectral component appears to be $\gtrsim 100$\,K.

\subsubsection{Spectral Component G4}
\label{NGC660G4}

\begin{figure}
\centering
\includegraphics[scale=0.45]{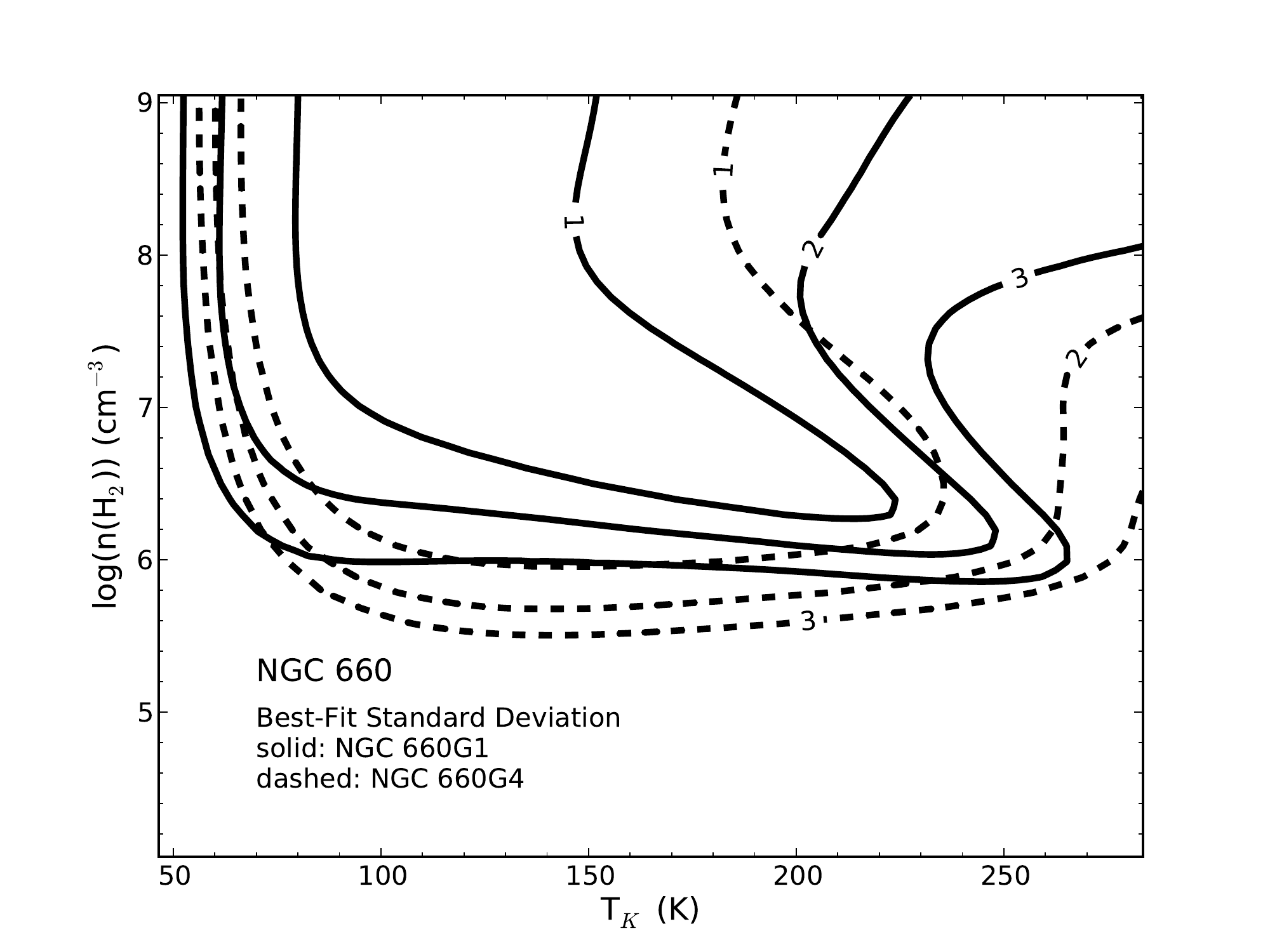}
\caption{NH$_3$ LVG model best-fit kinetic temperature toward the G1
  and G4 spectral components of NGC\,660 at the representative
  N(para-NH$_3$)/$\Delta$v = $10^{16.5}$\,cm$^{-2}$/\kms.  Shown are
  the metastable para-NH$_3$ best-fit 1, 2, and $3\sigma$ standard
  deviation to NGC\,660G1 (solid) and G4 (dashed) assuming a
  background continuum temperature of 300\,K.}
\label{fig:NGC660G1G4TkFit}
\end{figure}

Best-fit solutions constrained by all para-NH$_3$ transitions are
surprisingly good, yielding T$_K = 174\pm82$\,K.  Constraining only
the metastable para-NH$_3$ transitions results in a fit to the kinetic
temperature of T$_K = 149\pm84$\,K for n(H$_2$) $\gtrsim
10^6$\,cm$^{-3}$ and N(para-NH$_3$)/$\Delta$v $<
10^{16.75}$\,cm$^{-2}$/km\,s$^{-1}$
(Figure~\ref{fig:NGC660G1G4TkFit}).  For the LVG model fit using only
the metastable ortho-NH$_3$ (3,3)/(6,6) ratio we derive T$_K =
160\pm90$\,K.  Taking all three kinetic temperature
measurements together we estimate T$_K = 163\pm93$\,K for the G4
spectral component.

\subsection{NGC\,660 Spectral Component Summary}
\label{NGC660SpecSummary}

Table~\ref{tab:ngc660comps} summarizes the NH$_3$ spectral components,
their association with spatially-identified HI and/or OH spectral
components, and their derived kinetic temperatures.  Our NH$_3$
measurements suggest that both the disk and starburst ring components
harbor dense, high kinetic temperature gas.

%
\begin{deluxetable*}{lccclc} 
\tablewidth{0pt}
\tablecolumns{6}
\tablecaption{NGC\,660 Spectral/Spatial Component Correspondence\label{tab:ngc660comps}}
\tablehead{
\colhead{Component} & \colhead{$v_{hel}$(NH$_3$)} &
\colhead{$v_{hel}$(OH)} & \colhead{$v_{hel}$(HI)} & \colhead{Spatial
  Assoc} & \colhead{T$_K$} \\
& \colhead{(\kms)} & \colhead{(\kms)} & \colhead{(\kms)} &&
\colhead{(K)} 
}
\startdata
G1 & 765--769 & 765 & 761 & disk & $149\pm72$ \\
G2 & 785--800 & \ldots & \ldots & indeterminate & $\gtrsim 150$ \\
G3 & 828--839 & 861 & 851,862,870 & starburst ring & $\gtrsim 100$ \\
G4 & 970--989 & 992,1008 & 965,974,1002 & disk & $163\pm93$ \\
\enddata
\end{deluxetable*}


\section{Spatial Extent of the NH$_3$ Emitting Gas}
\label{NH3Spatial}

An estimate to the size of the NH$_3$ emitting region within each of
the galaxies in our sample can be derived from the measured NH$_3$
transition intensities ($T_{mb} = \frac{T^*_A}{\eta_{mb}}$, where $T_{mb}$
is the main beam brightness temperature of the measured transition), an
estimate of the optical depth ($\tau$), our derived kinetic
temperatures ($T_{rot} = T_K$), and the radiative transfer equation in
a uniform medium:
\begin{eqnarray}
\label{eq:radtrans}
T_{mb} &=& f T_{rot} \left(1 - \exp(-\tau)\right) \nonumber \\
\frac{T^*_A}{\eta_{mb}} &\simeq& f T_K \tau \nonumber \\
f &\simeq& \frac{T^*_A}{\eta_{mb} \tau T_K}
\end{eqnarray}
where $f$ is the areal filling factor of the NH$_3$
emission in our primary beam ($\theta_b \simeq 30^{\prime\prime}$).
In Equation~\ref{eq:radtrans} we have assumed that all measured
transitions are optically thin ($\tau \ll 1$) and that the NH$_3$
rotation temperature (T$_{rot}$) is equal to the kinetic temperature
in the gas (T$_K$).  Measured T$^*_A$ values are listed in
Table~\ref{tab:nh3results}, derived T$_K$ values are given in
Table~\ref{tab:lvg}, and $\eta_{mb} = 0.87$ (\S\ref{Observations}).
NH$_3$ transition optical depths can be derived from our LVG models,
but since the optical depth for a given transition is dependent upon
the intensity of that transition, which is, as
Equation~\ref{eq:radtrans} shows, dependent upon the areal filling
factor $f$, we only have an upper limit to $\tau$ at our disposal.
For non-masing (para-NH$_3$) transitions $\tau \lesssim 0.1$ with only
a minor dependence on spatial density.  Note too that for galaxies
with lower limits to their dense gas kinetic temperatures, coupled to
the upper limit to $\tau$, that $f$ is indeterminant.  With these
limitations in mind, and using the measured T$^*_A$, $\tau$, and T$_K$
(excluding lower limits) values for our NH$_3$ measurements, we
derive the lower-limits to the areal filling factor $f$ listed in
Table~\ref{tab:fareal}.  As the areal filling factor $f$ is related to
the source emission extent ($\theta_s$ or d$_s$) and the measurement
beam size ($\theta_b$ or d$_b$):
\begin{eqnarray}
\label{eq:f}
f &=& \frac{\theta^2_s}{\theta^2_s + \theta^2_b} \nonumber \\
&=& \frac{d^2_s}{d^2_s + d^2_b} \nonumber \\
&\simeq& \left(\frac{d_s}{d_b}\right)^2~\mathrm{for}~d_s \ll d_b
\end{eqnarray}
and our GBT primary beam diameter $\theta_b =
30^{\prime\prime}$ has a physical size ($d_b$) of 145\,pc at a distance of
1\,Mpc, we can derive lower-limits to the physical size of the NH$_3$
emitting regions ($d_s$; Table~\ref{tab:fareal}).  The lower limits to
$d_s$ derived correspond to the analogs of molecular and giant
molecular clouds in our own Galaxy.

\begin{deluxetable}{llll}
\tablewidth{0pt}
\tablecolumns{4}
\tablecaption{Galaxy Ammonia Areal Filling Factor Limits\label{tab:fareal}}
\tablehead{
\colhead{Galaxy} & 
\colhead{d$_b$\tablenotemark{a}} &
\colhead{f} &
\colhead{d$_s$} \\
& \colhead{(pc)} && \colhead{(pc)} 
}
\startdata
NGC\,253 & 499 & $\gtrsim 0.01$ & $\gtrsim 50$ \\
NGC\,660 & 1769 & $\gtrsim 0.01$ & $\gtrsim 177$ \\
NGC\,891\tablenotemark{b} & 1367 & $\gtrsim 0.0002$ & $\gtrsim 19$ \\
Maffei\,2 & 451 & $\gtrsim 0.001$ & $\gtrsim 15$ \\
NGC\,1365 & 3118 & $\gtrsim 0.001$ & $\gtrsim 98$ \\
IC\,342 & 554 & $\gtrsim 0.01$ & $\gtrsim 55$ \\
M\,82SW & 858 & $\gtrsim 0.001$ & $\gtrsim 27$ \\
NGC\,3079\tablenotemark{c} & 3002 & \nodata & \nodata \\
M\,83 & 584 & $\gtrsim 0.001$ & $\gtrsim 18$ \\
IC\,860 & 7801 & $\gtrsim 0.001$ & $\gtrsim 247$ \\
IR\,15107+0724 & 8967 & $\gtrsim 0.001$ & $\gtrsim 284$ \\
Arp\,220 & 12021 & $\gtrsim 0.001$ & $\gtrsim 380$ \\
NGC\,6946 & 796 & $\gtrsim 0.001$ & $\gtrsim 25$ \\
\enddata
\tablenotetext{a}{Using the angular size distance D$_A =
  \frac{D_L}{\left(1+z\right)^2}$ with D$_L$ from Table~\ref{tab:galaxies}.}
\tablenotetext{b}{The extremely low lower-limit for $f$ and d$_s$ for
  NGC\,891 is due to the very low upper-limit to T$_K$ of $< 30$\,K
  for this galaxy.}
\tablenotetext{c}{As all T$_K$ derived for NGC\,3079 are lower-limits,
  $f$ and d$_s$ are indeterminant.}
\end{deluxetable}

\section{How Can Galaxies Support Such High Kinetic Temperatures?}
\label{HighTK}

With few exceptions the kinetic temperatures we derive range from 50
to $>250$\,K.  The physical size scales over which our $\theta_b
\simeq 30^{\prime\prime}$ NH$_3$ measurements are sensitive range
from $\sim 10$\,pc to $\sim 30$\,kpc (\S\ref{NH3Spatial}).
As these physical size scales are potentially quite large, one has to
wonder how such high kinetic temperatures can be maintained in these
regions.

The dense gas environments and high star formation rates in
star forming galaxies have been modelled by several groups
using adaptations of 
chemical Photon Dominated Region (PDR) models which incorporate cosmic
ray \citep[CR;][]{Bayet2011}, CR plus UV, X-ray, and mechanical
\citep{Loenen2008}, and CR plus mechanical \citep[][ Meijerink, R.,
  private communication]{Meijerink2011} heating.  These models predict 
T$_K = 30 - 500$\,K, with higher kinetic temperatures corresponding
to higher CR and/or mechanical heating rates.  For reference, star
formation rates (SFR in $M_\odot$/yr), supernova rates (SN per year),
and mechanical heating rates ($\Gamma_{mech}$ in
erg\,cm$^{-2}$\,s$^{-1}$) correspond to (1, 0.01, $2\times10^{-20}$)
for disks of quiescent galaxies (\ie\ the Milky Way), (50, 0.3,
$1\times10^{-18}$) for star formation regions, and (1000, 6.4,
$2\times10^{-18}$) for extreme starbursts \citep{Kazandjian2012}.
This suggests that an active star formation process in galaxies can
generate the necessary mechanical heating input that is required to
in-turn generate the high kinetic temperatures measured in galaxies
with high star formation rates.

In the CR
plus mechanical heating models of \cite{Meijerink2011}, T$_K \simeq
100-200$\,K for a mechanical heating rate of $3\times
10^{-18}$\,erg\,cm$^{-2}$\,s$^{-1}$ in a high density (n(H$_2$) =
$10^{5.5}$\,cm$^{-3}$) high column density (N(H$_2$)
$>10^{22}$\,cm$^{-2}$) environment with CR rates ranging from $5\times 10^{-17} -
5\times 10^{-14}$\,s$^{-1}$.  In the
\cite{Bayet2011} models which include only CR heating the NH$_3$
abundance declines rapidly with increasing CR rate.  In contrast, the
modified PDR models of \cite{Meijerink2011} predict that the addition
of mechanical heating with energy density $\Gamma =
2\times10^{-17}$\,erg cm$^{-3}$ s$^{-1}$ will result in an
order-of-magnitude increase in the NH$_3$ abundance at the highest
column densities (N(H$_2$) $\simeq 3\times10^{22}$\,cm$^{-2}$)
and a CR rate of $5\times10^{-14}$\,s$^{-1}$.  We
should also point out that a similar heating scenario, involving cosmic
ray and/or turbulent energy dissipation (mechanical) heating, is
suggested based on measurements of the dense molecular clouds in the
central molecular zone (CMZ) of our Galaxy \citep{Ao2013}.  It appears
then that the high kinetic temperatures that we derive for the dense
star formation regions in our galaxy sample, and in the CMZ of our own
Galaxy, can be produced by both cosmic ray and mechanical heating
processes.

To investigate how the high kinetic temperatures we measure
might manifest themselves as traditional star formation indicators,
we have investigated the correlation between our measured gas
kinetic temperatures and star formation rate, as indicated by
infrared luminosity (L$_{IR}$) or radio flux at 5\,GHz.  Kinetic
temperature is uncorrelated to either of these quantities.  We have
also looked for correlation between T$_K$ and the relative
proportion of dust emission at 12\,$\mu$m, 25\,$\mu$m, and
60\,$\mu$m, again finding no correlation.  This lack of correlation
with global star formation indicators suggests that the high kinetic
temperatures we measure are associated with more localized
processes.  Unfortunately, physical processes that are more
localized in nature and that are also associated with star formation,
including shocks, outflows, or merger state, are not well
characterized in our galaxy sample.

A quantification of the spatial distribution of the dense hot
gas we measure within the galaxies in our sample would certainly
shed some light on the exact physical process that is producing the
high kinetic temperatures we measure.  NH$_3$ enhancement has
  been associated with shock excitation in outflows (\ie\ L1157:
  \cite{Umemoto1992, Tafalla1995, Viti2011}) within our Galaxy,
  allowing for the possibility that the actual distribution of dense
hot gas we measure is actually tracing the
interfaces between shocks and dense molecular clouds.  As noted
above, the mechanical heating rates implied by star formation and
starbursts are consistent with dense gas at T$_K \simeq 100 -
200$\,K.  A plausible model of the spatial distribution of this
dense hot gas would suggest that it emanates from numerous active
star formation regions that are concentrated within the dense gas
components, primarily nuclei, of the galaxies in our sample.  One
could envision this as a ``plum pudding'' model whereby the hot gas
we measure traces compact knots of active star formation within a
more extended, yet still dense, star-forming environment.

\section{Why Are Dust Temperatures a Poor Proxy for Gas Temperatures?}
\label{DustandGas}

\begin{figure}
\centering
\includegraphics[scale=0.35]{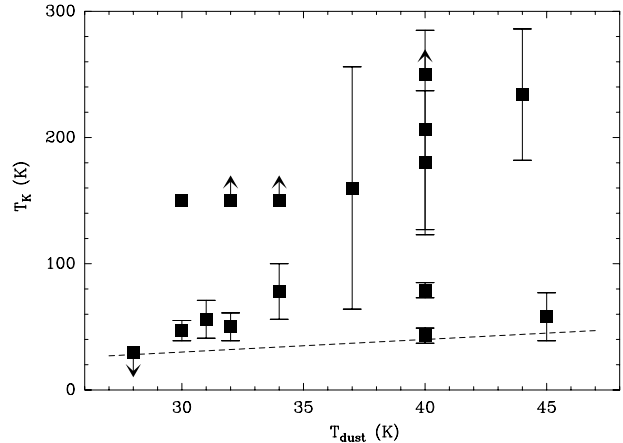}
\caption{Galaxy sample dust temperature (T$_{dust}$) versus kinetic
  temperature (T$_K$).  For the kinetic temperature values we have used
  the ``whole galaxy'' values listed in Table~\ref{tab:lvg}.  The
  dashed line represents T$_{dust}$ = T$_K$.}
\label{fig:TdTk}
\end{figure}

Related to the issue of the high kinetic temperatures in our
star forming
galaxy sample is the use of dust temperatures as a proxy for gas
kinetic temperatures in these galaxies.  As noted in
\cite{Mangum2013}, the common practice of using the dust temperature
derived from IRAS 60 and 100\,$\mu$m fluxes as a proxy for gas kinetic
temperature was applied to the star forming galaxy sample presented in
\cite{Mangum2008}.  Table~\ref{tab:lvg} lists the dust and gas kinetic
temperatures derived for all of the galaxies in our sample, while
Figure~\ref{fig:TdTk} displays these temperatures.  While
dust temperatures range from 28 to 45\,K, NH$_3$-based gas kinetic
temperatures range from $\sim 25$ to $> 250$\,K, with no
apparent correlation between T$_{dust}$ and T$_K$
(Figure~\ref{fig:TdTk}).  Clearly the assumption that $T_{dust} =
T_K$ is not correct.

Issues with the veracity of the application of dust temperatures as a proxy
for kinetic temperatures have been noted before.  \cite{Helou1986}
noted that dust emission at 12 and 25\,$\mu$m traces primarily
active star formation while 60 and 100\,$\mu$m emission traces cooler
``cirrus'' emission from ambient dust heated by existing stellar
populations.  A similar analysis of \textit{Spitzer} Space Telescope
and \textit{Herschel} Space Observatory measurements of the dust
emission from M\,81, M\,83, and 
NGC\,2403 \citep{Bendo2012} concluded that dust emission at
wavelengths less than 160\,$\mu$m generally originates from active star
formation regions, while dust emission at wavelengths greater than
250\,$\mu$m originates primarily from colder dust components which are
relatively unassociated with star formation activity.  A similar
\textit{Herschel} Space Observatory study of the dust continuum
emission from M\,31 \citep{Groves2012} found that the heating source
for the submillimeter dust emission from this galaxy originates in the
older ($\gtrsim 10^9$ years) stellar populations that inhabit the
bulge of M\,31.  It appears then that longer-wavelength infrared
measurements do not trace active star formation in galaxies,
and that long-wavelength infrared measurements should not be used as a
proxy for dust and gas kinetic temperatures in such regions.

Nowhere is this discrepancy between dust and gas kinetic temperatures
more apparent than in our own Galactic 
center.  \citet{Ao2013} used para-H$_2$CO $3_{03}-2_{02}$, $3_{22}-2_{21}$,
and $3_{21}-2_{20}$ measurements of the Galactic plane and the
Galactic center to measure the kinetic temperature of the dense
molecular clouds in these regions over spatial scales of 1.2\,pc.
\citet{Ao2013} found that T$_K$ ranged from 50 to $>100$\,K in the
Galactic center, with the molecular clouds possessing the highest
kinetic temperatures located closest to the Galactic center.  These
H$_2$CO-derived gas kinetic temperatures are much higher than the
fairly uniform dust temperatures of $\sim 14 - 20$\,K measured toward
the dense molecular clouds of the Galactic center
\citep{PiercePrice2000, GarciaMarin2011, Molinari2011}.
\citet{Ao2013} argue that the high gas kinetic temperatures measured
may be caused by turbulent and/or cosmic-ray heating, but cannot be
caused by photon (irradiative) processes.

\section{Conclusions}
\label{Conclusions}

We have used measurements of 15 NH$_3$ transitions to derive the
kinetic temperature in a sample of 23 star forming low-redshift
galaxies and one galaxy offset position.  A summary of our conclusions
are:
\begin{itemize}
\item Of the 23 galaxies and one galaxy offset position measured, we
  detect at least one NH$_3$ transition toward 13 of these
  galaxies.
\item Using our NH$_3$ measurements to derive the kinetic temperature,
  we find that 9 of the 13 galaxies possess multiple kinetic
  temperature and/or velocity components.
\item The kinetic temperatures derived toward our galaxy sample are in
  many cases at least a factor of two larger than kinetic temperatures
  derived previously from mainly lower-excitation NH$_3$ measurements.
\item  The derived kinetic temperatures in our galaxy sample, which
  are in many cases at least a factor of two larger than derived dust
  temperatures, point to a problem with the common assumption that
  dust and gas kinetic temperatures are equivalent.  As previously
  suggested, the use of dust emission at wavelengths greater than
  160\,$\mu$m to derive dust temperatures, or dust heating from older
  stellar populations, may be skewing derived dust temperatures in
  these galaxies to lower values.
\item The non-metastable NH$_3$ (2,1) transition has been detected
  toward 4 galaxies: two in emission and two in absorption.
  Non-metastable NH$_3$ (3,1), (3,2), (4,3), (5,4) (toward NGC\,660)
  and (10,9) (toward Arp\,220) have also been detected, all in
  absorption.
\item The polar ring galaxy NGC\,660 possesses the most remarkably
  complex NH$_3$ absorption profile in our sample.
  \begin{itemize}
  \item The NH$_3$ and H$_2$CO spectral components from this galaxy
    overlap in velocity but differ by almost an order-of-magnitude in
    velocity width (FWZI(NH$_3$) $\simeq 150$ to 325\,\kms;
    FWZI(H$_2$CO)) $\simeq 1000$\,\kms).  This suggests that NH$_3$
    traces a more quiescent environment in the star formation regions
    of NGC\,660 than that traced by H$_2$CO.  NGC\,660 is the only
    galaxy which shows this disparity between NH$_3$ and H$_2$CO
    spectral component widths.
  \item The four NH$_3$ velocity components which comprise NGC\,660
    have been associated with previously-measured disk, and possibly
    ring, components.
  \item Derived kinetic temperatures within the four spectral
    components of NGC\,660 range from $>100$ to $163\pm93$\,K.
  \end{itemize}
\item As originally shown by \cite{Mangum1994} for the
  ortho-NH$_3$ (3,3) transition, the ortho-NH$_3$ (3,3) and (6,6)
  transitions mase over a wide range of physical conditions.  As our
  LVG models include a proper treatment of the collisional excitation
  process which leads to maser emission in these transitions, the
  NH$_3$ $\frac{(3,3)}{(6,6)}$ ratio is a viable probe of the kinetic
  temperature when properly modelled.
\item Unlike all other galaxies in our NH$_3$ sample which exhibit
  high-excitation NH$_3$ (6,6) (and in the case of NGC\,253, NH$_3$
  (9,9)) emission, we do not detect the NH$_3$ (3,3) transition toward
  NGC\,3079.  As the lower-excitation ortho-NH$_3$ transitions can be
  excited into maser emission states under a variety of conditions, a
  two-spatial component model which includes overlapping NH$_3$ (3,3)
  emission and absorption components can explain this lack of NH$_3$
  (3,3) emission or absorption.  This physical scenario is supported
  by the complex emission and absorption observed in H$_2$CO
  $1_{10}-1_{11}$ \citep{Mangum2008,Mangum2013}.
\item The merger system Arp\,220, in addition to possessing high
  kinetic temperatures within its two velocity component structure,
  exhibits unusually broad NH$_3$ (2,2) and comparatively narrow and
  weak NH$_3$ (1,1) absorption.  This unusual NH$_3$ (1,1) and (2,2)
  spectral structure has been previously explained \citep{Ott2011} as
  due to an overlapping cold (T$_K \lesssim 20$\,K) component.  An
  alternative explanation which appears to be consistent with our
  NH$_3$ measurements is that there is a broad velocity component
  shifted by approximately $\sim 200$\,\kms\ relative to the nominal
  velocity components in Arp\,220 which modifies the relative widths
  and intensities of the NH$_3$ (1,1) and (2,2) transitions.  This
  scenario is supported by the measurement of a similar broad velocity 
  component in the \textit{Herschel}/PACS spectra of OH, [OI], and
  H$_2$O \citep{GonzalezAlfonso2012}.
\item We confirm the tentative detection of OH $^2\Pi_{3/2}$\,J=9/2
  absorption toward Arp\,220 by \citet{Ott2011}.  This
  confirmation is consistent with far infrared high-excitation OH
  $^2\Pi_{1/2}$ and $^2\Pi_{3/2}$ absorption measurements toward
  Arp\,220 \citep{GonzalezAlfonso2012}.  A significant contributor to this
  high-excitation OH absorption appears to originate from a dense
  outflow component in Arp\,220.
\end{itemize}

\acknowledgments

The GBT staff, especially Frank Ghigo, were characteristically helpful
and contributed significantly to the success of our observing
program.  Our anonymous referee provided very useful suggestions which
improved the quality of this work.  JGM thanks Rowin Meijerink for
providing calculations of NH$_3$ and H$_2$CO abundance from his
chemical models.  JGM also thanks Brian Kent, who was instrumental in
providing the python script that was used to extrapolate the NH$_3$
collisional excitation rates.  Support for this work was provided by
NASA through Hubble Fellowship grant \#HST-HF-01183.01-A awarded by
the Space Telescope Science Institute, which is operated by the
Association of Universities for Research in Astronomy, Incorporated,
under NASA contract NAS5-26555.  This research has made use of the
NASA/IPAC Extragalactic Database (NED) which is operated by the Jet
Propulsion Laboratory, California Institute of Technology, under
contract with the National Aeronautics and Space Administration.

\textit{Facilities:} \facility{GBT}

\bibliographystyle{apj}
\bibliography{Mangum}

\end{document}